\documentclass[12pt]{article}
\usepackage{jheppub}
\pdfoutput=1

\usepackage{amsmath,bbm,array,amsfonts,graphicx,wrapfig,lscape,float,mathtools,multirow,longtable}
\usepackage[dvipsnames]{xcolor}

\newcommand{\be}{\begin{equation}}
\newcommand{\ee}{\end{equation}}
\newcommand{\beq}{\begin{equation}}
\newcommand{\beql}[1]{\begin{equation}\label{#1}}
\newcommand{\eeq}{\end{equation}}
\newcommand{\ba}{\begin{array}}
\newcommand{\ea}{\end{array}}
\newcommand{\bea}{\begin{eqnarray}}
\newcommand{\beal}[1]{\begin{eqnarray}\label{#1}}
\newcommand{\eea}{\end{eqnarray}}
\newcommand{\ben}{\begin{enumerate}}
\newcommand{\een}{\end{enumerate}}
\newcommand{\bean}{\begin{eqnarray*}}
\newcommand{\eean}{\end{eqnarray*}}
\newcommand{\eref}[1]{(\ref{#1})}
\newcommand{\sref}[1]{\S\ref{#1}}
\newcommand{\tref}[1]{Table~\ref{#1}}

\newcommand{\fref}[1]{Figure \ref{#1}}
\newcommand{\btab}[1]{\begin{tabular}{#1}}
\newcommand{\etab}{\end{tabular}}

\def \Black {\pdfsetcolor{0 0 0 1}}

\newcommand{\comment}[1]{}

\newcommand{\qed}{\nobreak \ifvmode \relax \else
      \ifdim\lastskip<1.5em \hskip-\lastskip
      \hskip1.5em plus0em minus0.5em \fi \nobreak
      \vrule height0.75em width0.5em depth0.25em\fi}

\definecolor{darkspringgreen}{rgb}{0.09, 0.45, 0.27}
\definecolor{forestgreen}{rgb}{0.13, 0.55, 0.13}

\title{Brane Brick Models in the Mirror}

\author[a,b]{Sebasti\'an Franco,} 
\author[c,d,e]{Sangmin Lee,}
\author[f]{Rak-Kyeong Seong,}
\author[g]{Cumrun Vafa}

\affiliation[a]{
Physics Department, The City College of the CUNY \\
160 Convent Avenue, New York, NY 10031, USA}

\affiliation[b]{The Graduate School and University Center, The City University of New York  \\
365 Fifth Avenue, New York NY 10016, USA }

\affiliation[c]{
Center for Theoretical Physics, Seoul National University, Seoul 08826, Korea
}

\affiliation[d]{
Department of Physics and Astronomy, Seoul National University, Seoul 08826, Korea
}

\affiliation[e]{
College of Liberal Studies, Seoul National University, Seoul 08826, Korea
}

\affiliation[f]{
School of Physics, Korea Institute for Advanced Study, Seoul 02455, Korea
}

\affiliation[g]{Jefferson Physical Laboratory, Harvard University, Cambridge, MA 02138, USA}

\emailAdd{sfranco@ccny.cuny.edu}
\emailAdd{sangmin@snu.ac.kr}
\emailAdd{rakkyeongseong@gmail.com}
\emailAdd{vafa@physics.harvard.edu}

\preprint{
\begin{flushright}
CCNY-HEP-16-07 \\
SNUTP16-004 \\
KIAS-P16062
\end{flushright}
}

\abstract{Brane brick models are Type IIA brane configurations that encode the $2d$ $\mathcal{N}=(0,2)$ gauge theories on the worldvolume of D1-branes probing toric Calabi-Yau 4-folds. We use mirror symmetry to improve our understanding of this correspondence and to provide a systematic approach for constructing brane brick models starting from geometry. The mirror configuration consists of D5-branes wrapping 4-spheres and the gauge theory is determined by how they intersect. We also explain how $2d$ $(0,2)$ triality is realized in terms of geometric transitions in the mirror geometry. Mirror symmetry leads to a geometric unification of dualities in different dimensions, where the order of duality is $n-1$ for a Calabi-Yau $n$-fold. This makes us conjecture the existence of a quadrality symmetry in $0d$. Finally, we comment on how the M-theory lift of brane brick models connects to the classification of $2d$ $(0,2)$ theories in terms of 4-manifolds.
}

\begin{document}

\maketitle

\section{Introduction}

The interplay between Calabi-Yau (CY) geometry and branes probing it has played a key role in understanding duality symmetries of field theories that emerge in string theory.  More specifically D-branes probing CY singularities have given rise to an interesting class of SCFT's.
In particular, D3-branes probing CY 3-folds lead to $4d$ $\mathcal{N}=1$ theories (see e.g. \cite{Kachru:1998ys,Klebanov:1998hh, Morrison:1998cs, Beasley:1999uz, Feng:2000mi,Benvenuti:2004dy,Franco:2005sm} and references therein).  It was shown in \cite{Cachazo:2001sg} that one can use the mirror symmetry of CY 3-folds to not only understand what the corresponding quiver theory is, but also to understand Seiberg dualities between them as a continuous change of parameters in the mirror geometry.

More recently, D1-branes probing CY 4-fold singularities were considered in \cite{Franco:2015tna,Franco:2015tya,Franco:2016nwv}, where the corresponding gauge theories they give rise to were proposed. These theories lead to $2d$ $(0,2)$ SCFT's.  Moreover, in this context it was proposed in \cite{Gadde:2013lxa} that these theories enjoy triality symmetries, which is rather novel. The main goal of this paper is to extend the observation about applying mirror symmetry in the context of D3-branes probing CY 3-folds to demystify those theories: we use mirror symmetry to not only explain what the prescription of the resulting $2d$ $(0,2)$ quiver is, but also explain triality using the mirror geometry.  In a subsequent paper we show that this extends to the
case of D(-1) instantons probing CY 5-folds, but now mirror symmetry leads to a quadrality symmetry \cite{quadrality}.

Let us sketch the basic idea for a Calabi-Yau $n$-fold.  Consider a D$(9-2n)$-brane probing the CY singularity.
Let us first move the brane away from the singular locus.  Note that the position of the D-brane in the transverse space is a point in the CY.  Now we apply mirror symmetry, which converts the D-brane to wrap in addition an $n$-dimensional torus $T^n$, as in the SYZ picture of mirror symmetry \cite{Strominger:1996it}.  Now we move the position of the brane to the singular point.  At the singular point, the $T^n$ breaks up to subspaces.  From the geometry of the subspaces we can infer the resulting quiver gauge theory. 

The simplest example is when $n=2$, and we are probing an $A_{N-1}$ singularity with D5-branes. In this case, as the D5-brane approaches the singular locus, the mirror $T^2$ breaks up to a necklace of $N$ spheres touching one another at a point.  This gives rise the usual quiver description. Namely, for $k$ D5-branes probing it, we get each sphere being wrapped by $k$ D-branes giving the gauge group $\prod_{i=1}^N U_i(k)$ with bifundamental matter from their neighboring intersections,
leading to an affine quiver theory. We view this geometry as an $S^1$ fibration over a base $S^1$. The base will be depicted by a circle, broken by $N$ points which denotes the loci where the $S^1$ fiber shrinks.  So each of the intervals on $S^1$ corresponds to a sphere and the neighboring intervals will have a bifundamental matter field in common.

The same story repeats for any $n$-fold.  In the general case, the $T^n$ mirror fiber will be viewed as an $S^1$ fibration over a $T^{n-1}$ geometry.  Loci where $S^1$ shrinks break up the $T^{n-1}$ into regions, each of which will correspond to
a gauge factor.  The neighboring regions will lead to matter bifundamentals.  The codimension-2 interfaces where the codimension-1 faces meet lead to loci where interactions take place between matter multiplets.  From this structure we can read off the quiver theory, its matter content and its interactions.

To read off dualities, we use the complex mirror geometry and consider changing the complex structure.  The inequivalent geometries we obtain correspond to complex deformations and passing through vanishing cycles.  However the mirror geometry unifies the inequivalent geometries of the Calabi-Yau into a single Calabi-Yau manifold as is familiar from various examples of mirror symmetry.  In this way we can read off dualities by the uniqueness of the Calabi-Yau mirror.
We find that for the $n$-fold case we get generalized duality symmetries which return to the original theory after $n-1$ steps. 
So for 2-fold case, we get the usual Weyl reflection which is the self-duality of $4d$ $\mathcal{N}=2$ $SU(N)$ with $2N$ flavors \cite{Seiberg:1994rs}, for the 3-fold case we get Seiberg duality \cite{Seiberg:1994pq}, for the 4-fold case we get the triality of Gadde, Gukov and Putrov \cite{Gadde:2013lxa}, and for the 5-fold case we get a quadrality \cite{quadrality}.  The order of the symmetry is easiest to see in the context of the local Calabi-Yau given by $O(-n)$ bundle
over $\mathbb{CP}^{n-1}$.

The organization of this paper is as follows. Section \sref{section_2d_theories_and_BBMs} reviews $2d$ $(0,2)$ theories, D1-branes over toric CY 4-folds and brane brick models. Section \sref{section_mirror_general} presents a general discussion of the mirror of D$(9-2n)$-branes probing CY $n$-folds. Section \sref{section_mirror_CY3} specializes on D3-branes probing toric CY 3-folds. Section \sref{section_mirror_CY4} discusses the application of mirror symmetry to D1-branes probing toric CY 4-folds in detail and explains how to use it for constructing the corresponding brane brick models. Additional examples are presented in section \sref{section_examples}. Section \sref{section_triality} explains how triality arises from geometric transitions in the mirror. Section \sref{section_4-manifolds} connects to the classification of $2d$ $(0,2)$ theories in terms of 4-manifolds in the M-theory lift of brane brick models. We present our conclusions in section \sref{section_conclusions}. In two appendices we present additional examples and details about the open string spectrum at D-brane intersections.

\section{Brane Brick Models and $2d$ $(0,2)$ Theories}

\label{section_2d_theories_and_BBMs}

This paper is mainly devoted to the $2d$ $(0,2)$ gauge theories that arise on the worldvolume of D1-branes probing singular toric CY 4-folds. This section contains a brief review of general $2d$ $(0,2)$ theories, the special structure of the theories on D1-branes on toric CY 4-folds and brane brick models. We refer the reader to \cite{Franco:2015tna,Franco:2015tya,Franco:2016nwv}, where the ideas presented below were originally introduced.

\subsection{$2d$ $(0,2)$ Gauge Theories}

In order to set up the language, let us quickly review some basic aspects of $2d$ $(0,2)$ gauge theories. Thorough introductions to the subject can be found in \cite{Witten:1993yc,GarciaCompean:1998kh,Gadde:2013lxa,Kutasov:2013ffl}. These theories can be efficiently formulated in terms of $2d$ $(0,2)$ superspace $(x^\alpha,\theta^+,\bar{\theta}^+)$, $\alpha=0,1$. Their elementary building blocks are three types of superfields: 

\begin{itemize}

\item{\bf Vector:} it contains a gauge boson $v_\alpha$ $(\alpha =0,1)$, adjoint chiral fermions $\chi_-$, $\bar{\chi}_-$ and an auxiliary field $D$. Here and in what follows, $\pm$ subindices indicate the chirality of the corresponding fermions.

\item{\bf Chiral:} the component expansion and chirality condition of a chiral field take the form
\beal{chiralm}
\Phi = \phi + \theta^+ \psi_+ -i \theta^+ \bar{\theta}^+ D_+ \phi \,, 
\quad \overline{\mathcal{D}}_+ \Phi = 0 \,.
\eea
The on-shell degrees of freedom are a complex scalar $\phi$ and a chiral fermion $\psi_+$, and $\overline{\mathcal{D}}_+$ is a supercovariant derivative. 

\item{\bf Fermi:} the chirality condition of a Fermi field may be deformed by a holomorphic function of chiral fields $E(\Phi_i)$, which gives rise to interactions among matter fields. Fermi fields have the following component expansion
\beal{fermim}
\Lambda = \lambda_- - \theta^+ G -i \theta^+ \bar{\theta}^+ D_+ \lambda_- 
- \bar{\theta}^+ E \,, 
\quad \overline{\mathcal{D}}_+ \Lambda = E(\Phi_i)\,.
\eea
$G$ is an auxiliary field and the chiral fermion $\lambda_-$ is the only on-shell degree of freedom. 

\end{itemize}

Let us now discuss some important building blocks of the Lagrangian. The kinetic terms for the Fermi multiplets plus some interactions of matter fields arise from
\beal{LF}
L_F= \int d^2y \, d^2\theta \sum_a \left( \bar{\Lambda}_a \Lambda_a\right) \,,
\eea
where $a$ runs over the Fermi fields of the theory.

Interactions among matter fields can also be incorporated via the couplings
\beal{LJ}
L_J = -\int d^2y \, d\theta^+ \sum_a \left( \Lambda_a J_a(\Phi_i)|_{\bar{\theta}^+=0}\right)-h.c. \,,
\eea
where the $J_a(\Phi_i)$ are holomorphic functions of chiral fields. In summary, every Fermi field $\Lambda_a$ is associated to a pair of holomorphic functions of chiral fields $J_a$ and $E_a$. 
Consistency of the theory requires $J$- and $E$-terms satisfy the following constraint
\beq
\sum_a \mathrm{tr} \left[ E_a(\Phi_i) J_a(\Phi_i) \right]= 0 \,.
\eeq

Integrating out the auxiliary components $G_a$ in the Fermi fields, $L_F$ and $L_J$ give rise to the following contributions to the scalar potential\footnote{The scalar potential contains additional positive definite contributions from the D-terms in vector multiplets.}
\beq
V \supset \sum_a \left( \mathrm{tr}|E_a(\phi)|^2 +  \mathrm{tr}|J_a(\phi)|^2 \right)\,,
\label{V_JE}
\eeq
and to interactions between scalars and pairs of fermions
\beq
V_Y = - \sum_{a,i} \mathrm{tr} \left( \bar{\lambda}_{-a}\frac{\partial E_a}{\partial \phi_j}\psi_{+j} + \lambda_{-a}\frac{\partial J_a}{\partial \phi_j}\psi_{+j} +   \mathrm{h.c.} \right) \,.
\label{VY}
\eeq 

$2d$ $(0,2)$ theories are invariant under the swap $\Lambda_a \leftrightarrow \bar{\Lambda}_a$ for any  $\Lambda_a$, accompanied with the exchange $J_a \leftrightarrow E_a$.

\begin{figure}[htbp]
	\centering
	\includegraphics[width=12cm]{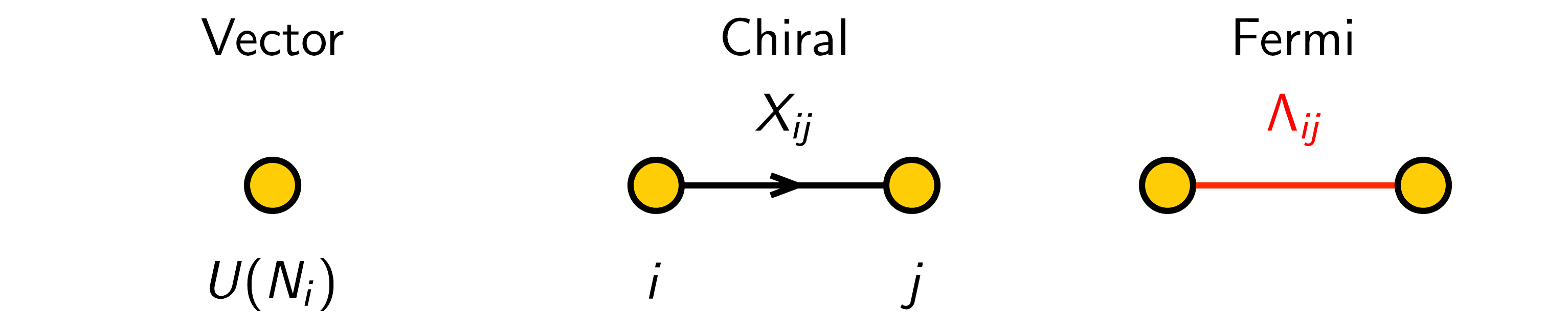}
\caption{Quiver representation of vector, chiral and Fermi superfields. Here we label fields with a pair of subindices indicating the gauge nodes under which they transform. Fermi fields are not assigned an orientation, in order to emphasize the $\Lambda_{ij}\leftrightarrow \bar{\Lambda}_{ji}$ symmetry.}
	\label{superfields_quiver}
\end{figure}

In this paper, we focus on theories in which all fields transform in either bifundamental or adjoint representations of a $\prod_i U(N_i)$ gauge group and can hence be represented by quiver diagrams as shown in \fref{superfields_quiver}.

\subsection{D1-Branes over Toric Calabi-Yau 4-Folds and Brane Brick Models}

\label{section_BBMs}

We will focus on the $2d$ $(0,2)$ theories that arise on the worldvolume of Type IIB D1-branes probing toric CY 4-folds. The probed CY$_4$ arises as the {\it classical mesonic moduli space} of the gauge theory. More precisely, for $N$ D1-branes, this moduli space is the algebraic variety
\beal{es0a56}
\mathcal{M}_N = 
\left(
\mathbb{C}[X_{\mu}] /
\langle
J_{a}= 0 ,
E_{a}= 0 
\rangle
\right)// U(N)^G
~,~
\eea
where $\mu$ and $a$ run over the chiral and Fermi fields, respectively, and $G$ is the number of $U(N)$ gauge groups. $G$ is equal to the volume of the toric diagram of the CY$_4$, normalized with respect to a minimal tetrahedron. Vanishing of the scalar potential requires the individual vanishing of $J$-, $E$- and $D$-terms. $\mathcal{M}_N$ is the $N^{th}$ symmetric product of the probed CY$_4$, $\mathcal{M}_N= \text{Sym}^{N}(\text{CY}_4)$. For $U(1)$ gauge groups, the variety in \eref{es0a56} becomes exactly the probed CY$_4$.

A new class of Type IIA brane configuration, denoted {\it brane brick models} was introduced in \cite{Franco:2015tya}. Brane brick models are related to D1-branes over toric CY$_4$ singularities by T-duality along three directions. Brane brick models substantially simplify and offer a new perspective on the connection between the CY$_4$ geometry and the corresponding $2d$ gauge theories. By doing so, they also provide a powerful tool for studying the dynamics of $2d$ $(0,2)$ theories.

A brane brick model consists of D4-branes suspended from an NS5-brane. The NS5-brane extends along the $(01)$ directions and wraps a holomorphic surface $\Sigma$ embedded into the $(234567)$ directions. The $(246)$ directions are periodically identified to form a 3-dimensional torus. The coordinates $(23)$, $(45)$ and $(67)$ are pairwise combined into three complex variables $x$, $y$ and $z$. The $T^3$ corresponds to the arguments of these complex variables, $(2,4,6)=(\arg(x),\arg(y),\arg(z))$. $\Sigma$ is defined as the zero locus of the Newton polynomial associated to the toric diagram of the CY$_4$, $P(x,y,z)=0$. Stacks of D4-branes extend along $(01)$ and are suspended inside the holes cut out by $\Sigma$ on the $(246)$ torus. The $U(1)$ R-symmetry of the gauge theories corresponds to rotations on the $(8,9)$ plane, on which all the branes sit at a point. \tref{tbconfig} summarizes the structure of a brane brick model.

\begin{table}[ht!!]
\centering
\begin{tabular}{c|cccccccccc}
\; & 0 & 1 & 2 & 3 & 4 & 5 & 6 & 7 & 8 & 9\\
\hline
\text{D4} & $\times$ & $\times$ & $\times$ & $\cdot$ & $\times$ & $\cdot$ & $\times$ & $\cdot$ & $\cdot$ & $\cdot$ \\
\text{NS5} & $\times$ & $\times$ & \multicolumn{6}{c}{----------- \ $\Sigma$ \ ------------} & $\cdot$ & $\cdot$\\
\end{tabular}
\caption{Brane brick models consist of D4-branes suspended from an NS5-brane wrapping a holomorphic surface $\Sigma$.}
\label{tbconfig}
\end{table}

\begin{figure}[h]
\begin{center}
\resizebox{0.96\hsize}{!}{
\includegraphics[width=8cm]{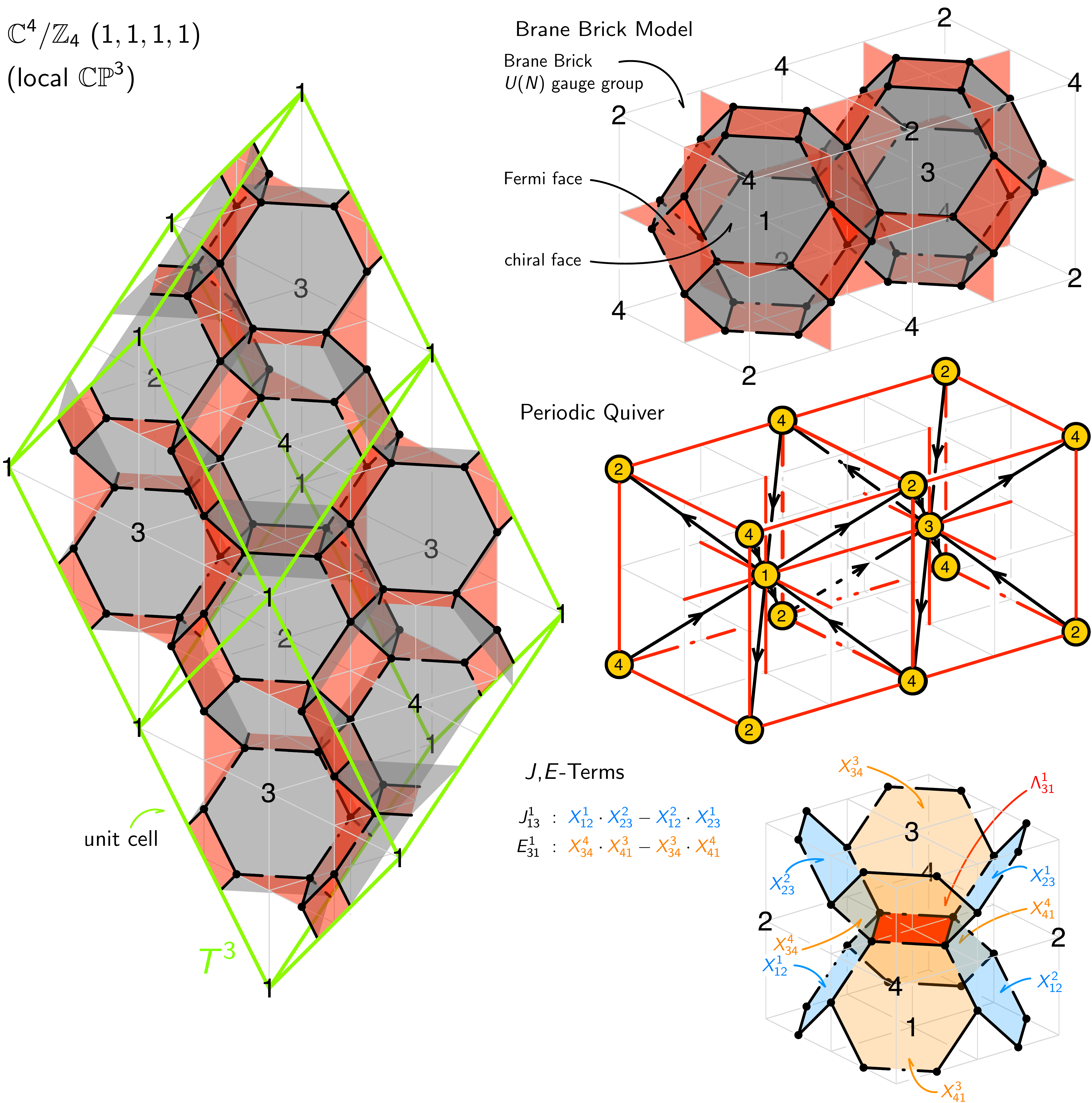}
}
\caption{Brane brick model for local $\mathbb{CP}^3$, i.e. the $\mathbb{C}^4/\mathbb{Z}_4$ orbifold with $(1,1,1,1)$ action. The figure shows two unit cells, which are indicated in green. The figures on the right focus on pieces of the unit cell and summarize the dictionary relating brane brick models to $2d$ gauge theories and the associated periodic quiver in $T^3$. The periodic quiver is related to the brane brick model by graph dualization.
\label{fbranebricksummary}}
 \end{center}
 \end{figure}

It is convenient to represent a brane brick model by its ``skeleton'' on $T^3$. We will often refer to this simplified object also as the brane brick model. Every brane brick model defines a $2d$ $(0,2)$ gauge theory according to the rules in \tref{tbrick}. Bricks correspond to $U(N_i)$ gauge groups.\footnote{Having different ranks is possible if fractional D1-branes are introduced in the T-dual configuration of branes at a CY$_4$ singularity.} There are two types of faces, representing to the two types of matter superfields present in $2d$ $(0,2)$ theories. Every oriented face corresponds to chiral field and every unoriented face represents a Fermi field $\Lambda$ and its conjugate $\bar{\Lambda}$. Throughout this paper, we will distinguish chiral and Fermi faces by coloring them grey and red, respectively. Fermi faces are 4-sided, this follows from the special structure of $J$- and E-terms in toric theories, as explained below. \fref{fbranebricksummary} illustrates the correspondence between brane brick models and gauge theories using the local $\mathbb{CP}^3$ example.

\begin{table}[h]
\centering
\resizebox{0.9\hsize}{!}{
\begin{tabular}{|l|l|}
\hline
\ \ \ \ \ \ {\bf Brane Brick Model} \ \ \ \ \ & \ \ \ \ \ \ \ \ \ \ \ \ \ \ \ \ \ \ \ \ {\bf $2d$ $(0,2)$ Theory} \ \ \ \ \ \ \ \ \ \ \ \ 
\\
\hline\hline
Brick  & Gauge group \\
\hline
Oriented face between bricks & Chiral field in the bifundamental representation \\
$i$ and $j$ & of nodes $i$ and $j$ (adjoint for $i=j$) \\
\hline
Unoriented square face between & Fermi field in the bifundamental representation \\
bricks $i$ and $j$ & of nodes $i$ and $j$ (adjoint for $i=j$) \\
\hline
Edge  & Plaquette encoding a monomial in a \\ 
& $J$- or $E$-term \\
\hline
\end{tabular}
}
\caption{Dictionary relating brane brick models to $2d$ $(0,2)$ gauge theories.
\label{tbrick}
}
\end{table}

In the $2d$ theories dual to toric CY$_4$'s, $J$- and $E$-terms have a special structure, which was dubbed the \textit{toric condition} in \cite{Franco:2015tna}. In these theories, the $J$- and $E$-terms take the form
\beal{esb1a1}
J_{ji} = J_{ji}^{+} - J_{ji}^{-} ~,~ E_{ij} = E_{ij}^{+} - E_{ij}^{-} ~,~
\eea
with $J_{ji}^{\pm}$ and $E_{ij}^{\pm}$ holomorphic monomials in chiral fields.\footnote{More precisely, the toric condition holds in those phases of the gauge theory that are described by brane brick models, which are often referred to as {\it toric phases}. Non-toric phases can be reached by general triality transformations, as explained in section \sref{section_triality}. In such phases, the ranks of all gauge nodes are no longer equal and the $J$- and $E$-terms do not necessarily obey the toric condition.} The origin of the toric condition can be understood geometrically. Toric CY$_4$'s are defined by monomial relations, and these are precisely the type of relations that arise from vanishing $J$- and $E$-terms when they are of the form \eref{esb1a1}. The toric condition has an alternative, more physical, derivation in terms of classical higgsing \cite{Franco:2015tna}. Higgsing corresponds to partial resolution of the probed geometries and can be systematically exploited for obtaining the gauge theory for an arbitrary toric CY$_4$'s starting from the one for an abelian orbifold of $\mathbb{C}^4$. The identification of edges in the brane brick model with monomials in $J$- and $E$-terms, together with the toric condition, imply that the faces associated to Fermi fields are 4-sided, as already anticipated.

\begin{figure}[h]
\begin{center}
\resizebox{0.75\hsize}{!}{
\includegraphics[width=8cm]{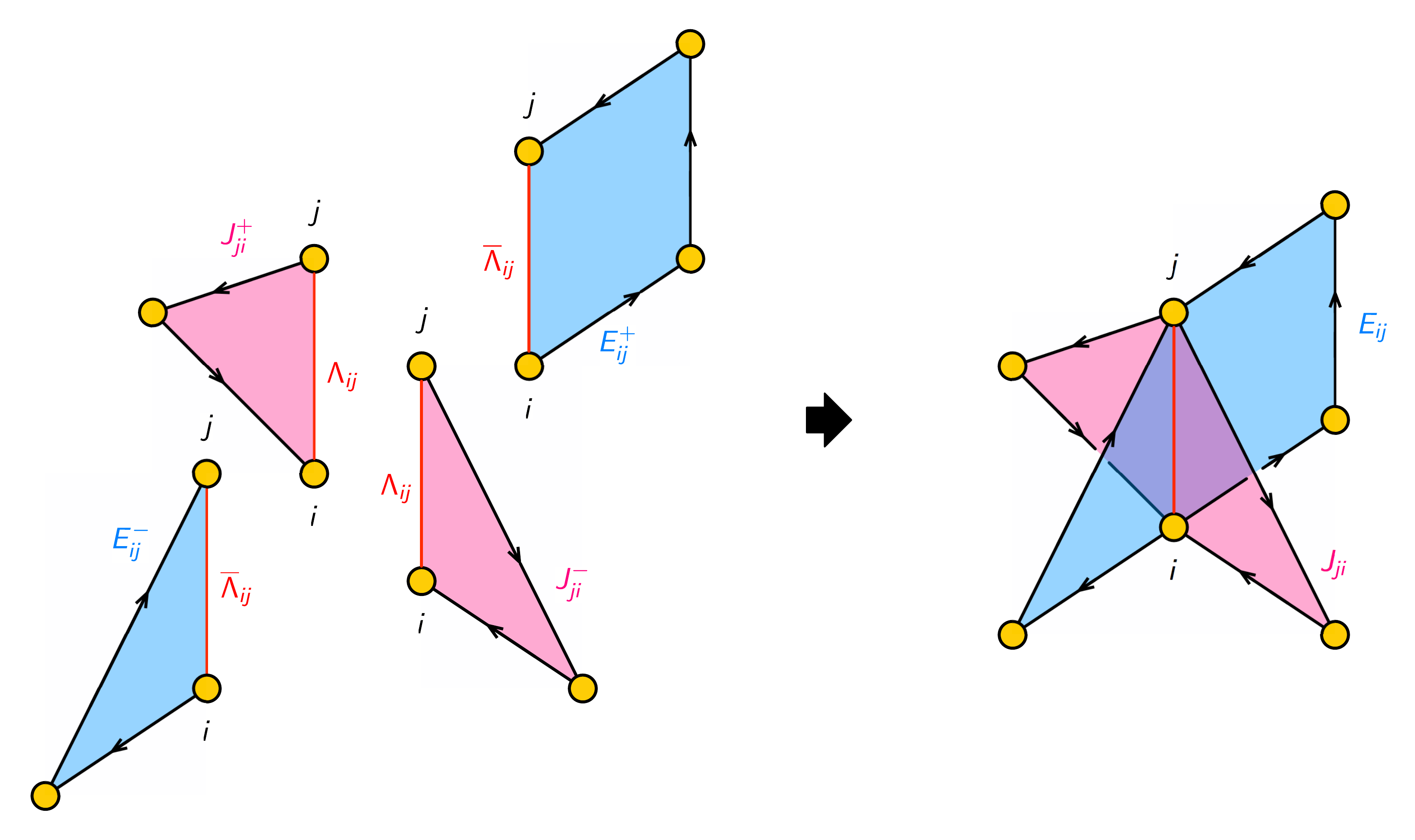}
}
\caption{The four plaquettes $(\Lambda_{ij}, J^{\pm}_{ji})$ and $(\overline\Lambda_{ij}, E^\pm_{ij})$ for a Fermi field $\Lambda_{ij}$. The $J$- and $E$-terms are $J_{ji}=J^{+}_{ji}- J^{-}_{ji}=0$ and $E_{ij}=E_{ij}^{+}- E_{ij}^{-}=0$, with $J_{ji}^\pm$ and $E_{ij}^{\pm}$ holomorphic monomials in chiral fields.
\label{fplaquettes}}
 \end{center}
 \end{figure}

Brane brick models are in one-to-one correspondence with {\it periodic quivers} on $T^3$, which automatically incorporate the toric condition. Periodic quivers not only encode the gauge symmetry and matter content of $2d$ toric theories, but also their $J$- and $E$-terms \cite{Franco:2015tna}. The latter are represented by \textit{minimal plaquettes}. A plaquette is defined as a gauge invariant closed loop in the quiver consisting of an oriented path of chiral fields and a single Fermi field. The toric condition implies that every Fermi is associated to four minimal plaquettes as shown in \fref{fplaquettes}. The full periodic quiver is constructed by assembling together all fields according to their minimal plaquettes. Brane brick models and periodic quivers are simply related by graph dualization, as shown in \fref{fbranebricksummary}.

\section{Mirror Approach to Brane Tilings and Brane Brick Models}

\label{section_mirror_general}

A central goal of this paper is to develop the mirror description for configurations of D1-branes probing toric CY 4-folds. The use of mirror symmetry for this purpose was pioneered in  \cite{Feng:2005gw}, in the context of D3-branes over toric CY 3-folds. There results were later generalized to toric CY$_n$ singularities with arbitrary $n$ in a beautiful paper by Futaki and Ueda \cite{Futaki:2014mpa}. Here we present the basics of this construction. A detailed analysis of the $n=3$ and $4$ cases is given in sections \sref{section_mirror_CY3} and \sref{section_mirror_CY4}.

Every toric CY$_n$ $\mathcal{M}$ is specified by its toric diagram $V$, which is a convex set of points $\mathbb{Z}^{n-1}$. Its mirror geometry \cite{Hori:2000kt,Hori:2000ck} is another $n$-fold $\mathcal{W}$ given by a double fibration over the complex $W$ plane
\beq
\begin{array}{rl}
W = & P(x_1,\ldots, x_{n-1}) \\[.1cm]
W = & uv
\end{array}
\label{double_fibration}
\eeq 
with $u,v \in \mathbb {C}$ and $x_\mu\in \mathbb{C^*}$, $\mu=1,\ldots,n-1$. Here $P(x_1,\ldots, x_{n-1})$ is the Newton polynomial associated to the toric diagram
\beq
P(x_1,\ldots, x_{n-1})=\sum_{\vec{v} \in V} c_{\vec{v}} \, x_1^{v_1} \ldots x_{n-1}^{v_{n-1}} , 
\eeq
where the $c_{\vec{v}}$ are complex coefficients and the sum runs over points $\vec{v}$ in the toric diagram. It is possible to scale $n$ of the coefficients to 1. 

The critical points of $P$ are defined as $(x_1^*,\ldots,x_{n-1}^*)$ such that
\beq
\left. {\partial \over \partial x_\mu} P(x_1,\ldots, x_{n-1})\right |_{(x_1^*,\ldots,x_{n-1}^*)}=0 \ \ \ \ \forall \, \mu 
\eeq
and, on the $W$-plane correspond to the critical values $W^*=P(x_1^*,\ldots,x_{n-1}^*)$. In \cite{Feng:2005gw} it was proved that for arbitrary $n$, when the toric diagram contains at least one internal point, the number of critical points of $P$ matches the normalized volume of the toric diagram.\footnote{Throughout the paper we will focus on toric diagrams that satisfy this condition. We are confident that our ideas can be extended to toric diagrams without internal points.} The number of critical points is precisely the one required for a basis of wrapped D$(9-n)$-branes in the mirror that accounts for the gauge nodes in the field theory, for any $n$.

\begin{figure}[htbp]
	\centering
	\includegraphics[width=14cm]{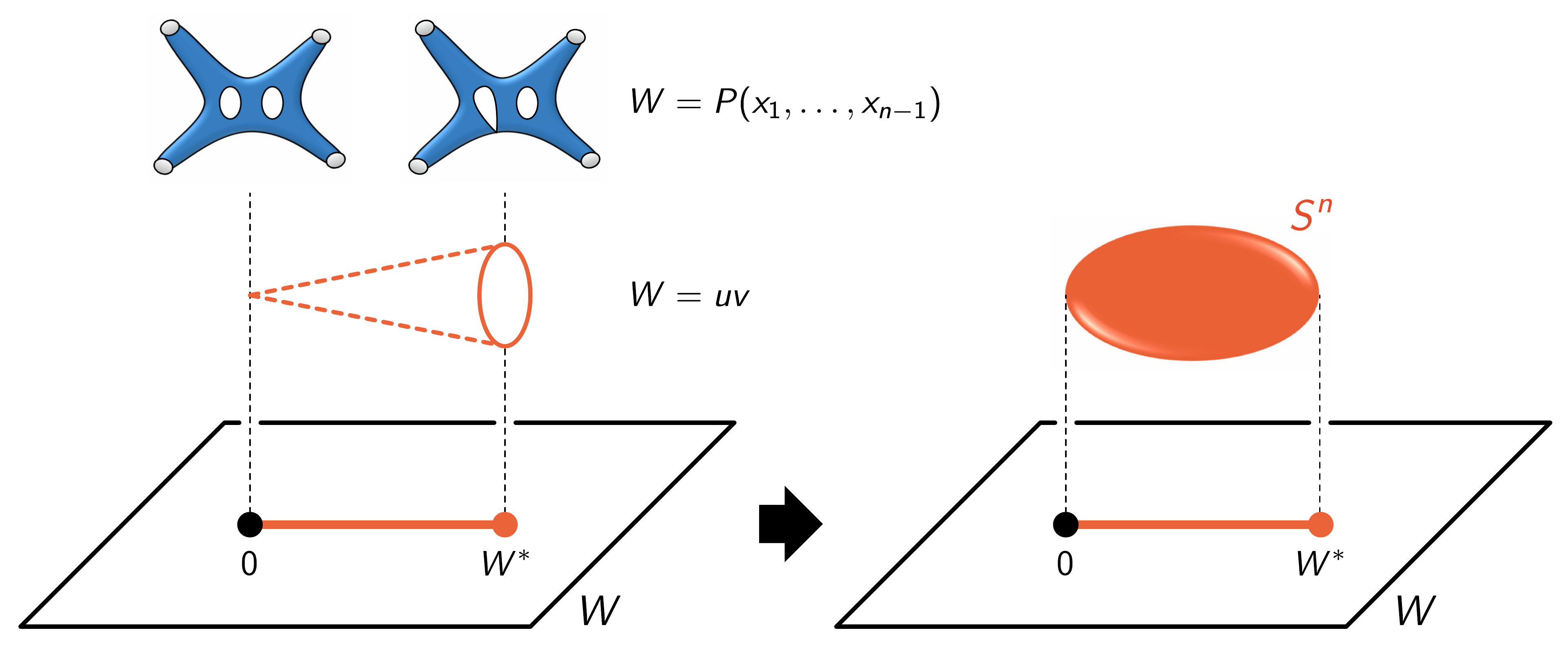}
\caption{Geometry of the mirror of a toric CY$_n$. It is a double fibration over the $W$-plane: one fiber is an $(n-2)$-complex dimensional surface $\Sigma_W$ containing an $S^{n-2}$ that degenerates at critical points $W^*$ and the other one is a $C^*$, with an $S^1$ that degenerates at the origin. The $S^{n-2} \times S^1$ fibered over an interval connecting the origin to $W^*$ gives rise to an $S^n$.}
	\label{n-sphere_mirror}
\end{figure}

The fiber associated to $P(x_1,\ldots, x_{n-1})$ corresponds to a holomorphic $(n-2)$-complex dimensional surface $\Sigma_W$, while the $uv$ one is a $\mathbb{C}^*$ fibration. For generic values of the $c_{\vec{v}}$ coefficients, an $S^{n-2} \subset \Sigma_W$ shrinks to zero size at each critical value $W^*$. In addition, the $S^1$ from the $uv$ fibration vanishes at $W=0$. Considering this $S^{n-2}\times S^1$ fibration over a straight {\it vanishing path} connecting $W=0$ and $W=W^*$, we obtain an $S^n$, as illustrated in \fref{n-sphere_mirror}.\footnote{In section \sref{section_triality} and appendix \sref{section_Q111_phases} we will comment on the possibility of non-straight vanishing paths on the $W$-plane.}

\begin{figure}[htbp]
	\centering
	\includegraphics[width=12cm]{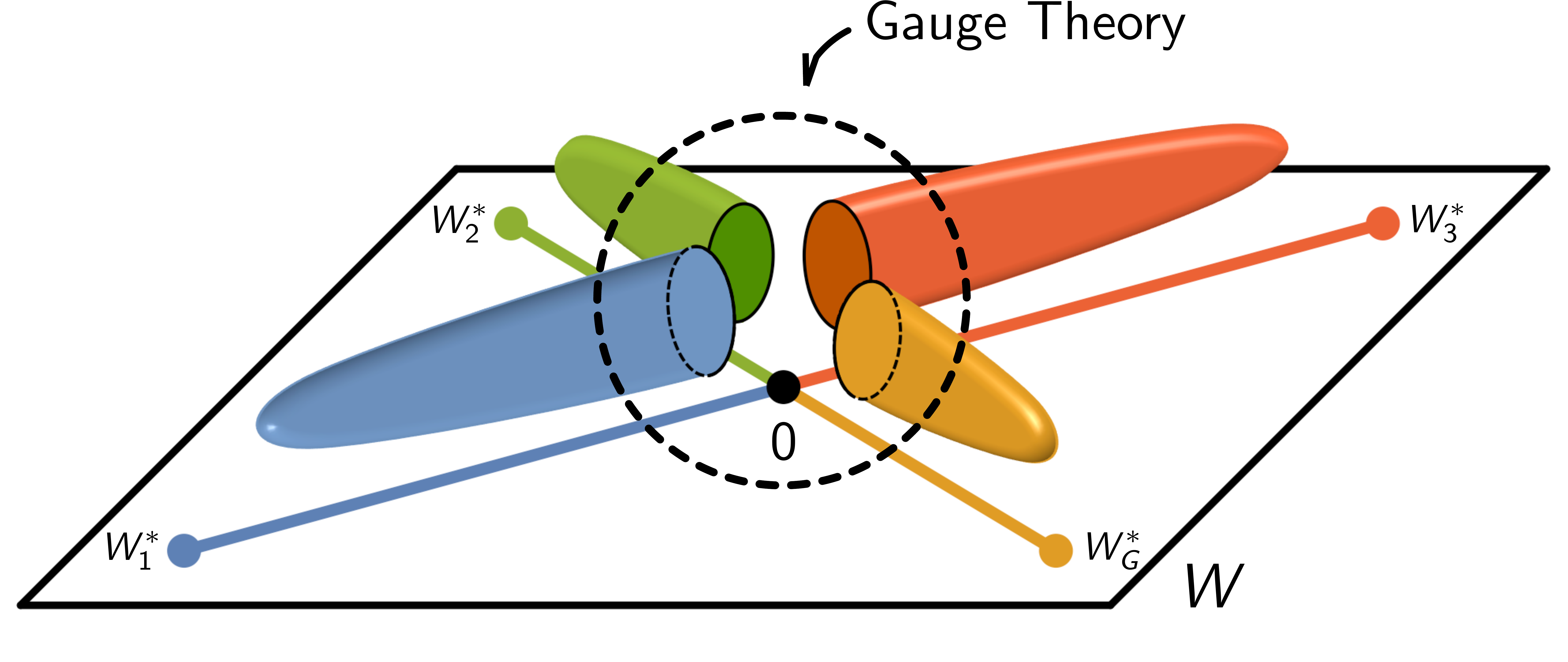}
\caption{The gauge theory is encoded in the way the $S^{n-2}$'s intersect on $\Sigma$ at $W=0$. For $n=3,4$, these intersections give rise to the corresponding brane tiling or brane brick model.}
	\label{basic_intersecting_Sn}
\end{figure}

All the $S^n$'s meet at $W=0$, where the $S^1$ fiber vanishes. The gauge theory is encoded in the way the $S^{n-2}$'s intersect on the vanishing locus $W^{-1}(0):P(x_\mu)=W=0$, as illustrated in \fref{basic_intersecting_Sn}. For $n=3,4$ this is precisely the holomorphic surface $\Sigma$ that underlies the brane tiling \cite{Franco:2005rj} and the brane brick model, as we explained in section \sref{section_BBMs}. This implies that these objects can be reconstructed from the intersections of the $S^{n-2}$'s. We will refer to each $S^{n}$ as $\mathcal{C}_i$, $i=1,\ldots,G$, and to the corresponding $S^{n-2}$ on $\Sigma$ as $C_i$. When studying the intersections of the $C_i$, it is often useful to consider two standard projections: the {\it amoeba}, which projects $\Sigma$ onto the $\log|x_\mu|$-plane, and the {\it coamoeba}, which projects it on the $\arg(x_\mu)$-torus.

\paragraph{Tomography.}

A useful tool for analyzing the mirror, to which we refer as {\it tomography}, was introduced in \cite{Futaki:2014mpa}. The $x_\mu$-tomography is the projection of the configuration of $S^{n-2}$ spheres at $W=0$ on the $x_\mu$-plane. The coamoeba projection of the $x_\mu$-tomographies provides a powerful systematic algorithm for constructing brane tilings and brane brick models. An appealing feature of tomography is its scalability. Every time the CY dimension $n$ is increased by one, we simply need to include an additional $x_\mu$ complex plane.
In sections \sref{section_mirror_CY3} and \sref{section_mirror_CY4}, we will discuss tomography in detail with various explicit examples.

\section{Calabi-Yau 3-Folds and Brane Tilings}

\label{section_mirror_CY3}

D3-branes over toric CY$_3$ singularities map to a collection D6-branes wrapped over 3-spheres $\mathcal{C}_i$ in the mirror \cite{Feng:2005gw}. Here we revisit the CY$_3$ case in order to set up the stage for CY$_4$'s, to be considered in section \ref{section_mirror_CY4}, and to illustrate the refined analysis of \cite{Futaki:2014mpa}.

\subsection{$\mathbb{C}^{3}/\mathbb{Z}_{3}$}

\begin{figure}[htbp]
	\centering
	\includegraphics[width=4cm]{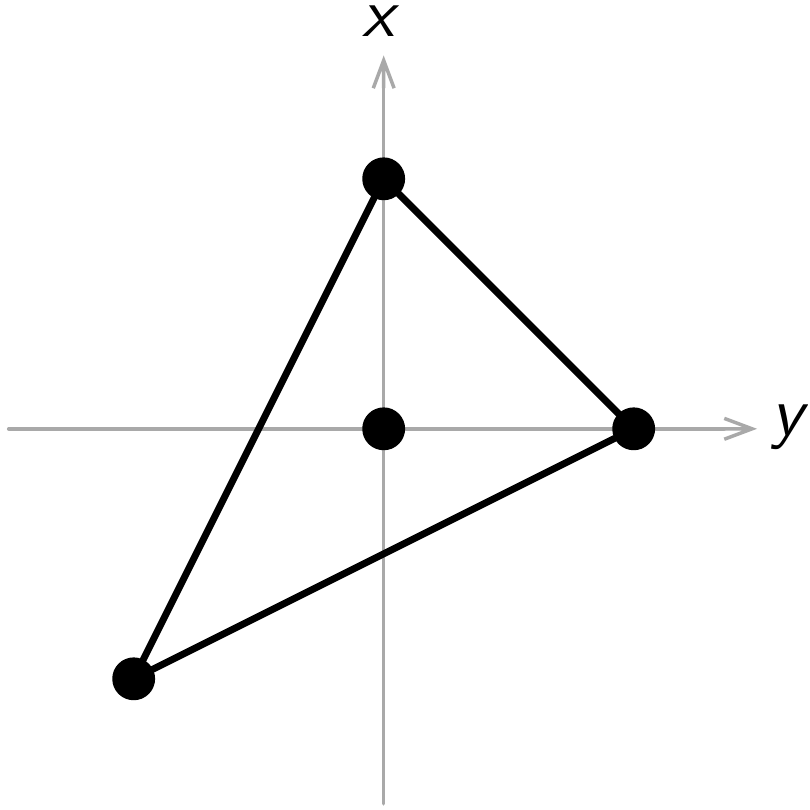}
\caption{Toric diagram for local $\mathbb{CP}^2$.}
	\label{toric_P2}
\end{figure}

Following \cite{Futaki:2014mpa}, we first consider the simplest example: local $\mathbb{CP}^2$, i.e. the $\mathbb{C}^3/\mathbb{Z}_3$ orbifold with action $(1,1,1)$. The toric diagram for this geometry is shown in \fref{toric_P2}.

\begin{figure}[htbp]
	\centering
	\includegraphics[height=7cm]{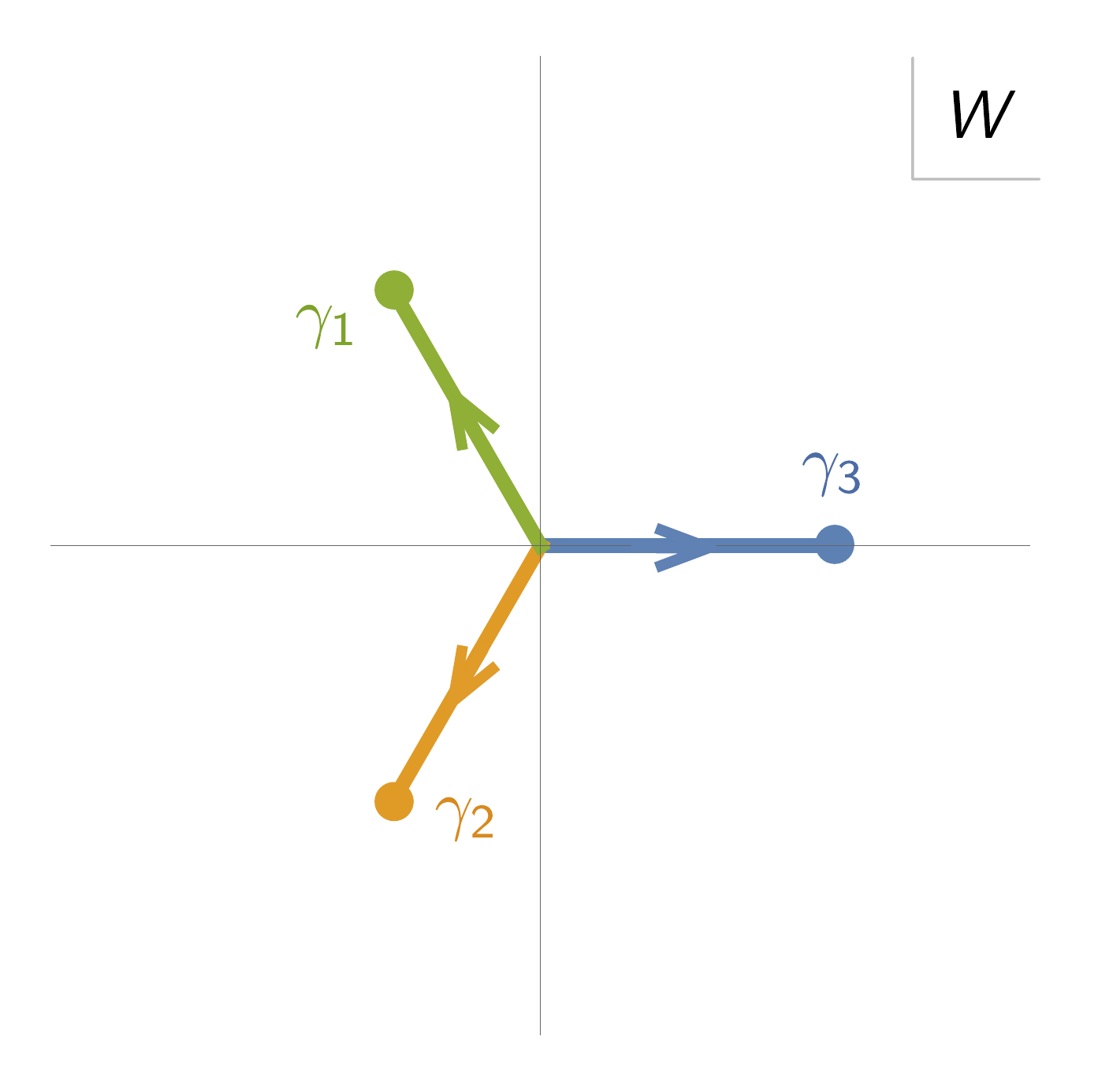}
\caption{Vanishing paths for local $\mathbb{CP}^2$.}
	\label{fig:P2-1a}
\end{figure}

The Newton polynomial has in general four terms. Three of the coefficients can be set to 1, leaving a single tunable parameter. Let us analyze the case in which we set $c_{(0,0)}=0$, i.e.  
\begin{align}
P(x,y) = x + y + \frac{1}{xy} \,.
\label{P_P2}
\end{align}
As expected, there are three critical points:
\begin{align}
(x^*_i,y^*_i)= (1,1), (\omega,\omega), (\omega^2,\omega^2)\,,\qquad \omega = e^{2\pi i/3}\,.
\label{P2-crit}
\end{align}
The critical values are $W^*_i=3, 3\omega,3\omega^2$, respectively. 
To each critical point, we can assign a vanishing path. A vanishing path is a curve $\gamma_i(t)$ on the $W$-plane such that 
\begin{align}
\gamma_i(0) = 0 \,, \quad \gamma_i(1) = W^*_i \,.
\end{align}
We choose $\gamma_i$, $i=1,2,3$, to be straight lines connecting the origin to $W=3\omega^i$, as shown in \fref{fig:P2-1a}.

As explained in section \sref{section_mirror_general}, every $W_i^*$ is in one-to-one correspondence with an $S^3$ $\mathcal{C}_i$ and with an $S^1$ $C_i$. D6-branes wrapping the $\mathcal{C}_i$'s give rise to the $4d$ gauge theory. In \cite{Futaki:2014mpa}, tomography was proposed as a convenient tool for studying the embedding of the $C_i$ into the Riemann surface $W^{-1}(0)\equiv \Sigma$. Let us consider the $y$-tomography, i.e. let us examine the critical points of $y$ as a function of $x$ at $W=0$. We get
\begin{align}
P(x,y) = 0 \quad \Longrightarrow \quad y = f(x) \,, 
\;\; \left.\frac{dy}{dx}\right|_{x=x_0} = 0\,,
\end{align}
which we call $x_0$ in order to differentiate them from the critical points of $P(x,y)$, $(x^*,y^*)$. Near every critical point $x_0$, we can locally write 
\begin{align}
y - y _0 = c_0 (x-x_0)^2 \quad \Longrightarrow \quad 
x - x_0 = \sqrt{(y-y_0)/c_0} \,.
\label{branching}
\end{align}
Hence, if we consider the inverse function, $x = f^{-1}(y)$, $y_0$ becomes a branch point. Explicitly, solving $P(x,y)=0$ for $x$, we find
\begin{align}
x+y+\frac{1}{xy} = 0 
\quad \Longrightarrow \quad
x = -\frac{y}{2} \pm \sqrt{\frac{y^2}{4} - \frac{1}{y}} \,.
\end{align}
Thus, the branch points on the $y$-plane are the solutions to $y^3 = 4$. To avoid confusion between $y^*$ in \eqref{P2-crit} and $y_0$ in \eqref{branching}, 
from here on we will refer to the former as {\em critical points} and the latter as {\em branch points}. 

The surface $\Sigma$ is a branched double cover of the $y$-plane. We should also remember that since $y$ is a $\mathbb{C}^*$ variable, $y=0$ can also be regarded as a branch point. In fact, as we go around $y=0$, $x$ returns to itself after two turns. The three branch points, together with $y=0$, give rise to a branched double cover description of a torus with three punctures. This fact was indeed expected, since $\Sigma$ can be regarded as the ``dual" of the toric diagram, which for this example is shown in \fref{toric_P2}.

\begin{figure}[htbp]
	\centering
	\includegraphics[height=7cm]{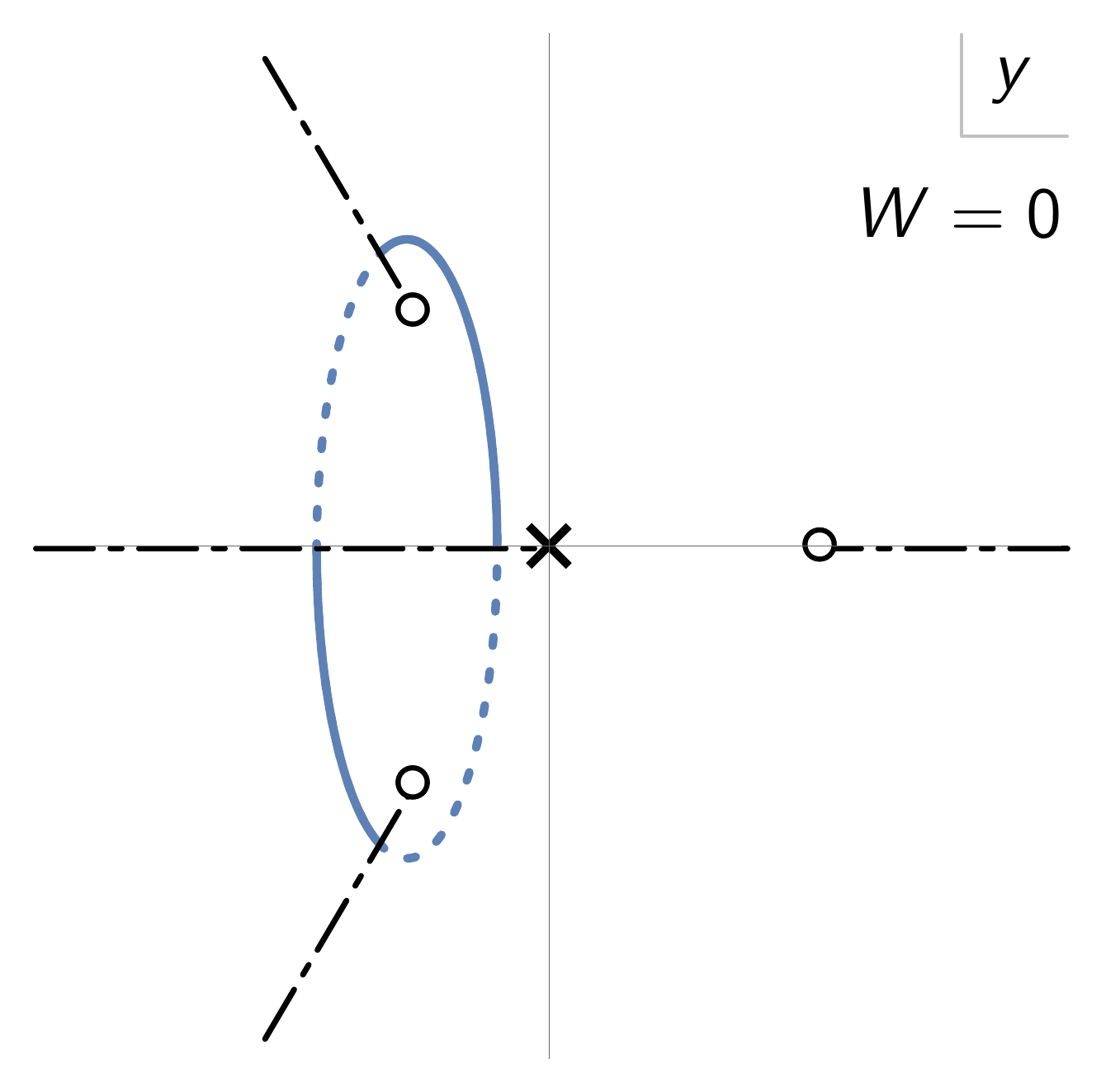}
\caption{Branch points, branch cuts, and an example of a cycle on the $y$-plane for local $\mathbb{CP}^2$.}
	\label{fig:P2-1b}
\end{figure}

\fref{fig:P2-1b} illustrates a choice of branch cuts and an example of a non-trivial cycle of the torus. In the figure, we marked the point $y=0$ to emphasize that it is a puncture. (Two other punctures are located at infinity and hence do not appear in the figure.) 

\begin{figure}[htbp]
	\centering
	\includegraphics[height=7cm]{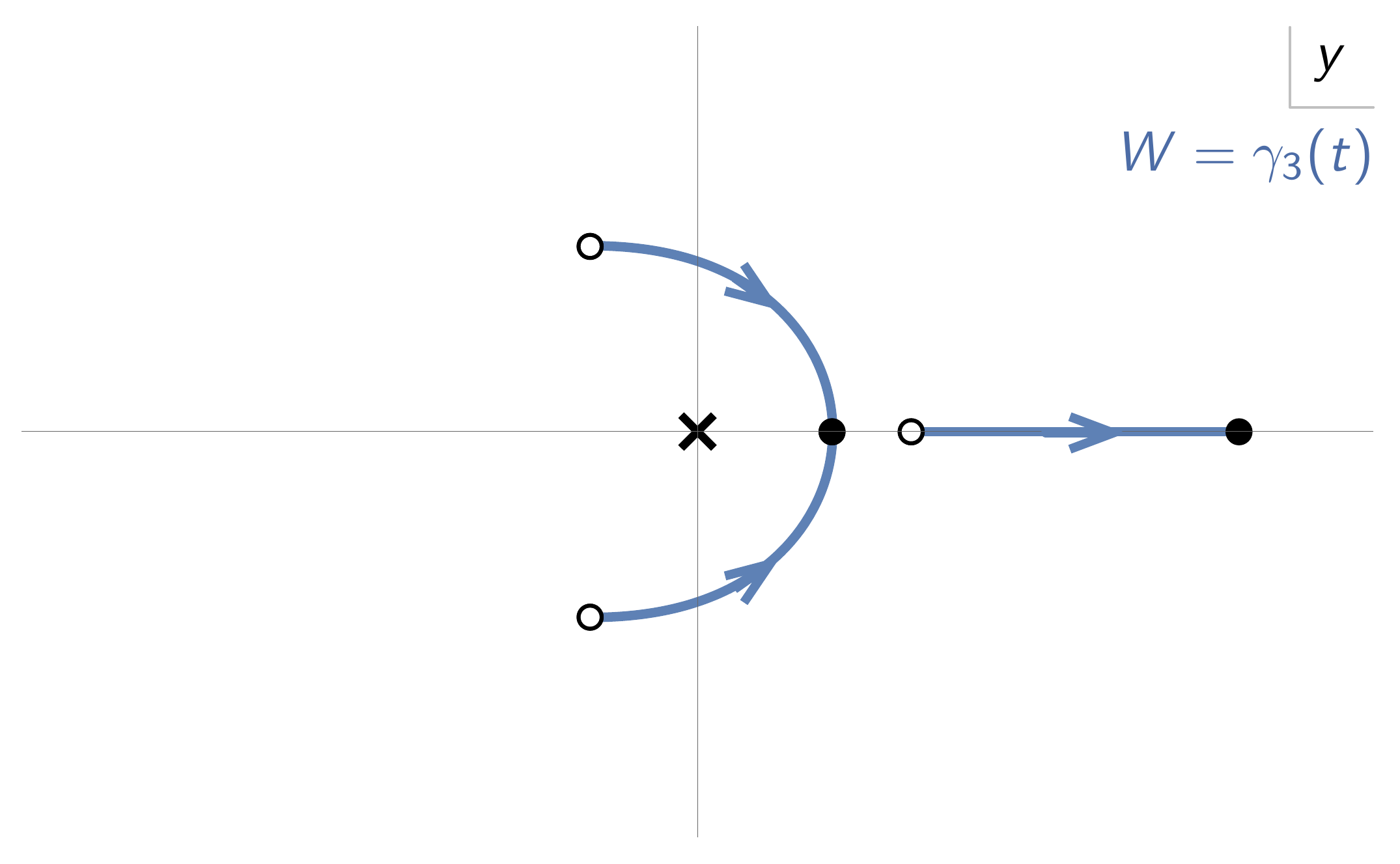}
\caption{Evolution of the branch points along $\gamma_3(t)$.}
	\label{fig:P2-1c}
\end{figure}

$\Sigma$ contains several 1-cycles. How do we identify $C_i$, i.e. the one pinching off at a given critical point $W_i^*$? Since the $C_i$'s are distinguished by their winding numbers around the branch points and the punctures, we can study what happens as we vary the value of $W$ along the corresponding vanishing path $\gamma_i(t)$. For example, let us consider $\gamma_3(t)$, along which we have
\begin{align}
x+y+\frac{1}{xy} = 3t 
\quad \Longrightarrow \quad x = -\frac{y-3t}{2} \pm \sqrt{\frac{(y-3t)^2}{4} - \frac{1}{y}} \,.
\end{align}
The branch points sit at the solutions to $y(y-3t)^2=4$. In particular, 
at $t=1$ we have a double root at $y=1$
and a simple root at $y=4$. Recall that $y=1$ is the critical value $y^*_3$. \fref{fig:P2-1c} shows the evolution of the branch points as $t$ goes from $0$ to $1$.

\begin{figure}[htbp]
	\centering
	\includegraphics[height=7cm]{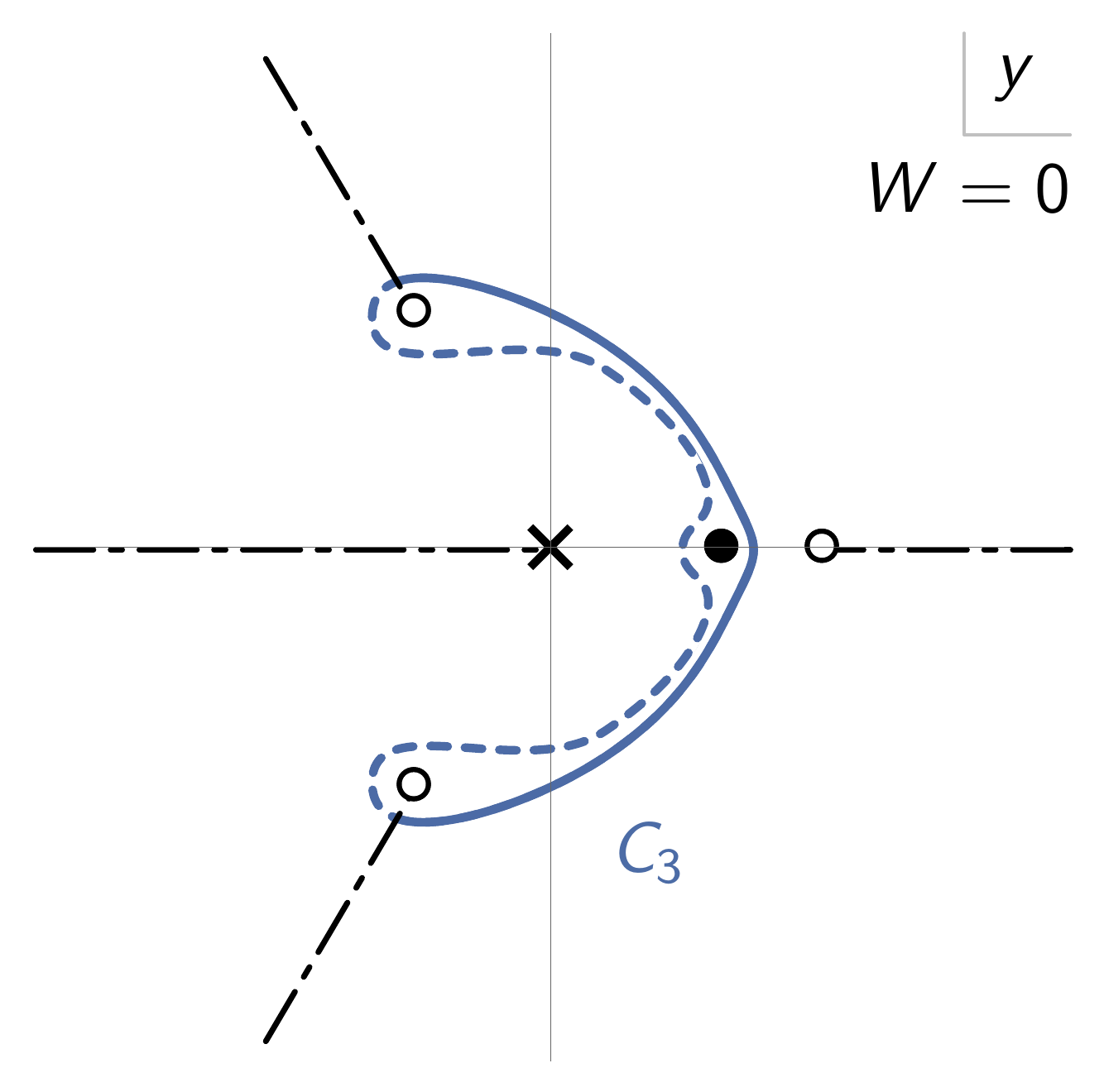}
\caption{Vanishing cycle $C_3$.}
	\label{fig:P2-1d}
\end{figure}

Two branch points coalescing into a single point indicate that a cycle is collapsing. It follows that the vanishing cycle $C_3$ corresponding to the vanishing path $\gamma_3(t)$ takes the form shown in \fref{fig:P2-1d}. The same argument can be repeated to determine $C_1$ and $C_2$. 

\begin{figure}[htbp]
	\centering
	\includegraphics[height=7cm]{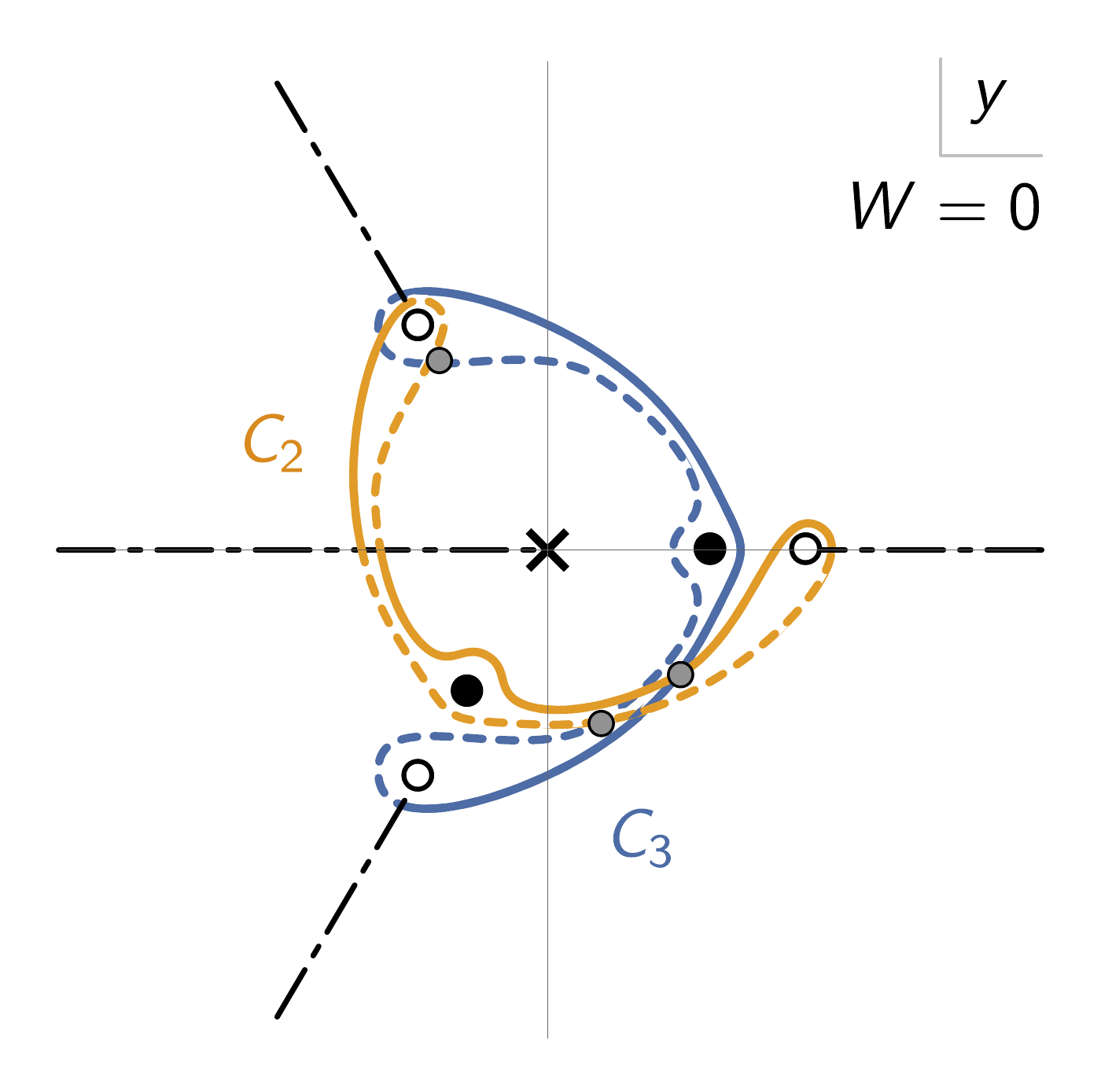}
\caption{Intersections (indicated by grey dots) between two vanishing cycles.}
	\label{fig:P2-1e}
\end{figure}

The next step is to study the intersections between pairs of vanishing cycles, since they give rise to bifundamental chiral fields in the gauge theory. Let us consider 
$C_2$ and $C_3$. In \fref{fig:P2-1e}, we have slightly deformed the two cycles in order to resolve the individual intersections. Solid and dotted segments lie on two different sheets. Only segments on the same sheet can intersect. We see that $C_2$ and $C_3$ intersect at three points: one near the common branch point and two away from the branch points.  Without the deformation, the two bulk intersections would coincide in the $y$-tomography. This is a general property in any dimension: intersections at branch points are always single, while bulk intersections might have higher multiplicity \cite{Futaki:2014mpa}. In fact, in order to get a precise understanding of intersections it is necessary to consider both the $x$- and $y$-tomographies (in general CY$_n$'s, we should take into account all the $x_i$-tomographies). Sometimes, a pair of cycles might seem to intersect when projected onto some of the planes but another tomography might reveal that they actually never meet. We will illustrate this phenomenon in some of the examples of section \sref{section_examples}.

The previous counting, one intersection at a branch point and two in the bulk, holds in any branched double cover descriptions of $\Sigma$. We should not, however, get the impression that branch point and bulk intersections are intrinsically different. A branch point on the $y$-plane is not a branch point on the $x$-plane. More generally, we can change the nature of the intersection points by switching among several $SL(2,\mathbb{Z})$ frames. 

\begin{figure}[htbp]
	\centering
	\includegraphics[height=7cm]{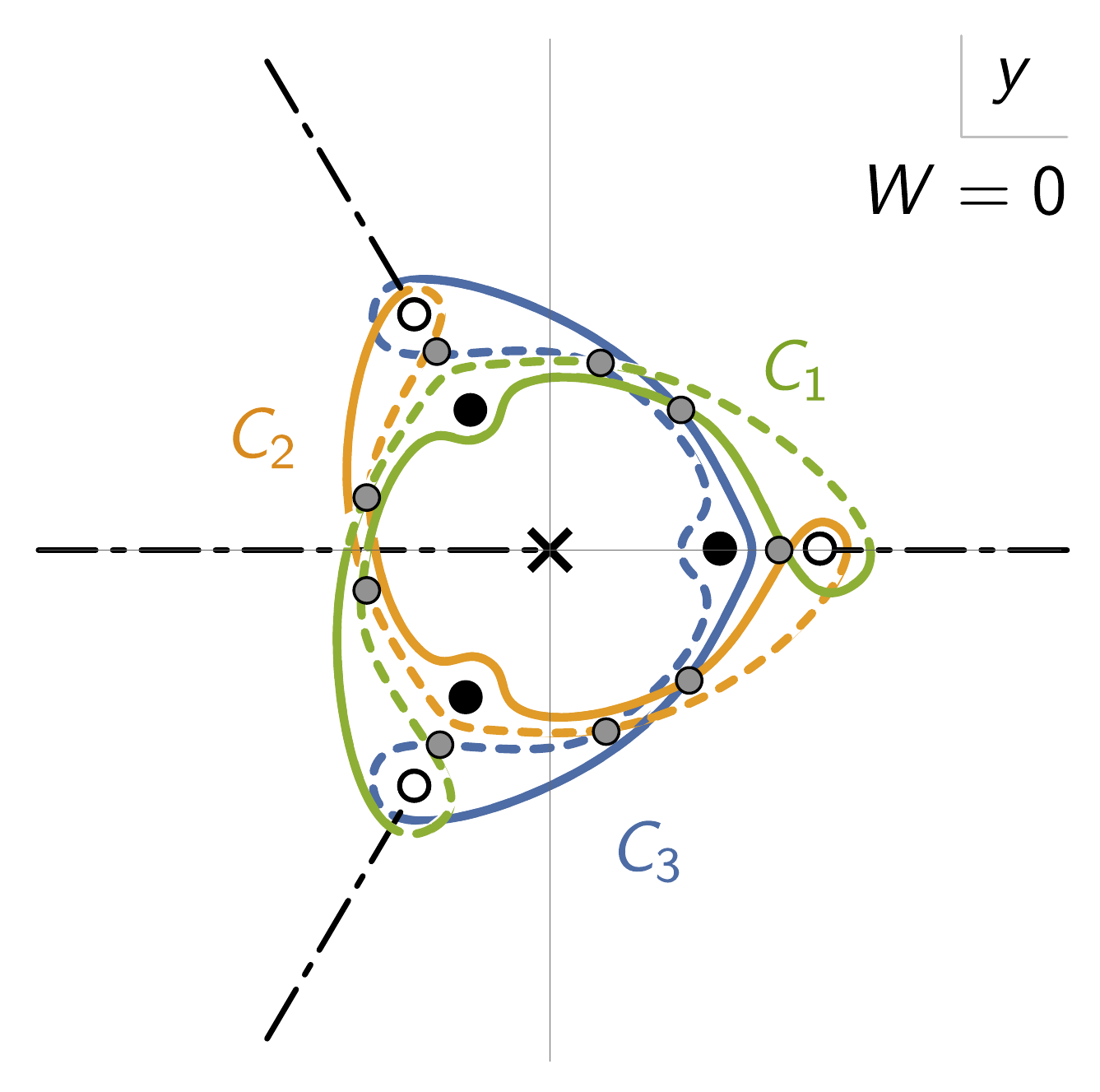}
\caption{The $y$-tomography for local $\mathbb{CP}^2$. The critical points are shown in black. Given the $x\leftrightarrow y$ symmetry of \eref{P_P2}, it is identical to the $x$-tomography.}
	\label{y-tomography_P2}
\end{figure}

Having presented a meticulous description of the geometry of vanishing cycles in the mirror, we can now streamline the discussion as follows. As noted in \cite{Futaki:2014mpa}, among the various figures we have discussed, the most important one is \fref{fig:P2-1c}. As we move on the $W$-plane along a vanishing path $\gamma_i(t)$, a pair of branch points from $t=0$ converges to $y^*_i$ at $t=1$. The union of the two trajectories of branch points, let us call it an {\it arc}, defines the vanishing cycle $C_i$. As shown in \fref{fig:P2-1d}, $C_i$ is topologically an $S^1$; two points fibered over the arc meet at the two $t=0$ branch points, closing the circle. The full $y$-tomography is shown in \fref{y-tomography_P2}. In principle, it is still necessary to construct the $x$-tomography. In this example, however, it is identical to the $y$-tomography due to the $x\leftrightarrow y$ symmetry of \eref{P_P2}.

\begin{figure}[htbp]
	\centering
	\includegraphics[width=11cm]{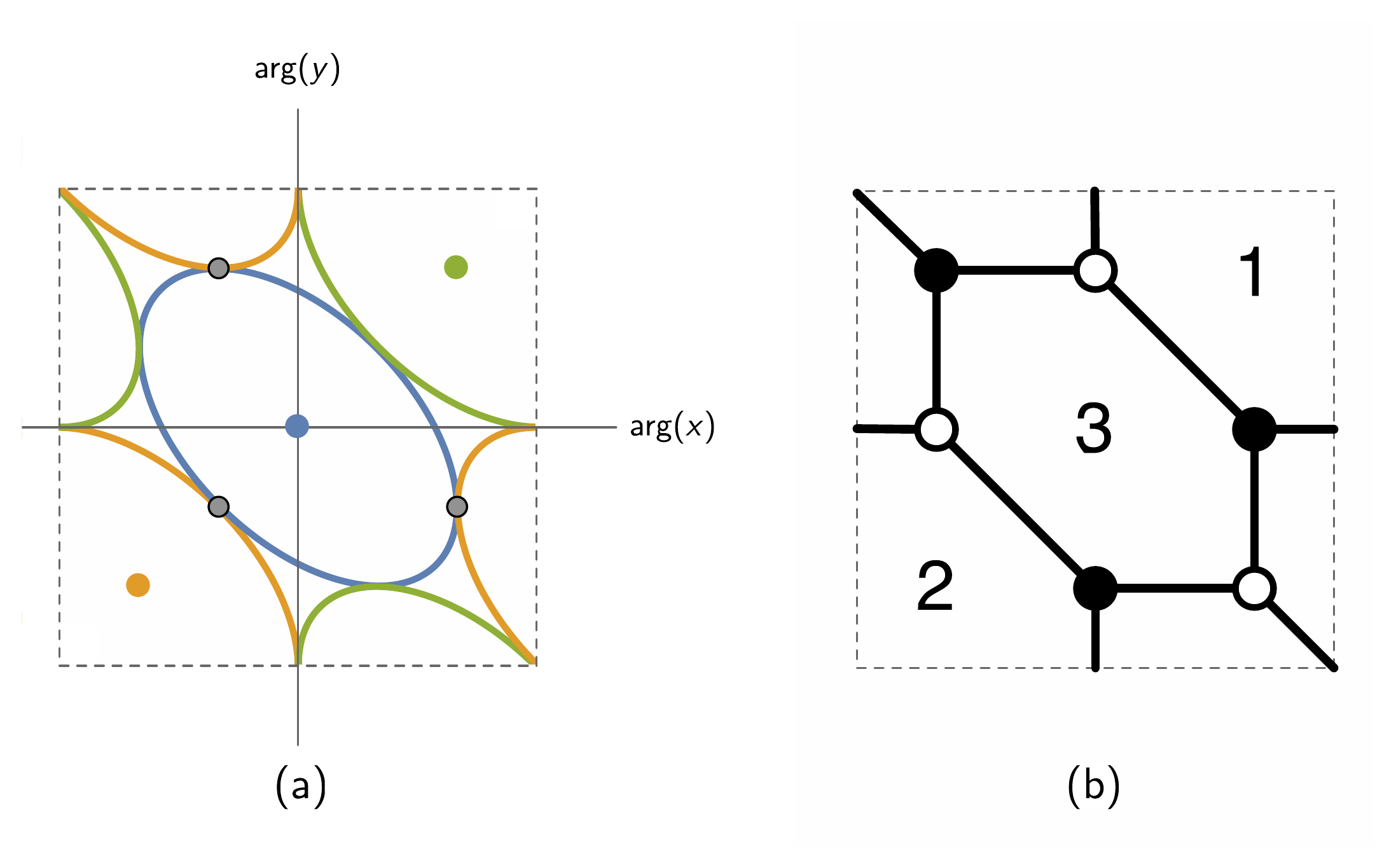}
\caption{a) Vanishing cycles and intersections on the coamoeba torus for local $\mathbb{CP}^2$. The three grey dots are the three intersections between $C_2$ and $C_3$ from \fref{fig:P2-1e}. b) The corresponding brane tiling.}
	\label{coamoeba-tiling_P2}
\end{figure}

It is now possible to systematically build the corresponding brane tiling. The $T^2$ of the brane tiling is precisely the {\it coamoeba torus} spanned by $(\mathrm{arg}(x),\mathrm{arg}(y))$. The coamoeba projection maps every fixed value of the argument in a given tomography to a planar slice of $T^2$. Every vanishing cycle $C_i$ at $W=0$ is mapped to a topologically trivial circle (i.e to the boundary of a disc) on the coameoba $T^2$. Its ``center" is located at the critical point $(\mathrm{arg}(x^*_i),\mathrm{arg}(y^*_i))$. The intersections between neighboring cycles give rise to chiral fields. The coamoeba diagram \cite{Feng:2005gw} is the complement of the union of all discs $D_i$, whose boundaries are the vanishing cycles $C_i$. We can immediately construct the brane tiling from the coamoeba, which has a segment for every intersection of $C_i$'s, as shown in \fref{coamoeba-tiling_P2}.

\subsection{$F_0$}

\label{section_F0}

Let us now consider the complex cone over $F_0$, which is a chiral $\mathbb{Z}_2$ orbifold of the conifold. For brevity, we will refer to it simply as $F_0$. This is an interesting example, since it is one of the simplest geometries that admit more than one toric Seiberg dual phases \cite{Feng:2001xr,Feng:2002zw}. In $4d$, we define a toric phase as one that can be encoded in terms of a brane tiling. Following the detailed presentation in the previous section, the discussion of this example will be more succinct. 

\begin{figure}[htbp]
	\centering
	\includegraphics[width=4cm]{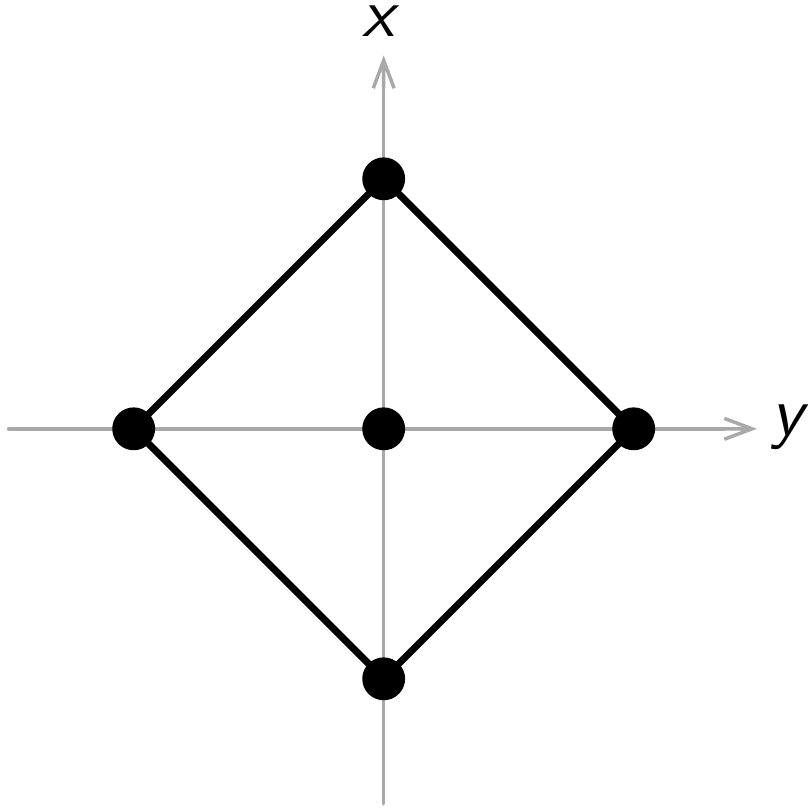}
\caption{Toric diagram for $F_0$.}
	\label{toric_F0}
\end{figure}

The toric diagram for $F_0$ is shown in \fref{toric_F0}. Let us first consider the following choice of coefficients in the Newton polynomial
\begin{align}
P(x,y) = x + \frac{1}{x} + i \left(y + \frac{1}{y} \right) \,.
\label{P-xy-F0-1}
\end{align} 
The four critical points and the corresponding critical values are
\begin{align}
(x^*,y^*) = (\pm 1, \pm 1) \,, \quad W^* = 2(x^* + i y^*) \,.
\end{align}

\begin{figure}[htbp]
	\centering
	\includegraphics[width=15.5cm]{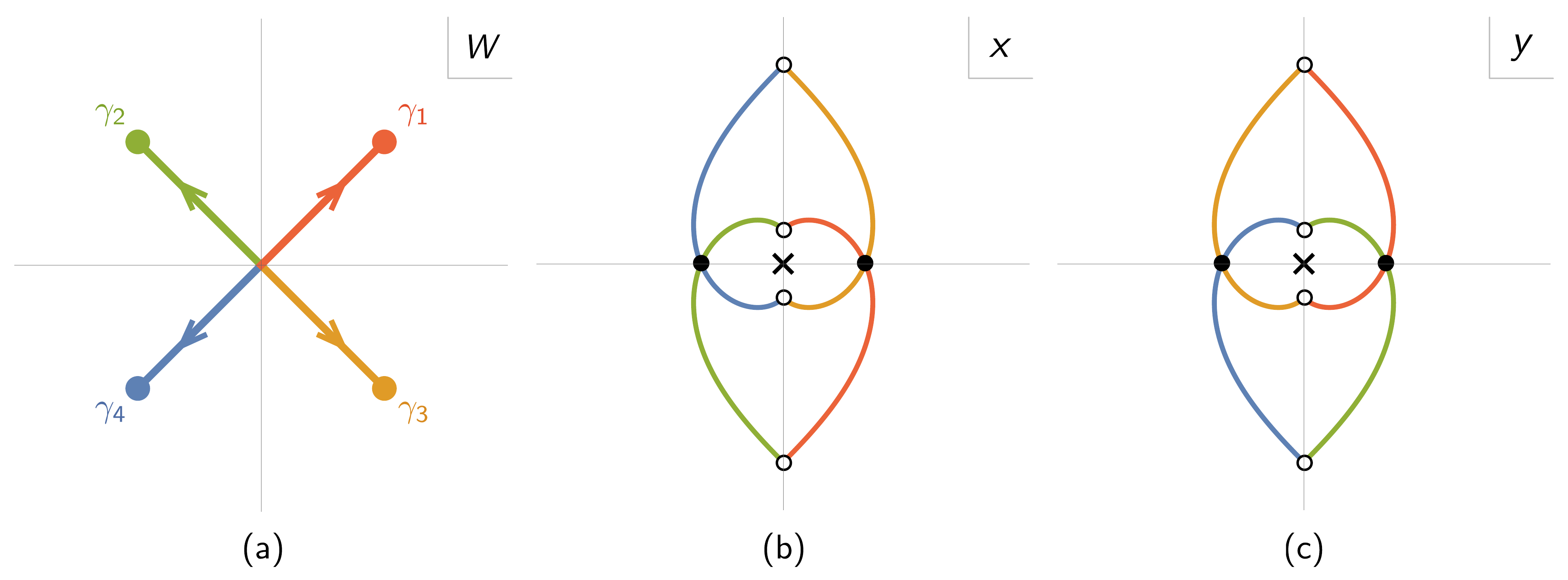}
\caption{a) Vanishing paths, b) $x$- and c) $y$-tomographies for phase 1 of $F_0$.}
	\label{mirror_F0_1}
\end{figure}

\begin{figure}[htbp]
	\centering
	\includegraphics[width=12cm]{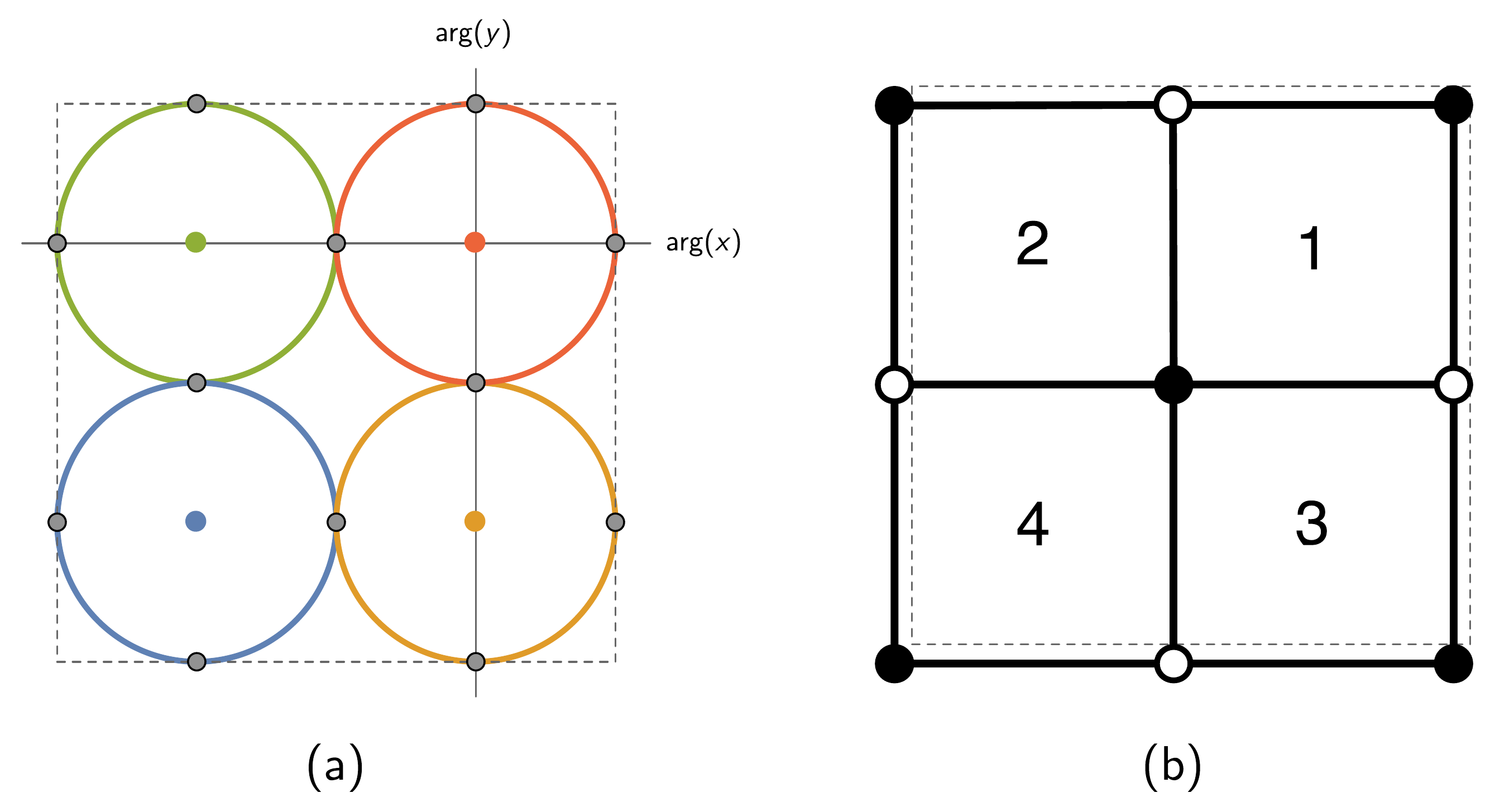}
\caption{a) Vanishing cycles and intersections on the coamoeba torus for phase 1 of $F_0$. The three grey dots indicate intersections between cycles. b) The corresponding brane tiling.}
	\label{coamoeba-tiling_F0_1}
\end{figure}

\fref{mirror_F0_1} shows the vanishing paths on the $W$-plane and the $x$- and $y$-tomographies. From them, we can construct the coamoeba diagram and brane tiling, which are shown in \fref{coamoeba-tiling_F0_1}. They correspond to one of the two toric phases of $F_0$ \cite{Feng:2001xr,Feng:2002zw}, which we call phase 1. 

\begin{figure}[htbp]
	\centering
	\includegraphics[width=15.5cm]{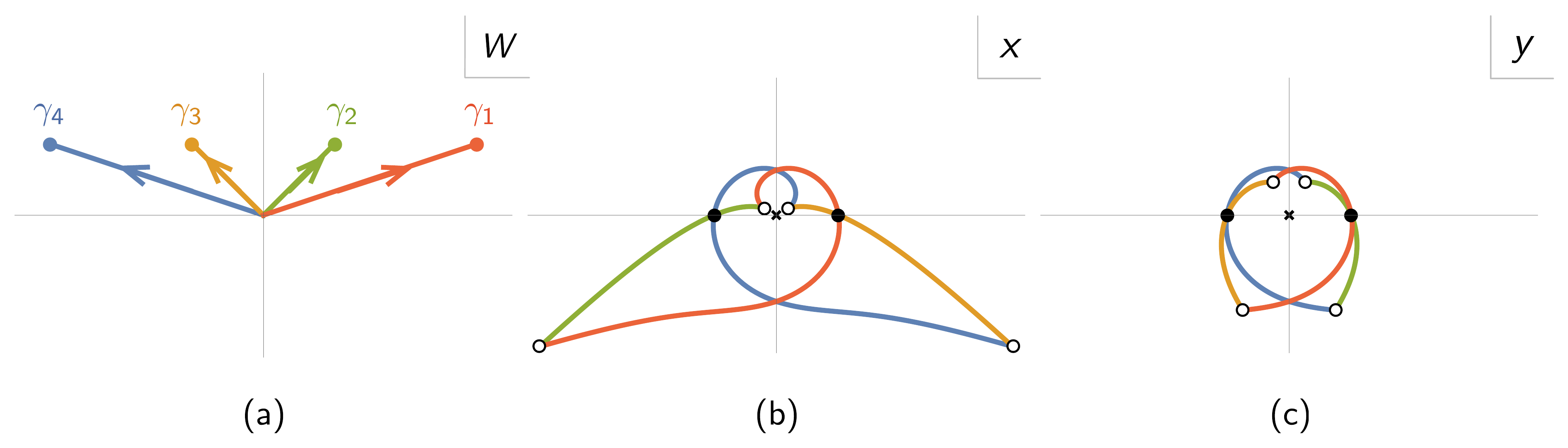}
\caption{a) Vanishing paths, b) $x$- and c) $y$-tomographies for phase 2 of $F_0$.}
	\label{mirror_F0_2}
\end{figure}

\begin{figure}[htbp]
	\centering
	\includegraphics[width=12cm]{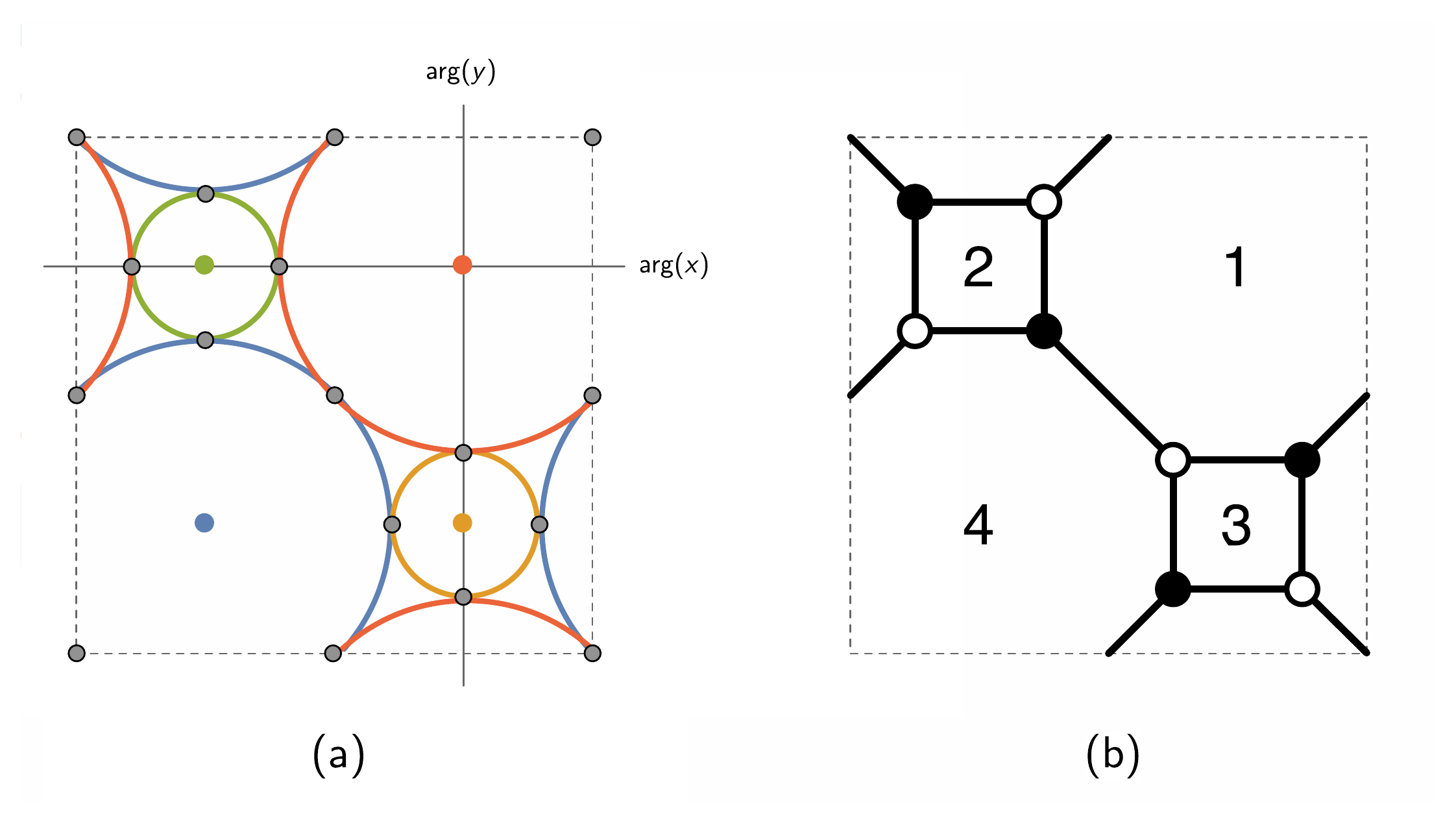}
\caption{a) Vanishing cycles and intersections on the coamoeba torus for phase 2 of $F_0$. The three grey dots indicate intersections between cycles. b) The corresponding brane tiling.}
	\label{coamoeba-tiling_F0_2}
\end{figure}

Interestingly, the second toric phase of $F_0$ can be generated by varying the coefficients in the Newton polynomial. Consider, for example,
\begin{align}
P(x,y) = x + \frac{1}{x} + 2 \left(y + \frac{1}{y} \right) +2i \,.
\label{P-xy-F0-2}
\end{align}
Comparing \eqref{P-xy-F0-1} and \eqref{P-xy-F0-2}, we note that the coefficient multiplying $(y+1/y)$ in the second phase is real. 
If we set it to 1, two of the critical points would coincide. As long as it is not 1, its absolute value is not important. Another novelty of \eqref{P-xy-F0-2} is 
that it contains a constant term. It prevents the vanishing paths from overlapping. The four critical points and their critical values are now
\begin{align}
(x^*,y^*) = (\pm 1, \pm 1) \,, \quad W^* = 2x^* + 4 y^*+2i \,.
\end{align}
The resulting vanishing paths and $x$- and $y$-tomographies are shown in \fref{mirror_F0_2}. The corresponding coamoeba and brane tiling are shown in \fref{coamoeba-tiling_F0_2}. They represent phase 2 of $F_0$.

A crucial distinction between the mirror configurations describing both phases is the different cyclic orderings of the vanishing paths on the $W$-plane, as shown in Figures \ref{coamoeba-tiling_F0_1} and \ref{coamoeba-tiling_F0_2}. When going from phase 1 to phase 2, $\gamma_3$, which corresponds to the dualized gauge group, moves over $\gamma_4$, associated to the node in the quiver from which the flavors going into node $3$ emanate. This geometric realization of Seiberg duality was introduced in \cite{Cachazo:2001sg} and will be revisited in section \sref{section_triality}, where we generalize it to $2d$ triality.

\newpage

\section{Calabi-Yau 4-Folds and Brane Brick Models}

\label{section_mirror_CY4}

Having extensively reviewed the application of mirror symmetry to brane tilings, we are ready to explain how a similar approach can be developed for brane brick models. The construction of brane brick model consists of three steps. 

First, we identify some compact 2-cycles $C_i$ on the surface $\Sigma = W^{-1}(0)$ as the gauge nodes in complete parallel with brane tilings. Upon double fibration over vanishing paths, they form the compact 4-cycles $\mathcal{C}_i$ of the mirror CY$_4$. D5-branes wrapping these 4-cycles give rise to the gauge nodes. 

Second, we identify the intersections among the cycles as bifundamental fields of the $2d$ gauge theory. The fields originate from string segments connecting two different gauge nodes. A main novelty here is that, unlike in brane tilings, there are two distinct types of supermultiplets: chiral and Fermi. We will show that they are distinguished by the
orientation of the intersections. A closer look at the oriented intersections can also determine the orientation of the chirals. 

Third, we construct the $J$- and $E$-terms that form plaquettes in the brane brick model. Given the coordinates of the gauge nodes and the intersection points in the coamoeba $T^3$, we will show how to graphically construct all the plaquettes. 

In this section, following \cite{Futaki:2014mpa}, we will illustrate the ideas with the simplest example: the local $\mathbb{CP}^3$ (namely the $\mathbb{C}^4/\mathbb{Z}_4$ orbifold with action $(1,1,1,1)$). We will present additional examples in the next section. 

\begin{figure}[ht]
\begin{center}
\includegraphics[width=12cm]{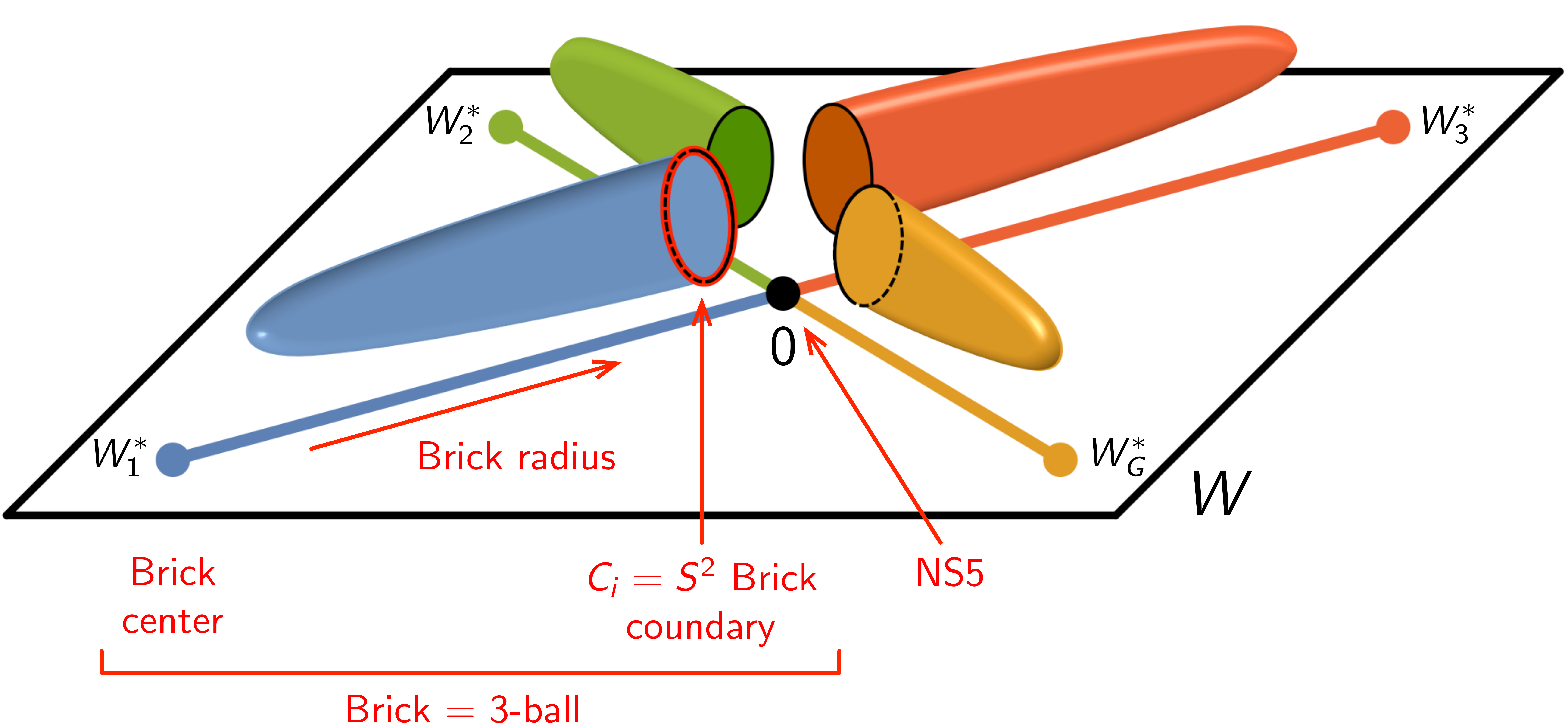}
\caption{The $\Sigma_W$ fibration in the mirror and the correspondence with some of the basic elements in the brane brick mode.
\label{mirror-bbm}}
 \end{center}
 \end{figure}

\fref{mirror-bbm} summarizes the basic ingredients in the correspondence between the mirror and brane brick models. We focus on the $\Sigma_W$ fibration, leaving the $\mathbb{C}^*$ $uv$ fibration implicit. In red we indicate the brane brick model objects associated with different parts of the mirror. Of course, an analogous picture applies to the connection between D6-branes in the mirror of toric CY 3-folds and brane tilings.

\subsection{Cycles as Gauge Nodes}

Most of the discussion from the previous section generalizes straightforwardly to CY$_4$. 
We begin by noting that $\Sigma = W^{-1}(0)$ now defines a four real dimensional hypersurface. As we will see shortly, the vanishing cycles $C_i$ are $S^2$'s intersecting with each other transversely. 

The main tool for studying $\Sigma$ and its vanishing cycles is the same as before; we take the tomography of the surface by expressing one of the variables, say $z$, as a function of the other two. Locally, we regard $\Sigma$ as a fibration over the $z$-plane. At every fixed value of $z=z_1$, the fiber $\mathcal{F}(z_1)$, defined by the equation $P(x,y,z_1)=0$, is a Riemann surface. For a generic value of $z_1$, $P(x,y,z_1)$ is nothing but the Newton polynomial of the projection of the CY$_4$ toric diagram onto the $(x,y)$-plane. 
Hence, we can study the vanishing 2-cycles of the four-dimensional $\Sigma$ and their intersections by first learning about the fibration structure and then examining how the 1-cycles of the Riemann surface fiber intersect.

\begin{figure}[ht]
\begin{center}
\includegraphics[width=6cm]{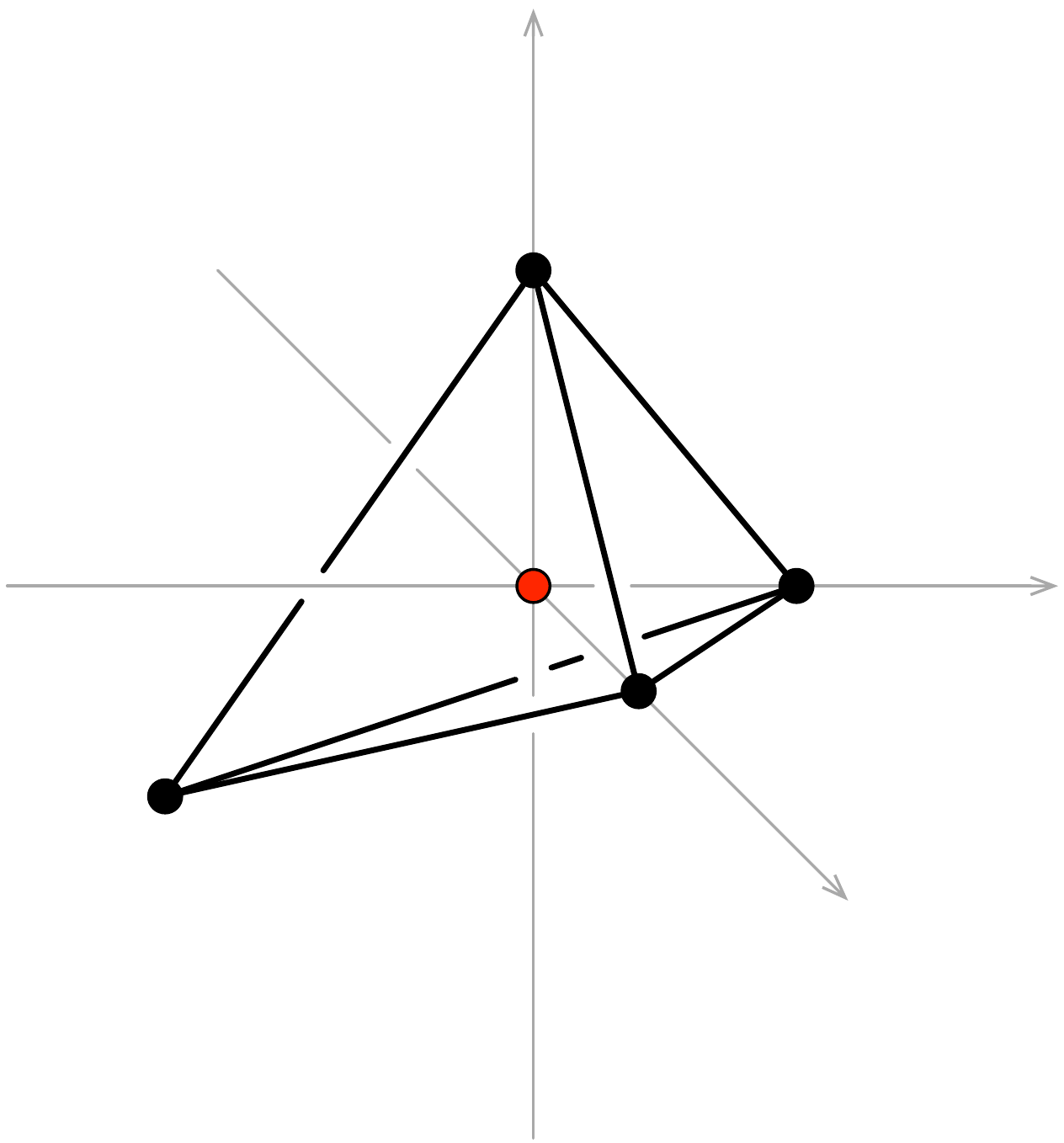}
\caption{Toric diagram for local $\mathbb{CP}^3$. 
\label{toric_CP3}}
 \end{center}
 \end{figure} 

\begin{figure}[htbp]
	\centering
	\includegraphics[height=6cm]{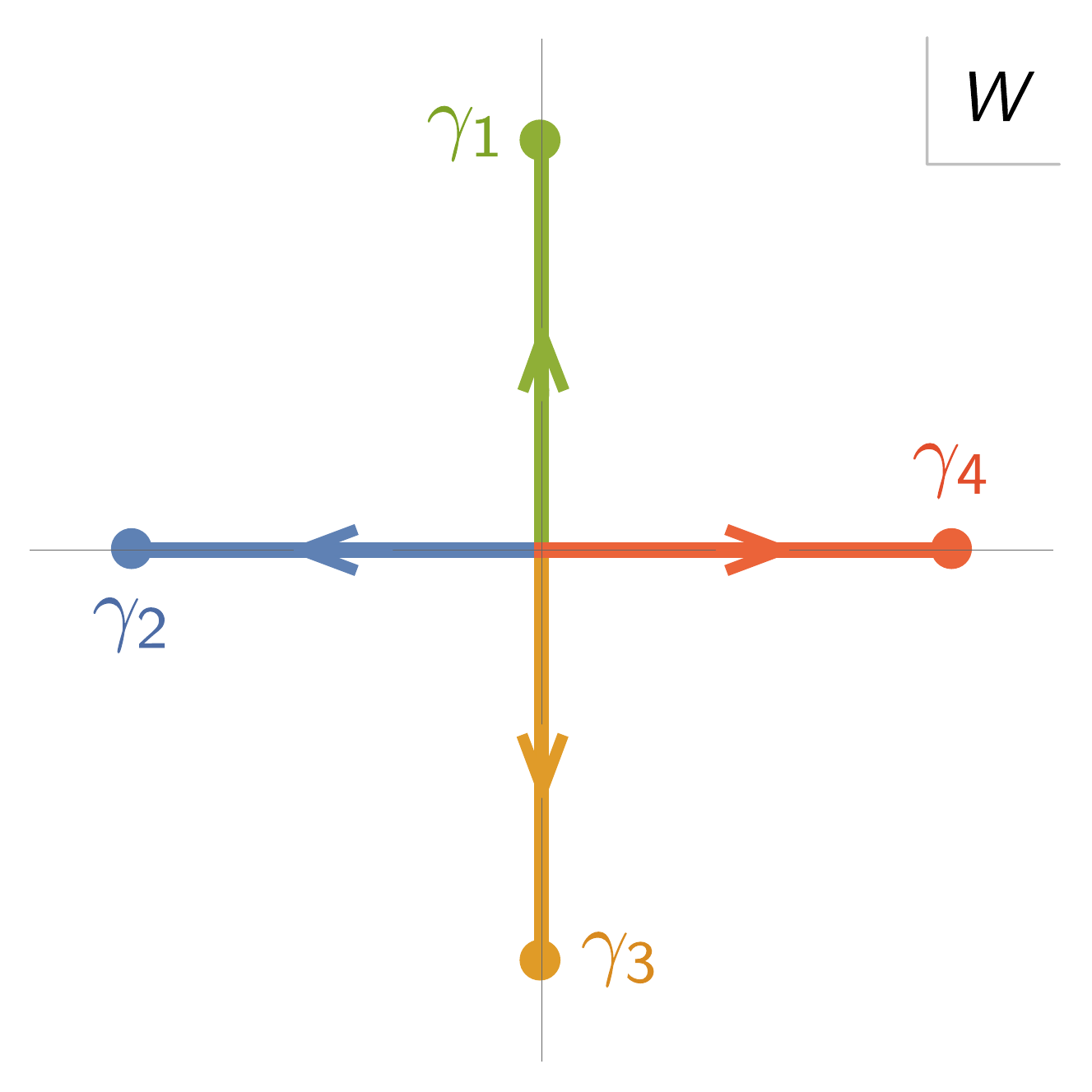}
\caption{Vanishing paths for local $\mathbb{CP}^3$. 
}
	\label{fig:P3-a}
\end{figure}

Let us consider local $\mathbb{CP}^3$, whose toric diagram is shown in \fref{toric_CP3}. The Newton polynomial has in general five terms. Four of the coefficients can be scaled to 1. Setting $c_{(0,0,0)}=0$, we obtain
\begin{align}
W = P(x,y,z) = x + y + z + \frac{1}{xyz} \,.
\label{P_CP3}
\end{align}
It has four critical points at\footnote{In this expression we momentarily use $a$ to index cycles in order to avoid confusion with $i=\sqrt{-1}$. We will soon return to the notation for cycles in terms of $i$ and $j$ indices.}
\begin{align}
x^* = y^* = z^* = i^a, \quad W^* = 4 i^a \quad (a=0,1,2,3) \,.
\label{P3-crit}
\end{align}
The vanishing paths, defined in the usual way, are shown in \fref{fig:P3-a}. 

\begin{figure}[htbp]
	\centering
	\includegraphics[height=8cm]{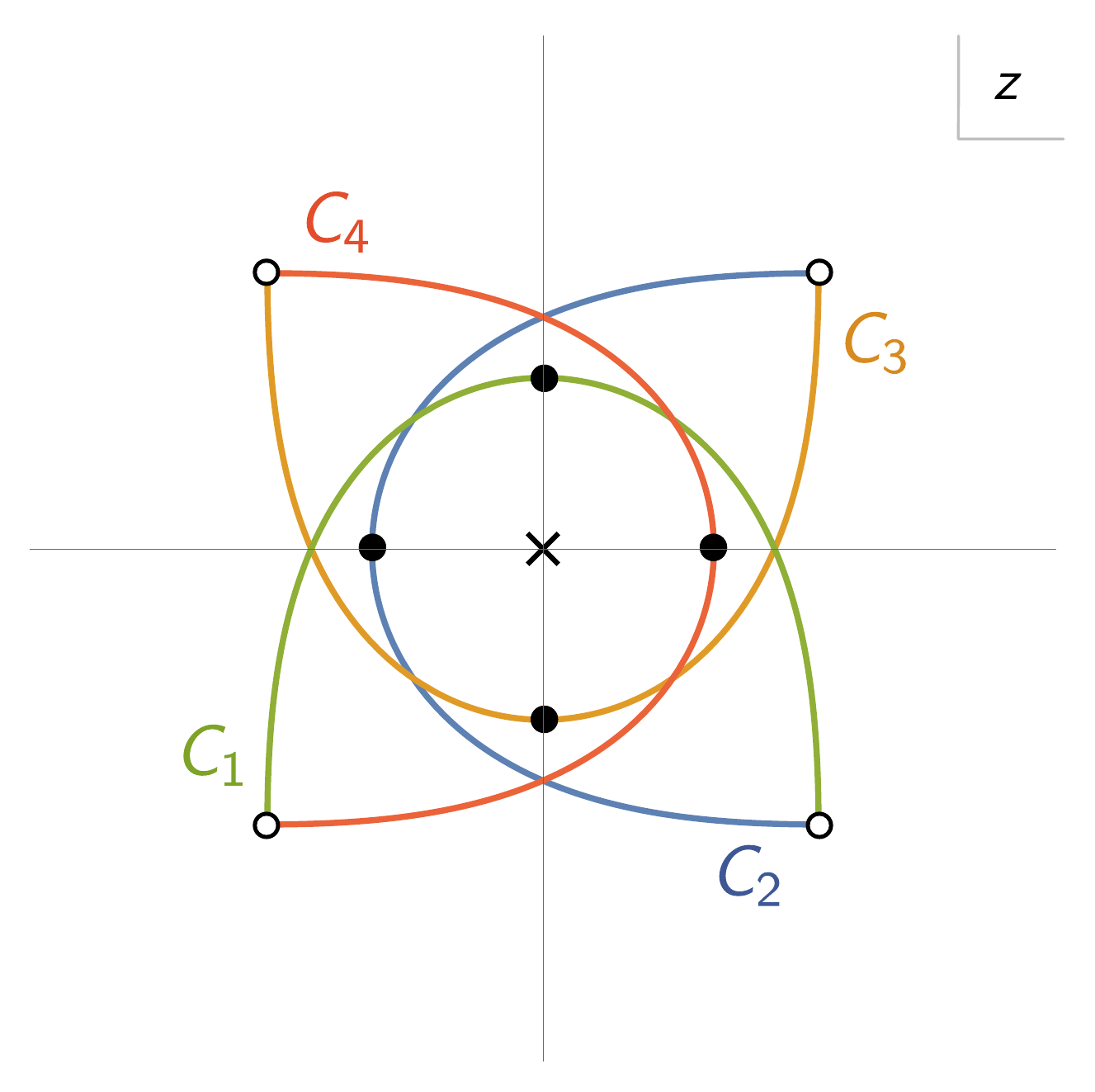}
\caption{The $z$-tomography for local $\mathbb{CP}^3$. The critical points are shown in black.
}
	\label{fig:P3-b}
\end{figure}

To construct the $z$-tomography, we study the critical points of $z$ as a function of $x$ and $y$
\begin{align}
P(x,y,z) = 0 \quad \Longrightarrow \quad z = f(x,y) \,, 
\;\;\frac{\partial z}{\partial x} = 0 = \frac{\partial z}{\partial y}\;\;
\mbox{at} \;\; (x_0,y_0)\,, \;\; z_0 = f(x_0,y_0)\, .
\end{align}
Although $z=f(x,y)$ cannot be inverted as \eqref{branching}, in order to keep consistency in nomenclature and avoid confusion with $(x^*,y^*,z^*)$ in \eqref{P3-crit}, we will call $(x_0,y_0,z_0)$ {\em branch points}. 
We find four branch points located at 
\begin{align}
z_0 = (-3)^{3/4}, \quad x_0 = y_0 = -z_0/3 \,.
\label{P3-branch}
\end{align}
As before, we study how these points move as we vary the value of $W$ along a vanishing path. We obtain four arcs, each of which is a union of two images of the corresponding vanishing path. The result is shown in \fref{fig:P3-b}.

For every value $z=z_1$ along an arc, we have the Riemann surface fiber $P(x,y,z_1)=0$, which contains several 1-cycles. One of them pinches off as we approach the two endpoints of the arc. The fibration of the vanishing 1-cycle over the arc defines an $S^2$. We conclude that each arc in \fref{fig:P3-b} represents an $S^2$.

How do we determine the intersection number between two of these $S^2$'s? We will first focus on the absolute values of the intersection numbers and consider their signs in the coming section. If the $S^2$'s meet at a branch point, since both of them shrink to a point there, the local intersection number must be one. If instead the corresponding arcs intersect in the bulk, i.e. away from the branch points, their local intersection number is inherited from that of the 1-cycle on the Riemann surface fiber. 

For local $\mathbb{CP}^3$, the Riemann surface fiber away from the branch points is always the one of local $\mathbb{CP}^2$. Hence, modulo signs, all the bulk intersection numbers are equal to 3. Adding up all contributions, we can summarize the (absolute value of) the intersection numbers as 
\begin{align}
|\langle C_i , C_{i+1} \rangle | = 4 \,,
\quad 
| \langle C_i , C_{i+2} \rangle | = 6 \,.
\label{P3-int-abs}
\end{align}
The same analysis applies to the $x$- and $y$-tomographies. In this case, due to the symmetry among $x$, $y$ and $z$ of the Newton polynomial \eref{P_CP3}, the $x$- and $y$ tomographies are identical to \fref{fig:P3-b}. The identity of the fields associated to each intersection might differ between tomographies, as we will show in \fref{xyz-tomographies_CP3}.

We have not explained how to distinguish between chiral and Fermi fields yet. If we compare the intersection numbers with the known quiver for local $\mathbb{CP}^3$ \cite{GarciaCompean:1998kh,Franco:2015tna}, we immediately recognize that the 4's correspond to chiral fields and the 6's correspond to Fermi fields.

\subsection{Chiral vs Fermi Fields from Oriented Intersections \label{sec:field-type}} 

Intersections between two vanishing cycles give rise to bifundamental fields between two gauge nodes. We will now explain how to distinguish between chiral and Fermi multiplets. In string theory, these fields arise from massless open string modes localized at the intersection. In the current context, a short answer to the field type question is that the signs of oriented intersections determine the types of fields at the intersections, as follows:
\beq
\begin{array}{lcccc}
\mbox{Fermi} & : & \langle \mathcal{C}_i, \mathcal{C}_j \rangle & > & 0  \\[.15cm]
\mbox{Chiral} & : & \langle \mathcal{C}_i, \mathcal{C}_j \rangle & < & 0 
\end{array}
\label{signs_field_type}
\eeq
In appendix \ref{appendix:angles} we present an explicit analysis of the open string spectrum in the simpler context of branes intersecting at $SU(n)$ angles in flat space, which leads to the above rule. 

The particular form of the mirror CY geometry, $uv - P(x,y,z) = 0$, allows us to simplify the discussion. The orientation of the intersection of the 4-cycles $\mathcal{C}_i$ in the mirror CY$_4$ is completely determined by that of the corresponding 2-cycles $C_i$ on the surface $\Sigma=W^{-1}(0)$.

\subsubsection*{Definition of the Intersection Number}

Consider an oriented plane and two oriented curves $C_1$ and $C_2$ on the plane intersecting transversely at a point $p$. We can measure the angle from the tangent vector of $C_1$ to that of $C_2$ in the counterclockwise direction such that the value lies in $(-\pi,0)\cup (0,+\pi)$. We define the oriented intersection number at the point, $\langle C_1,  C_2\rangle_p$ to be the sign of the angle. Equivalently, $\langle C_1,  C_2\rangle_p$ is $+1$ if $C_2$ crosses $C_1$ from the right to the left. For a symplectic manifold with a symplectic form $\omega_{mn}$, we can also write 
\begin{align}
\langle C_1,  C_2\rangle_p =\mathrm{sgn}[\omega_{mn} (v_1)^m (v_2)^n]_p \,, 
\label{ori-R2}
\end{align}
where $v_1$ and $v_2$ are the tangent vectors of $C_1$ and $C_2$ at the intersection, respectively. 

\begin{figure}[htbp]
	\centering
	\includegraphics[height=2.2cm]{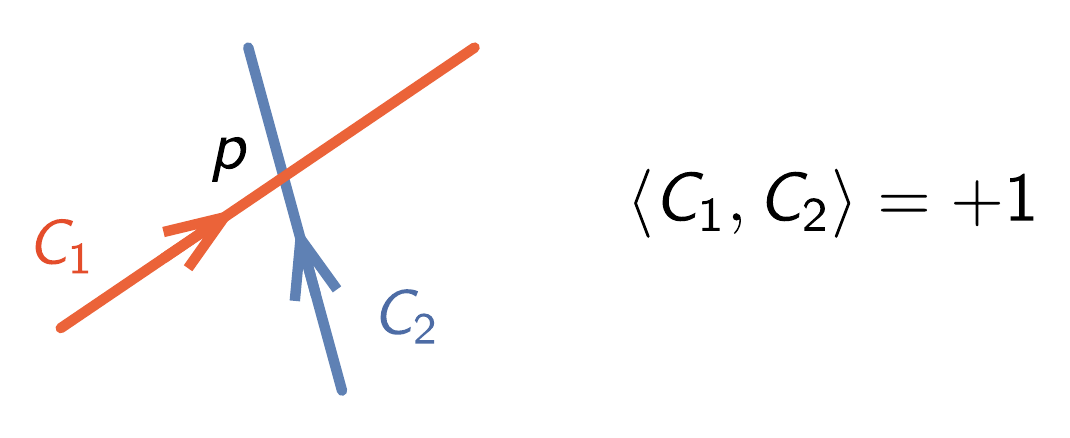}
\caption{Oriented intersection of two curves on a plane 
}
	\label{fig:Ori-1}
\end{figure}

Now consider oriented 2-cycles intersecting transversely in a symplectic 4-manifold. 
The tangent plane of each 2-cycle is given by a bi-vector $(v_i)^{mn} = - (v_i)^{nm}$. 
A natural generalization of \eqref{ori-R2} is to use the volume form $\mathrm{vol}=\omega^2/2$, 
\begin{align}
\langle C_1,  C_2\rangle_p =\mathrm{sgn}[\omega_{[mn} \omega_{pq]} (v_1)^{mn} (v_2)^{pq}]_p \,.
\label{ori-R4}
\end{align}

As an example, consider $\mathbb{C}^2$ with the standard K\"ahler form as the symplectic form. 
Let $(x,y)$ be the complex coordinates. Take $C_1$ the holomorphic cycle $y=0$ and 
$C_2$ the holomorphic cycle $x=0$. They intersect at the origin $O$. Clearly, $\langle C_1, C_2 \rangle_O = +1$. 
If instead we consider $C_3$ and $C_4$, the Lagrangian manifolds defined by the bi-vectors
\begin{align}
\partial_{\mathrm{Re}(x)} \wedge \partial_{\mathrm{Re}(y)}
\,, \quad 
\partial_{\mathrm{Im}(x)} \wedge \partial_{\mathrm{Im}(y)} \,,
\label{slag-ex}
\end{align}
then we find that $\langle C_3, C_4 \rangle_O = -1$.

In what follows, we will focus on Lagrangian 2-cycles of a K\"ahler 4-manifold. 
When the manifold is locally described by two complex coordinates, $(x,y)$, 
a Lagrangian 2-cycle $C_i$ is locally a product of a curve on the $x$-plane 
and a curve on the $y$-plane. Let us denote the curves by $c_i^x$ and $c_i^y$. The intersection number of a pair of 2-cycles follows from 
the intersection numbers of the component curves as 
\begin{align}
\langle C_1,  C_2\rangle = -\langle c_1^x,  c_2^x \rangle \langle c_1^y,  c_2^y \rangle  \,.
\label{ori-R4R2}
\end{align}
The overall minus sign in \eqref{ori-R4R2} comes from the fact that 
\begin{align}
\begin{split}
\frac{1}{2}\omega^2 &= 
d \mathrm{Re}(x) \wedge d \mathrm{Im}(x) \wedge  d \mathrm{Re}(y) \wedge d \mathrm{Im}(y) 
\\
&= - 
d \mathrm{Re}(x) \wedge d \mathrm{Re}(y) \wedge  d \mathrm{Im}(x) \wedge d \mathrm{Im}(y) \,.
\end{split}
\end{align}

\subsubsection{From $n$-cycles to $(n-2)$-cycles \label{cycle-reduction}} 

In section \sref{section_mirror_general}, we saw how, topologically, the $S^n$'s $\mathcal{C}_i$ can be viewed as a double fibration of $S^{n-2}$'s $C_i$ and an $S^1$ over vanishing paths $\gamma_i$ on the $W$-plane. 
In order to determine the orientation of intersections between a pair of $\mathcal{C}_i$'s, we need additional information on their geometry. 

Let us recall a few facts regarding Calabi-Yau manifolds and their special Lagrangian submanifolds (see e.g. \cite{joyce2007riemannian} and references therein). The coefficients of the Newton polynomial are complex structure moduli of the mirror Calabi-Yau. Up to an overall scaling, a unique Calabi-Yau metric exists for each value of the complex moduli. The holomorphic $n$-form $\Omega$ is unique up to an overall complex constant. The constant can be fixed by requiring that the Lagrangian $n$-cycles are calibrated by $\mathrm{Re}(\Omega)$. 

In principle, the precise loci of the Lagrangian $n$-cycles can be obtained by solving some partial differential equation. In practice, the explicit form of the metric, or the loci of the Lagragian cycles are out of reach. Even without these geometric data, we can deduce some general features of the intersections from the holomorphic $n$-form. In particular, we can see the relation between $\langle \mathcal{C}_i, \mathcal{C}_j \rangle$ in CY$_n$ and $\langle C_i, C_j \rangle$ in $\Sigma$. Let us discuss the CY$_3$ and CY$_4$ in some detail.

\subsubsection*{CY$_3$}

The coordinates $(u,v,x,y)$ span $\mathbb{C}^2 \times (\mathbb{C}^*)^2$. The mirror CY$_3$ is the vanishing locus of the function, 
\begin{align}
f = uv - P(x,y) \,.
\end{align}
Let us adopt a notation for logarithmic coordinates: 
\begin{align}
x = e^X\,, \quad 
y = e^Y\,, \quad 
u = e^U\,, \quad 
v = e^V\,. \quad 
\label{log-coord}
\end{align} 
The holomorphic 3-form $\Omega$ of the CY$_3$ is given by    
\begin{align}
\Omega = i \frac{du \wedge dv \wedge dX \wedge dY }{df} \,.
\label{CY3-Omega-0}
\end{align}
If we eliminate $u$, we obtain 
\begin{align}
\Omega =  i \frac{dv \wedge dX \wedge dY}{\partial f/\partial u} 
= i\, \frac{dv}{v} \wedge dX \wedge dY 
= i\, dV \wedge dX \wedge dY  \,.
\label{CY3-Omega-1}
\end{align}
The overall factor of $i$ is included such that 
\begin{align}
\mathrm{Re}{(\Omega)} = d\mathrm{Im}(V) \wedge d\mathrm{Im}(X)  \wedge d\mathrm{Im}(Y) + \mbox{(other terms)}\,.
\end{align}

The representation \eqref{CY3-Omega-1} is convenient near a critical point, where the local geometry of an $S^3$ is described by a disc embedded in the $(X,Y)$-space times a circle in the $v$-plane. In the $S^1 \times S^1$ fibration of the $S^3$ over a vanishing path on the $W$-plane, the $S^1$ from the $(u,v)$ space can be parametrized by 
\begin{align}
u = \sqrt{W} e^{-it} \,, 
\quad 
v = \sqrt{W} e^{it} \,, 
\quad t \in \mathbb{R}\,.
\end{align}
The factor $\sqrt{W}$ depends on the vanishing path, but $d\mathrm{Im}(V) = dt$ is common to all 3-cycles. The remaining part of $\mathrm{Re}(\Omega)$ is precisely the area 2-form on the coamoeba torus. Since all the 3-cycles are calibrated by $\mathrm{Re}(\Omega)$, it follows that all the corresponding vanishing cycles on the coamoeba torus, as illustrated in \fref{coamoeba-tiling_P2} and \fref{coamoeba-tiling_F0_1}, should have the same orientation. 

Alternatively, if we eliminate $x$, we obtain 
\begin{align}
\Omega = 
\frac{du \wedge dv \wedge dY}{\partial f/\partial X} 
= du \wedge dv \wedge d\zeta \,.
\label{CY3-Omega-2}
\end{align}
Here, $d \zeta$ is the holomorphic 1-form on the Riemann surface $W^{-1}(\gamma(t))$.
This representation is convenient near $W=0$, where the local geometry of the $S^3$ is described by a disc embedded in the $(u,v)$-space times a 1-cycle on the Riemann surface. When we examine the intersection between a pair of cycles, the angles from three complex planes should add up to zero (mod $2\pi$). The expression \eqref{CY3-Omega-2} implies a correlation between the angle between two vanishing paths and the angle between the corresponding vanishing cycles on $\Sigma$. We will illustrate the idea with concrete examples shortly.

\subsubsection*{CY$_4$}

Again, we can write $\Omega$ in two different ways:
\begin{align}
\begin{split}
\Omega 
&= dV \wedge dX \wedge dY \wedge dZ 
\\
&= du \wedge dv \wedge \Omega_2 \,.
\end{split}
\end{align}
In the second line, $\Omega_2$ is the holomorphic 2-form of the surface $W^{-1}(\gamma(t))$. 
The overall factor is chosen such that 
\begin{align}
\mathrm{Re}{(\Omega)} = d\mathrm{Im}(V) \wedge d\mathrm{Im}(X)  \wedge d\mathrm{Im}(Y) \wedge d\mathrm{Im}(Z) + \mbox{(other terms)}\,.
\end{align}
Again, the first representation implies that the vanishing cycles should be oriented uniformly in the coamoeba $T^3$. The second representation implies a correlation between the angles on the $W$-plane and the angles on $\Sigma$. 

One notable consequence of the reduction is that there is an overall sign flip in the oriented intersection numbers 
\begin{align}
\langle \mathcal{C}_i , \mathcal{C}_j \rangle 
= - \langle C_i , C_j \rangle \,,
\end{align}
to be discussed below. Then, In terms of the intersections between 2-cycles, \eref{signs_field_type} becomes
\beq
\begin{array}{lcccc}
\mbox{Fermi} & : & \langle C_i, C_j \rangle & < & 0  \\[.15cm]
\mbox{Chiral} & : & \langle C_i, C_j \rangle & > & 0 
\end{array}
\label{signs_field_type2}
\eeq

\subsubsection{Brane Tilings Revisited \label{sec:dimer-ori}}

In brane tilings, the vanishing cycles should be oriented uniformly in the coamoeba torus. Let us choose a convention such that they have counterclockwise orientation. In the tomography plots, it is important to consider the multiple sheets of complex planes to describe the whole punctured Riemann surface. Along a cycle, we can determine the maxima and minima of $\mathrm{arg}(x)$ and $\mathrm{arg}(y)$. 
Then we can orient the cycle by going along the following sequence:
\begin{align}
\mathrm{max}(\mathrm{arg}(x)) \rightarrow 
\mathrm{max}(\mathrm{arg}(y)) \rightarrow 
\mathrm{min}(\mathrm{arg}(x)) \rightarrow
\mathrm{min}(\mathrm{arg}(y)) \rightarrow
\mathrm{max}(\mathrm{arg}(x)) \,.
\label{dimer-cycle-ori}
\end{align} 

\begin{figure}[htbp]
	\centering
	\includegraphics[height=6cm]{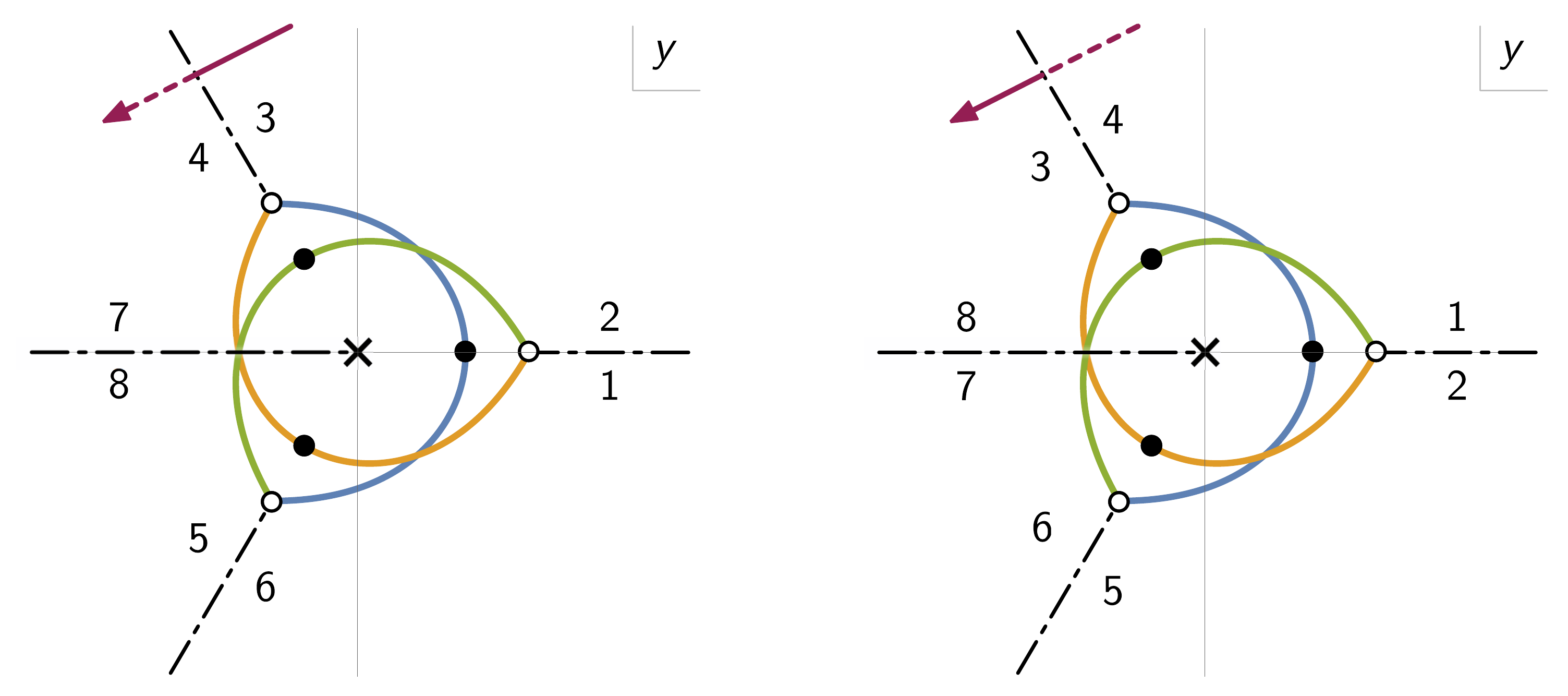}
\caption{Two sheets for the Riemann surface for local $\mathbb{CP}^2$ with its vanishing cycles.}
	\label{fig:P2-ori-1}
\end{figure}

Consider local $\mathbb{CP}^2$. The surface $P(x,y)=0$ is a genus-1 Riemann surface 
with three punctures. Two copies of the $y$-plane, reproduced in \fref{fig:P2-ori-1}, cover the whole surface. 
When gluing the two copies of the $y$-plane, care should be taken along the branch cuts. 
Naively, it seems that the angle around a branch point on the $y$-plane is $2\pi$, but it is actually $\pi$.   
In addition, we should remember that, since $y$ is a $\mathbb{C}^*$ variable, $y=0$ is also a branch point. As we go around $y=0$, $x$ comes back to itself after two turns. Hence, the correct angle around $y=0$ is also $\pi$. 

\begin{figure}[htbp]
	\centering
	\includegraphics[height=5cm]{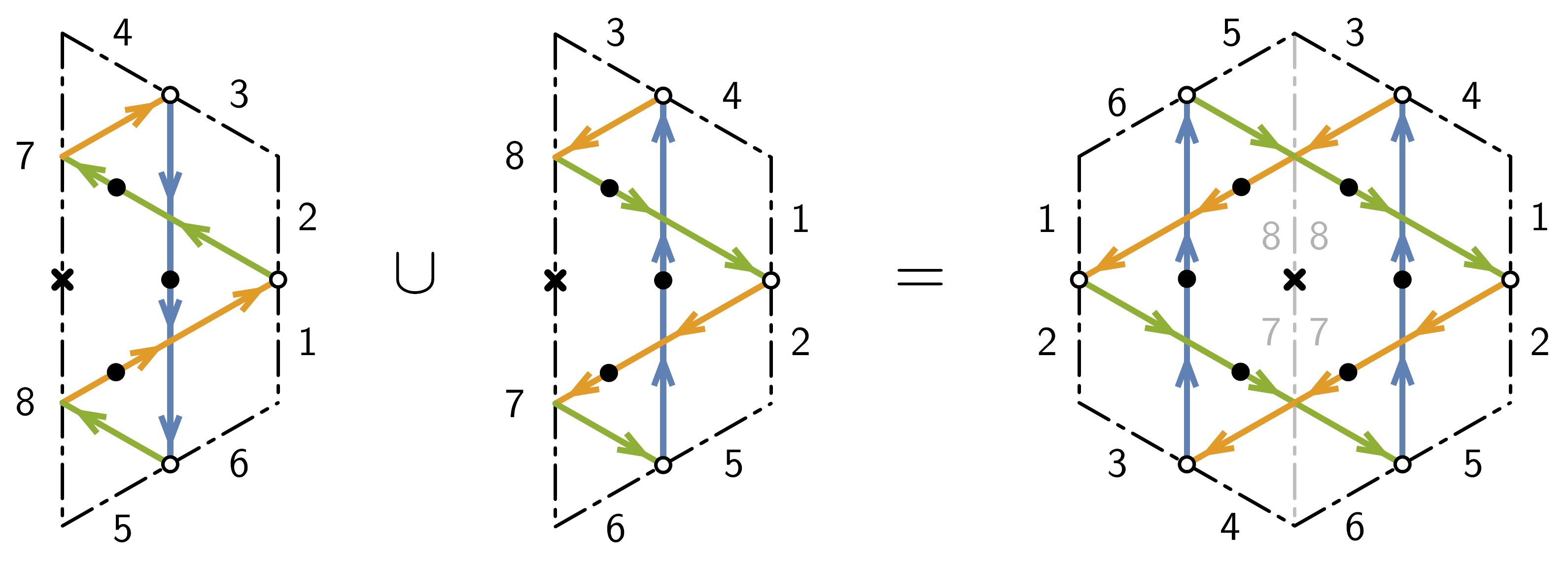}
\caption{Cutting and gluing two sheets to form a Riemann surface. The labels for the branch cuts conform to the convention of \fref{fig:P2-ori-1}. The vertices of the ``hexagon" in this figure, after identifying the edges, constitute the two punctures at $y=\infty$. }
	\label{fig:P2-ori-2}
\end{figure}

Taking into account the angles around branch points, we can cut the two copies of the $y$-plane and glue them again to form a genus-1 Riemann surface. The process is illustrated in \fref{fig:P2-ori-2}. In the figure, the cycles are oriented according to the rule \eqref{dimer-cycle-ori}. 

In the labeling convention of \fref{fig:P2-1a}, the signs of the intersections are given by 
\begin{align}
\mathrm{sgn} \langle C_3, C_2 \rangle = \mathrm{sgn} \langle C_2, C_1 \rangle = \mathrm{sgn}  \langle C_1, C_3 \rangle = + 1\,.
\label{RGB-int}
\end{align}
In brane tilings, this sign determines the orientation of chiral fields stretching between pairs of gauge groups. 

When an intersection occurs at a branch point, we can perform a simple computation to determine the sign. For example, consider the intersection of $C_1$ and $C_2$ on the real axis of the $y$-plane as shown in \fref{fig:P2-ori-1}. Recall that the branch point lies at $(x_0,y_0) = (-2^{-1/3},2^{2/3})$. We can examine a small neighborhood of the branch point by substituting\footnote{The log coordinates $X$ and $Y$ here are related to those in \eqref{log-coord} by constant shifts and rescalings.}
\begin{align}
x = x_0 e^{\epsilon X} \,, \quad y = y_0 e^{\epsilon^2 Y} \,.
\label{CY3-branch-local}
\end{align}
Then, $P(x,y)=0$ at the $\epsilon^2$ order gives
\begin{align}
Y = \frac{1}{3} X^2 \,.
\end{align}
In \fref{fig:P2-ori-1}, the two arcs approach the branch point along $\mathrm{arg}(Y) = \pm 2\pi/3$. 
Then, on the local $X$-plane, 
the two cycles lie along $\sqrt{3} \mathrm{Re}(X) \pm \mathrm{Im}(X) = 0$. Clearly, $C_1$ has $\mathrm{Im}(Y)\ge 0$ while $C_2$ has $\mathrm{Im}(Y)\le 0$. 
Then, the rule \eqref{dimer-cycle-ori} fixes the signs of $\partial_{\mathrm{Im}(X)}$ for the two tangent vectors $v_1$ and $v_2$. Altogether, these conditions fix the tangent vectors up to a rescaling by positive real numbers to be
\begin{align}
v_1 = \partial_{\mathrm{Re}(X)} + \sqrt{3} \partial_{\mathrm{Im}(X)}\,,
\quad 
v_2 = \partial_{\mathrm{Re}(X)} - \sqrt{3} \partial_{\mathrm{Im}(X)}\,.
\label{dimer-ori-ex}
\end{align}
It follows from \eqref{ori-R2} that $\langle C_1, C_2\rangle = -1$, in agreement with \eqref{RGB-int}.

\begin{figure}[htbp]
	\centering
	\includegraphics[height=6cm]{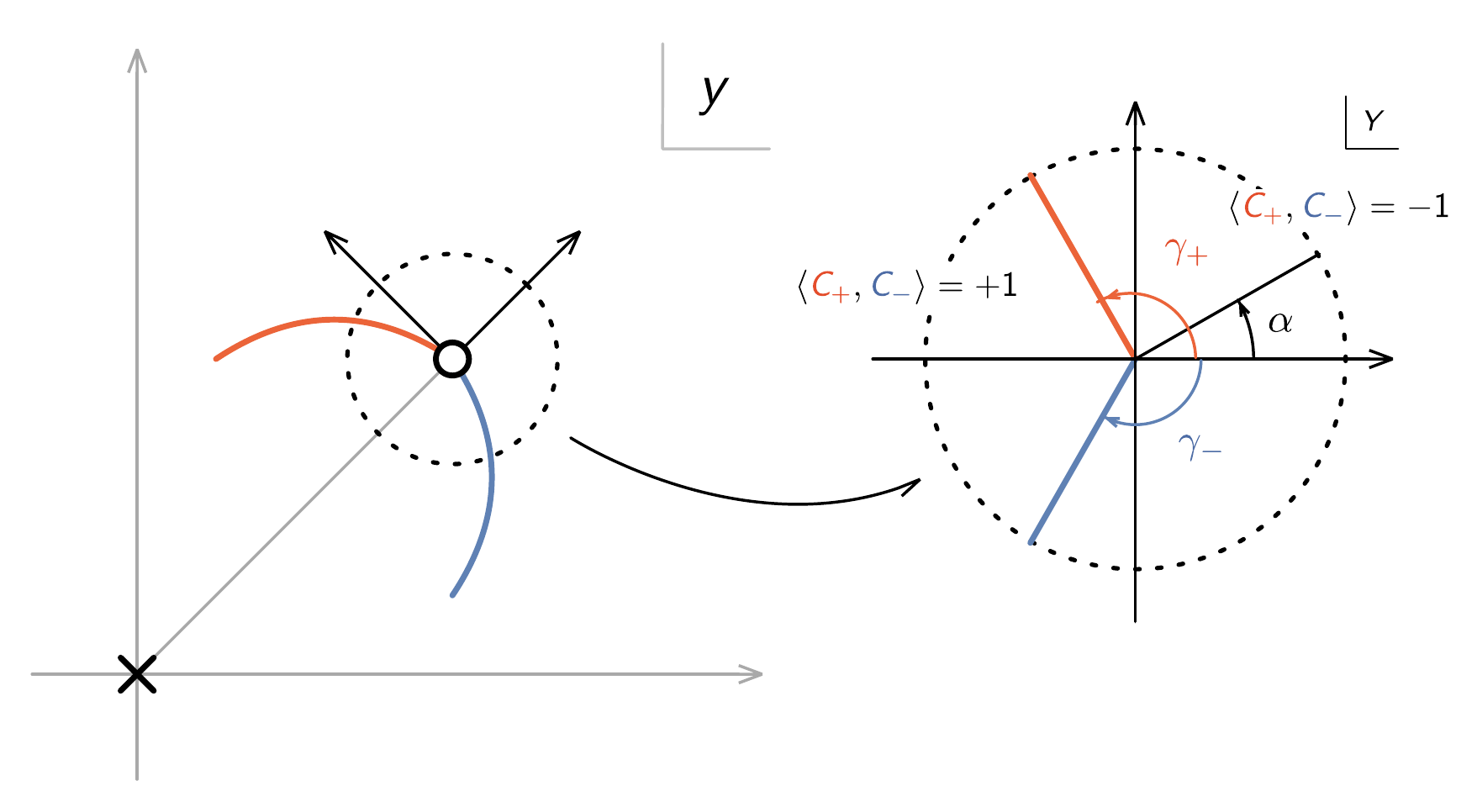}
\caption{Determination of $\langle C_+, C_- \rangle$ at a branch point. The sign depends on the relative position of $\alpha = \arg(A)$ from \eqref{dimer-A} with respect to the two angles for $C_\pm$.
}
	\label{dimer-sign}
\end{figure}

It is straightforward to generalize this analysis. At an arbitrary branch point on the $y$-plane, the substitution \eqref{CY3-branch-local} at the $\epsilon^2$ order will give
\begin{align}
Y = A X^2 \,, 
\label{dimer-A}
\end{align}
where $A$ is some complex number. The magnitude of $A$ is irrelevant. Let $\alpha$ be $\arg(A)$, taken in the range $-\pi < \alpha < \pi$. Focusing on the two cycles intersecting at the branch point, 
let us call $C_+$ the one with $\mathrm{Im}(Y)\ge 0$ and $C_-$ be the one with $\mathrm{Im}(Y)\le 0$. As we move along each of the cycles slightly away from the branch point, let the angles on the $Y$-plane be 
\begin{align}
\gamma_+ = \mathrm{arg}(\delta Y)_{C_+} \,, 
\quad 
\gamma_- = \mathrm{arg}(\delta Y)_{C_-}\,.
\end{align}
By definition, they are in the range $-\pi < \gamma_- < 0 < \gamma_+ < \pi$. Repeating the argument around \eqref{dimer-ori-ex}, one can show that the desired sign is given by 
\begin{align}
\mathrm{sgn} \langle C_+ , C_- \rangle_{y_0} = \mathrm{sgn}[(\gamma_+ - \alpha )(\gamma_- - \alpha)] \,.
\label{dimer-sign-final}
\end{align}
Pictorially, one can understand the sign as in \fref{dimer-sign}.

\subsubsection{Oriented Intersections in Brane Brick Models}

As we argued in \sref{cycle-reduction}, we want to orient the vanishing cycles such that, when projected onto the coamoeba, 
they become spheres with uniform orientation. We can use the $z$-tomography to introduce something similar to 
the polar angle in spherical coordinates. We orient every arc on the $z$-plane from the maximum (north pole) to the minimum (south pole) of $\mathrm{arg}(z)$ 
in analogy with the latitude on a sphere. At each point $z_1$ along the arc, we have the Riemann surface fiber $P(x,y,z_1)=0$. One of the 1-cycles of the Riemann surface shrinks at the two poles to form the sphere. Then, we orient the 1-cycle as we did in brane tilings.

\subsubsection*{Bulk Intersections}

\begin{figure}[htbp]
	\centering
	\includegraphics[height=8cm]{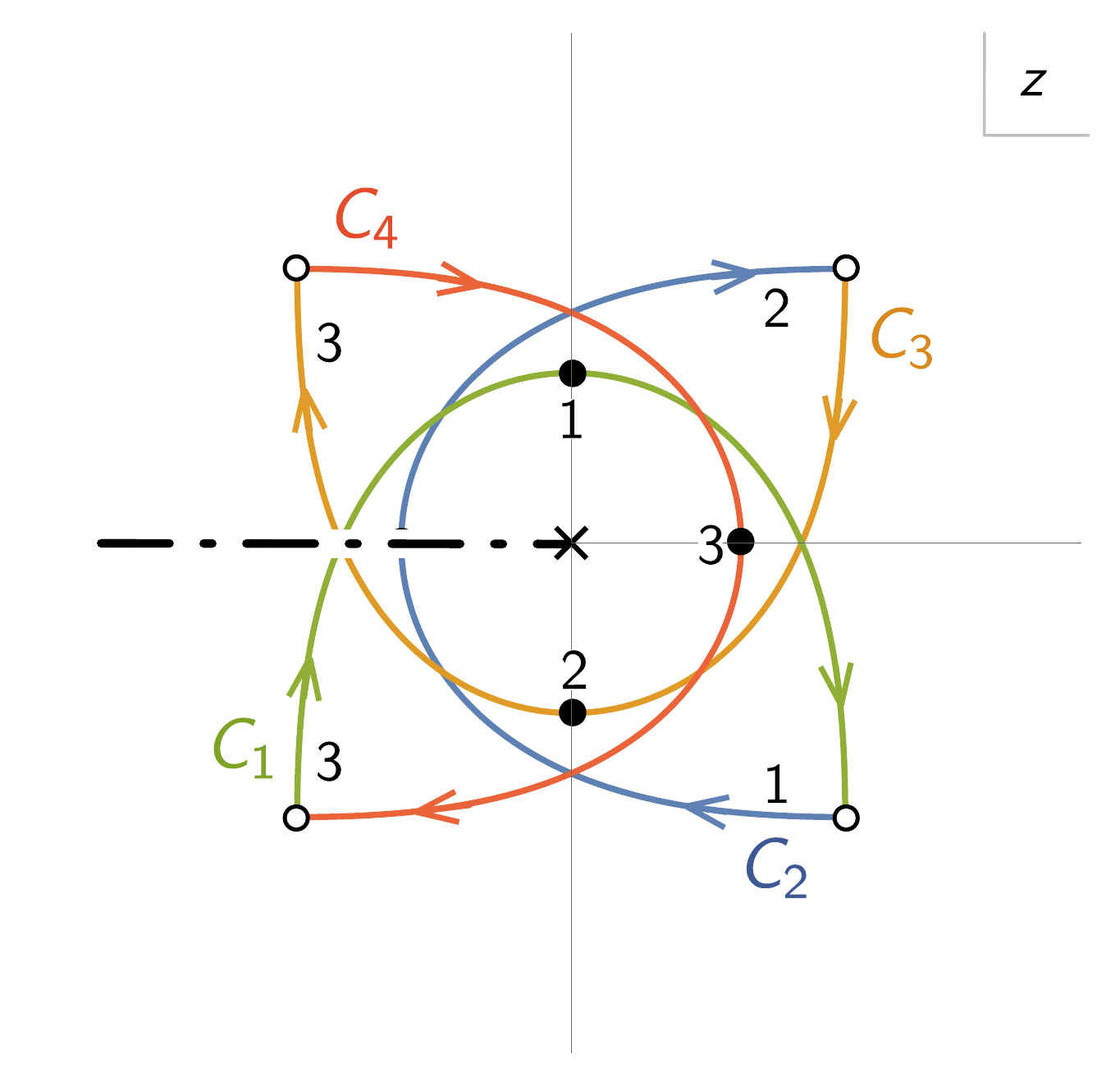}
\caption{Vanishing 2-cycles local $\mathbb{CP}^3$ as 1-cycles fibered over arcs on the $z$-plane.}
	\label{fig:P3-ori-1}
\end{figure}

When a pair of 2-cycles intersect away from a branch point on the $z$-plane, we can use the formula \eqref{ori-R4R2} (with $x$ replaced by $z$) to compute their intersection number. The $\langle c_1^z, c_2^z \rangle$ factor is manifest on the $z$-plane. 
To compute $\langle c_1^y, c_2^y \rangle$, we should figure out which cycle on the fiber corresponds to each of the arcs. The analysis gets complicated by the fact that the Riemann surface fiber undergoes a non-trivial monodromy as we go around $z=0$. The 1-cycles on the fiber get permuted. In the local $\mathbb{CP}^3$ example, the fibration of 1-cycles over arcs, their monodromy, and how the monodromy acts on the 1-cycles were explained in \cite{Futaki:2014mpa}. The result is summarized in \fref{fig:P3-ori-1}.

Let us, for example, compute $\langle C_1 , C_3 \rangle$ at the intersection at $\mathrm{arg}(z)=\pi/2$. We have 
\begin{align}
\langle C_1 , C_3 \rangle 
= - \langle c_1^z , c_3^z \rangle \langle c_1^y , c_3^y \rangle
= - (+1) \langle \hat{C}_3 , \hat{C}_2 \rangle = -1 \,.
\end{align}
We read off the first factor directly from the figure. For the second factor, we put hats on the 1-cycles of the Riemann surface fiber to distinguish them from 2-cycles, and applied \eqref{RGB-int} based on the information about the 1-cycle fibers from the figure. 

As another example, let us compute $\langle C_1 , C_2 \rangle$ at the intersection at $\mathrm{arg}(z)=\pi/4$. In this case, 
\begin{align}
\langle C_1 , C_2 \rangle 
= - \langle c_1^z , c_2^z \rangle \langle c_1^y , c_2^y \rangle
= - (+1) \langle \hat{C}_3 , \hat{C}_1 \rangle = +1 \,.
\end{align}
Again, we used the information on the fibration in \fref{fig:P3-ori-1} 
and applied \eqref{RGB-int}.

\subsubsection*{Branch Point Intersections}

When an intersection occurs at a branch point, we can compute the sign locally, without worrying about the Riemann surface fibration and monodromy. This method is a straightforward generalization of the discussion in the second half of section \sref{sec:dimer-ori}.

Take the $z$-tomography. Let $(x_0,y_0,z_0)$ be a branch point. We magnify a neighborhood of 
the branch point by setting 
\begin{align}
x = x_0 e^{\epsilon X} \,, \quad y = y_0 e^{\epsilon Y} 
\,, \quad z = z_0 e^{\epsilon^2 Z} \,,
\label{CY4-epsilon}
\end{align}
and taking the $\epsilon^2$ order term of $P(x,y,z)=0$. In general, the local geometry near a branch point takes the form
\begin{align}
Z = 
\begin{pmatrix}
X & Y
\end{pmatrix} 
H 
\begin{pmatrix}
X \\ Y
\end{pmatrix} \,.
\end{align}
where $H$ is a complex $2 \times 2$ symmetric matrix. 
We can diagonalize $H$ by an $SL(2,\mathbb{R})$ basis change in the $(X,Y)$ coordinates. For a local analysis, the distinction between $SL(n,\mathbb{R})$ and $SL(n,\mathbb{Z})$ is immaterial. Assume 
the diagonal form of the local geometry,
\begin{align}
Z = A X^2 + B Y^2 \,.
\label{local_branch_point}
\end{align}
Define $\alpha = \arg(A)$ and $\beta = \arg(B)$ in the range $-\pi < \alpha, \beta < \pi$. 

There are two cycles intersecting at the branch point: $C_+^{(z_0)}$ with $\mathrm{Im}(Z) \ge 0$ 
and $C_-^{(z_0)}$ with $\mathrm{Im}(Z) \le 0$. 
For instance, in the upper right quadrant of \fref{fig:P3-ori-1}, $C_2$ is identified with $C_+^{(z_0)}$ and $C_3$ is identified with $C_-^{(z_0)}$. 

As we move along each cycle slightly away from the branch point, let the angles on the $Z$-plane be 
\begin{align}
\gamma_+ = \mathrm{arg}(\delta Z)_{C_+} \,, 
\quad 
\gamma_- = \mathrm{arg}(\delta Z)_{C_-} 
\quad 
(-\gamma < \beta_- < 0 < \gamma_+ < \pi) \,.
\end{align}
In view of \eqref{dimer-sign-final}, we introduce
\begin{align}
\sigma(A) = \mathrm{sgn}[(\gamma_+ - \alpha)(\gamma_- - \alpha)] \,,
\quad  
\sigma(B) = \mathrm{sgn}[(\gamma_+ - \beta )(\gamma_- - \beta)]
\label{def-sigmaAB}
\end{align}

There are two bits of information we can extract from these quantities:
\begin{enumerate}

\item
The sign of the intersection, which distinguishes Fermi fields from chiral fields as in \eref{signs_field_type2}, is given by 
\begin{align}
\mathrm{sgn}\langle C_+ , C_- \rangle_{z_0} 
= \sigma(A) \sigma(B) \,.
\label{br-int-formula}
\end{align} 

\item 
When the field is chiral, its orientation is determined by
\begin{align}
[C_+,C_-]_{z_0} = \frac{1}{2} (\sigma(A) + \sigma(B))\,.
\label{br-int-chiral}
\end{align} 
We adopt a convention in which the chiral field in the quiver diagram is represented by an arrow from $C_+$ to $C_-$ when $[C_+,C_-]_{z_0}$ is positive. 

\end{enumerate}

\subsubsection*{Sample Computations}

Let us look at the branch point intersection between $C_2$ and $C_3$ in the upper-right quadrant of the $z$-plane in \fref{fig:P3-ori-1}. 
Recall that the coordinates of the intersection point are $(x_0,y_0,z_0)=   (-3)^{3/4} \times (-1/3,-1/3,1)$. The expansion \eqref{CY4-epsilon} yields
\begin{align}
4Z = X^2+ XY + Y^2 \,.
\label{P3-chiral-branch}
\end{align}
The two ``eigenvalues" ($A$ and $B$ of \eqref{local_branch_point}) are real and positive. 
According to \eqref{def-sigmaAB}, \eqref{br-int-formula} and \eqref{signs_field_type2}, 
this intersection gives a chiral field. Up to an $SL(3,\mathbb{Z})$ change of basis, all intersections between $C_i$ and $C_{i+1}$ can be brought to the form \eqref{P3-chiral-branch}. 
Thus, we conclude that all of them give chiral fields.

We can determine the sign of the intersection between $C_i$ and $C_{i+2}$ in a similar way. To begin with, we note that the coordinates of the intersections can be completely determined by symmetries. Recall from \eqref{P3-crit} that the centers of the cycles are located at $(x^*,y^*,z^*) = i^{n}(1,1,1)$ for $C_n$. There are eight ``mid-points" between $C_4=(1,1,1)$ and $C_2=(-1,-1,-1)$: $M_{42}^{\pm, \pm,\pm} = (\pm i, \pm i, \pm i)$. 
Two among them, $M_{42}^{+++}$ and $M_{42}^{---}$, are identified with $C_1$ and $C_3$. The other six are the intersection points between $C_4$ and $C_2$. Take $(x_0,y_0,z_0) = (i,i,-i)$ for example. Near this point, we can perform a basis change motivated by the root system of $SU(4)$ Lie algebra,  
\begin{align}
x= \frac{tu}{s} \,, \quad 
y = \frac{us}{t} \,, \quad 
z = \frac{st}{u} \,.
\end{align}
Expanding around $(s_0,t_0,u_0) = (1,1,i)$ by setting $s=e^{\epsilon S}$, $t=e^{\epsilon T}$, $u = e^{\epsilon^2 U}$, we find 
\begin{align}
U = ST \,. 
\end{align}
The two eigenvalues $A$ and $B$ are both real but one is positive and the other is negative. So, this intersection gives rise to a Fermi field. 

\begin{figure}[ht]
\centering
\includegraphics[height=12cm]{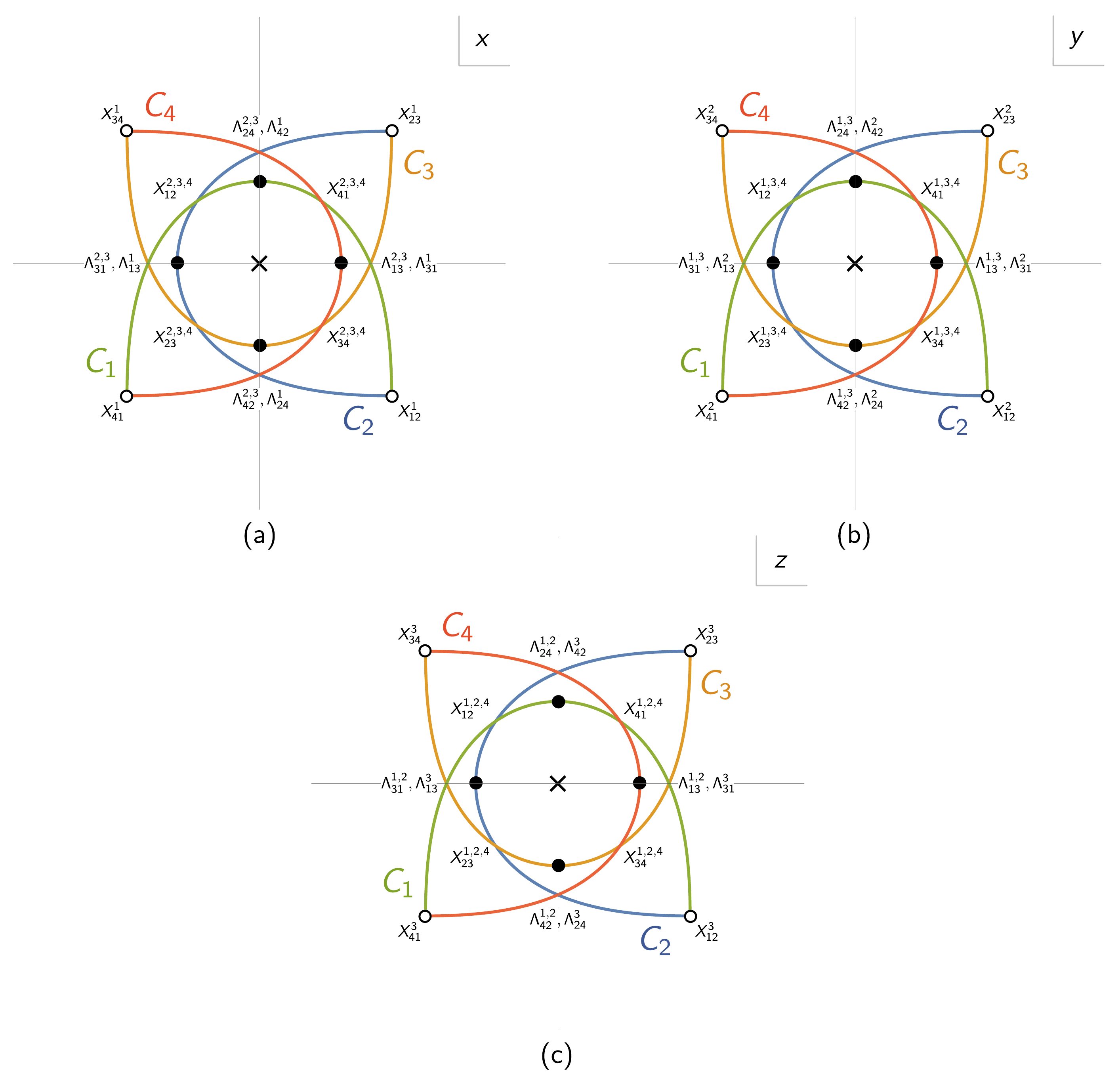}
 \label{fig:test1}  \label{fig:test2}
\caption{a) $x$-, b) $y$- and c) $z$-tomographies for local $\mathbb{CP}^3$. We indicate the fields associated with each intersection. 
}
\label{xyz-tomographies_CP3}
\end{figure}

Applying the techniques we just explained to all bulk and branch point intersections of the local $\mathbb{CP}^3$ model, we obtain
\beq
\begin{array}{ccccl}
\langle C_i , C_{i+1} \rangle & =&  +4 & \ \ \ \to \ \ \ & \mbox{Chiral} \\[.15 cm]
\langle C_i , C_{i+2} \rangle & = & -6 &\ \ \ \to \ \ \ & \mbox{Fermi}
\label{P3-ori-final}
\end{array}
\eeq
These results agree perfectly with the known matter content of the gauge theory \cite{GarciaCompean:1998kh,Franco:2015tna}. In fact, we can be more precise and identify the Fermi and chiral fields associated to each intersection in the $x$-, $y$- and $z$-tomographies, as shown in \fref{xyz-tomographies_CP3}. The notation is such that the superindices are labels identifying fields with the same gauge quantum numbers, which are indicated by the subindices. As anticipated, even though the cycles look identical in the three tomographies due to the symmetry between $x$, $y$ and $z$ in \eref{P_CP3}, the locations of the fields distinguish between them.  

\subsection{Interaction Terms}

We have identified all the gauge groups (cycles) and the oriented fields (intersections). Now we present a graphical method to construct all the $J$- and $E$-terms, thereby completing our prescription 
for deriving the gauge theory from the mirror geometry.  
 
In the coamoeba diagram, the 4-cycles become 3-balls with $S^2$ boundaries. 
The precise shape of each $S^2$ is not important. However, we know the coordinates of the center, given by the coamoeba projection of the corresponding critical point, and of the intersection points, given by the coamoeba projections of the intersections in the three tomographies shown in \fref{xyz-tomographies_CP3}. It suffices to draw an $S^2$ ``anchored" at all of its intersection points. 

For a fixed $S^2$, we can mark the fields (intersections) on the surface. Looking from the exterior of $S^2$, we can distinguish the field types by assigning the symbols $\otimes$, $\odot$, $\circ$ to incoming chiral, outgoing chiral and Fermi, respectively. 
Let $n_\mathrm{in}$, $n_\mathrm{out}$, $n_\mathrm{F}$ be the number of intersections of each type. For anomaly cancellation and consistency under triality, they should satisfy the constraints:
\begin{align}
n_\mathrm{in}, n_\mathrm{out}, n_\mathrm{F} \ge 2\,, \quad 
n_\mathrm{in} + n_\mathrm{out} - n_\mathrm{F} = 2\,.
\end{align}

\begin{figure}[H]
\begin{center}
\resizebox{0.3\hsize}{!}{
\includegraphics[trim=0cm 0cm 0cm 0cm,totalheight=6 cm]{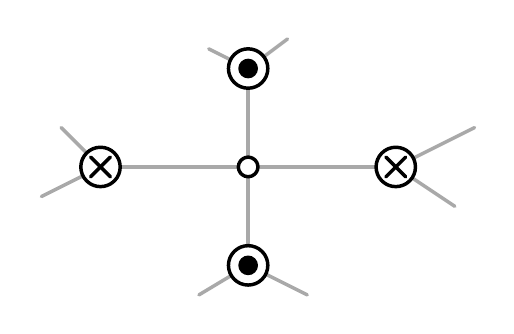}
}  
\caption{A Fermi point connected to nearby chiral points.
\label{Marking}}
 \end{center}
 \end{figure}

To determine the adjacency among the marked points, we draw a graph on the $S^2$ by connecting neighboring points. In generic theories, the following properties are satisfied:\footnote{Theories that do not satisfy these properties, e.g. with (Fermi)-(Fermi) connections, exist \cite{Franco:2016nwv}. Such theories are connected to generic ones by triality.}
\begin{enumerate}

\item 
Allowed connections: (chiral)-(Fermi) or (incoming chiral)-(outgoing chiral). Equivalently, connections between fields of the same type are forbidden. 

\item 
All Fermi fields are 4-valent. Their connections to incoming and outgoing chirals should alternate as shown in \fref{Marking}. 

\end{enumerate}

\begin{figure}[H]
\begin{center}
\resizebox{0.3\hsize}{!}{
\includegraphics[trim=0cm 0cm 0cm 0cm,totalheight=6 cm]{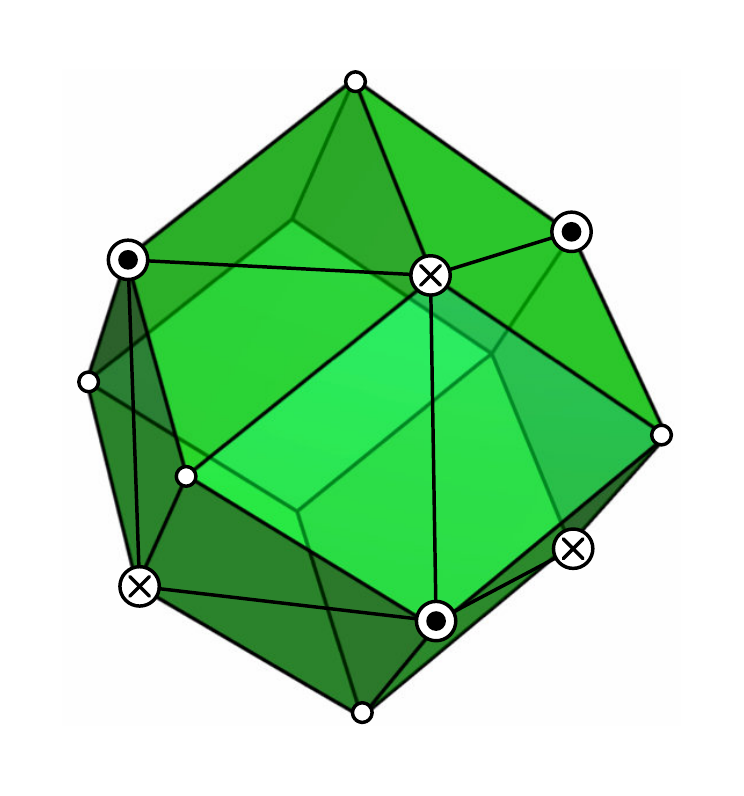}
}  
\caption{The graph on an $S^2$ for each of the four gauge groups in local $\mathbb{CP}^3$. 
To avoid clutter, we leave the points on the back unmarked.
\label{RD-marked}}
 \end{center}
 \end{figure}

For the local $\mathbb{CP}^3$ model, these rules determine that the each of the four gauge groups is associated to a rhombic dodecahedron, as shown in \fref{RD-marked}. This polyhedron has already appeared in the context of abelian orbifolds of $\mathbb{C}^4$, like local $\mathbb{CP}^3$, in the phase boundary approach to brane brick models \cite{Franco:2015tna,Franco:2015tya}. A new feature is that now its faces are triangulated by (incoming)-(outgoing) connections.

\begin{figure}[H]
\begin{center}
\resizebox{0.75\hsize}{!}{
\includegraphics[trim=0cm 0cm 0cm 0cm,totalheight=10 cm]{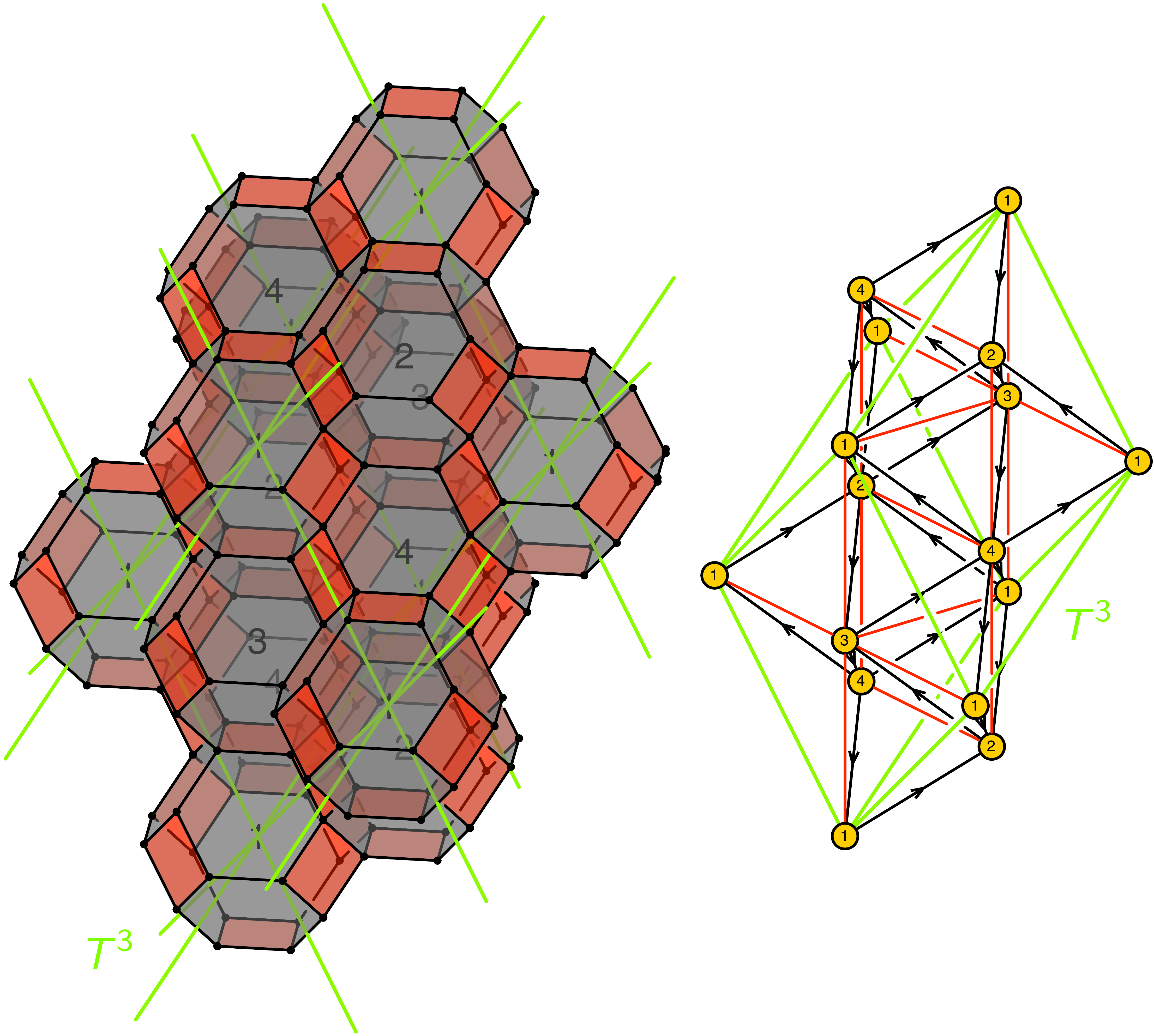}
}  
\vspace{-.3cm}\caption{Brane brick model and periodic quiver for local $\mathbb{CP}^3$.
\label{periodic_quiver_cp3}}
 \end{center}
 \end{figure}

To go from the marked $S^2$'s to brane bricks, we simply take the dual graph on the $S^2$. In taking the dual, we should ``inflate" the bricks such that there are no voids between them. For the local $\mathbb{CP}^3$ model, we obtain the brane brick model based on truncated octahedra that is shown in \fref{periodic_quiver_cp3}. 

Once the brane brick model is constructed, we can read the $J$- and $E$-terms of the gauge theory using the dictionary presented in section \sref{section_BBMs} \cite{Franco:2015tna,Franco:2015tya}.  All the chiral fields meeting at an edge form a plaquette. Every Fermi face has four edges. Two of these plaquettes become the corresponding $J$-term and the two others give rise to the $E$-term. They are given by:
\beq
\begin{array}{lcccc}
& &  J & &  E \\
\Lambda_{13}^{1} & :\ \ \ & X^2_{34} \cdot X^3_{41}-X^3_{34} \cdot X^2_{41} & \ \ \ \ & X^4_{12} \cdot X^1_{23}-X^1_{12} \cdot X^4_{23} \\
\Lambda_{13}^{2} & :\ \ \ & X^3_{34} \cdot X^1_{41}-X^1_{34} \cdot X^3_{41} & \ \ \ \ & X^4_{12} \cdot X^2_{23}-X^2_{12} \cdot X^4_{23} \\
\Lambda_{13}^{3} & :\ \ \ & X^1_{34} \cdot X^2_{41}-X^2_{34} \cdot X^1_{41} & \ \ \ \ & X^4_{12} \cdot X^3_{23}-X^3_{12} \cdot X^4_{23} \\
\Lambda_{31}^{1} & :\ \ \ & X^2_{12} \cdot X^3_{23}-X^3_{12} \cdot X^2_{23} & \ \ \ \ & X^4_{34} \cdot X^1_{41}-X^1_{34} \cdot X^4_{41} \\
\Lambda_{31}^{2} & :\ \ \ & X^3_{12} \cdot X^1_{23}-X^1_{12} \cdot X^3_{23} & \ \ \ \ & X^4_{34} \cdot X^2_{41}-X^2_{34} \cdot X^4_{41} \\
\Lambda_{31}^{3} & :\ \ \ & X^1_{12} \cdot X^2_{23}-X^2_{12} \cdot X^1_{23} & \ \ \ \ & X^4_{34} \cdot X^3_{41}-X^3_{34} \cdot X^4_{41} \\
\Lambda_{24}^{1} & :\ \ \ & X^2_{41} \cdot X^3_{12}-X^3_{41} \cdot X^2_{12} & \ \ \ \ & X^4_{23} \cdot X^1_{34}-X^1_{23} \cdot X^4_{34} \\
\Lambda_{24}^{2} & :\ \ \ & X^3_{41} \cdot X^1_{12}-X^1_{41} \cdot X^3_{12} & \ \ \ \ & X^4_{23} \cdot X^2_{34}-X^2_{23} \cdot X^4_{34} \\
\Lambda_{24}^{3} & :\ \ \ & X^1_{41} \cdot X^2_{12}-X^2_{41} \cdot X^1_{12} & \ \ \ \ & X^4_{23} \cdot X^3_{34}-X^3_{23} \cdot X^4_{34} \\
\Lambda_{42}^{1} & :\ \ \ & X^2_{23} \cdot X^3_{34}-X^3_{23} \cdot X^2_{34} & \ \ \ \ & X^4_{41} \cdot X^1_{12}-X^1_{41} \cdot X^4_{12} \\
\Lambda_{42}^{2} & :\ \ \ & X^3_{23} \cdot X^1_{34}-X^1_{23} \cdot X^3_{34} & \ \ \ \ & X^4_{41} \cdot X^2_{12}-X^2_{41} \cdot X^4_{12} \\
\Lambda_{42}^{3} & :\ \ \ & X^1_{23} \cdot X^2_{34}-X^2_{23} \cdot X^1_{34} & \ \ \ \ & X^4_{41} \cdot X^3_{12}-X^3_{41} \cdot X^4_{12} 
\end{array} 
\eeq

\section{Additional Examples \label{sec:examples}}

\label{section_examples}

In this section we present two additional examples illustrating the use of mirror symmetry to construct periodic quivers and, equivalently, brane brick models. Further examples are collected in appendix \ref{section_Q111_phases}.

\subsection{$M^{3,2}$}

\begin{figure}[H]
	\centering
	\includegraphics[width=5cm]{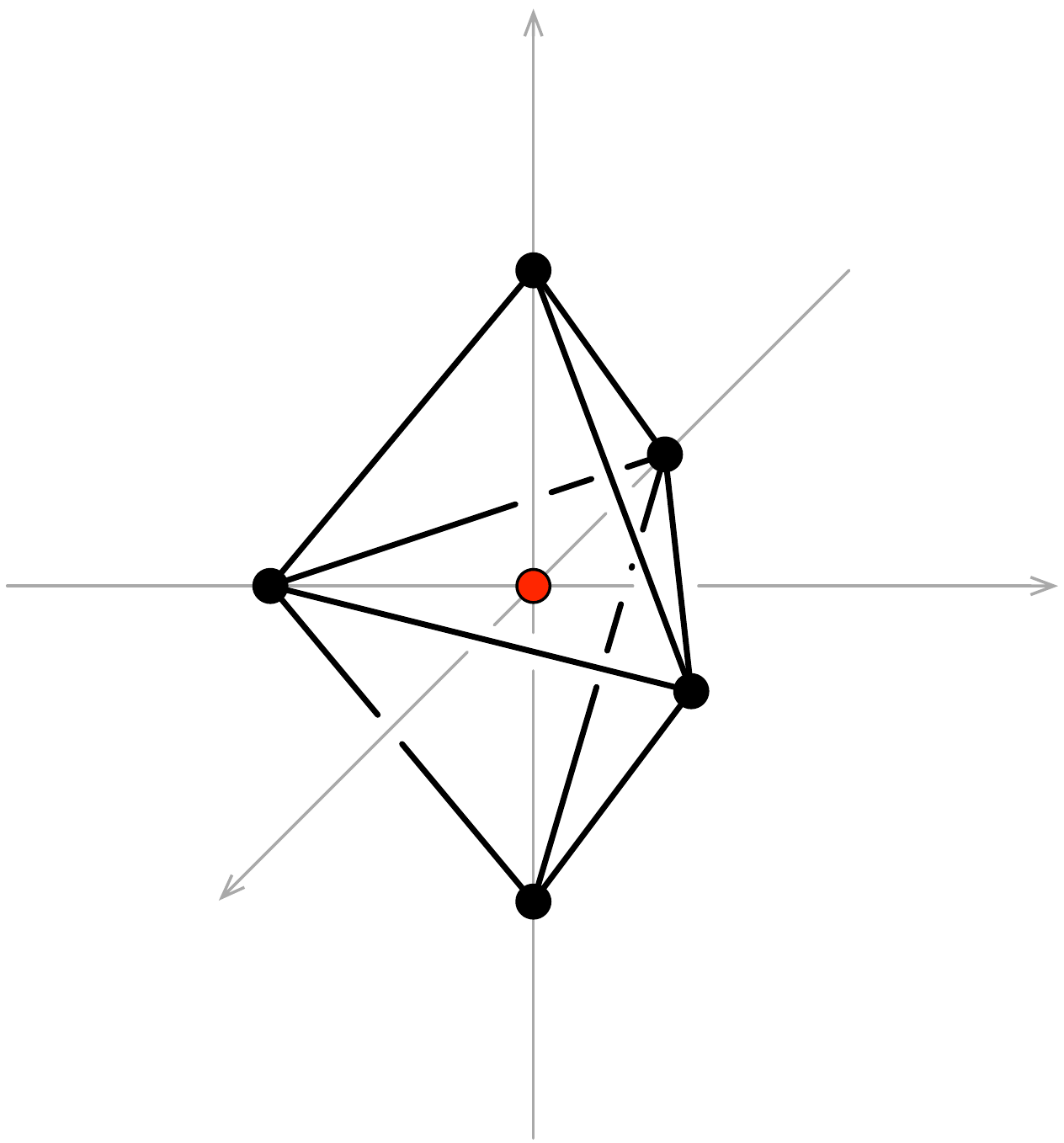}
\caption{Toric diagram for $M^{3,2}$.}
	\label{toric_M32}
\end{figure}

The toric diagram of $M^{3,2}$ is given in \fref{toric_M32}. The gauge theory for this geometry has not yet appeared in the literature and we will use mirror symmetry to construct it for the first time. 

\begin{figure}[ht]
	\centering
	\includegraphics[height=6cm]{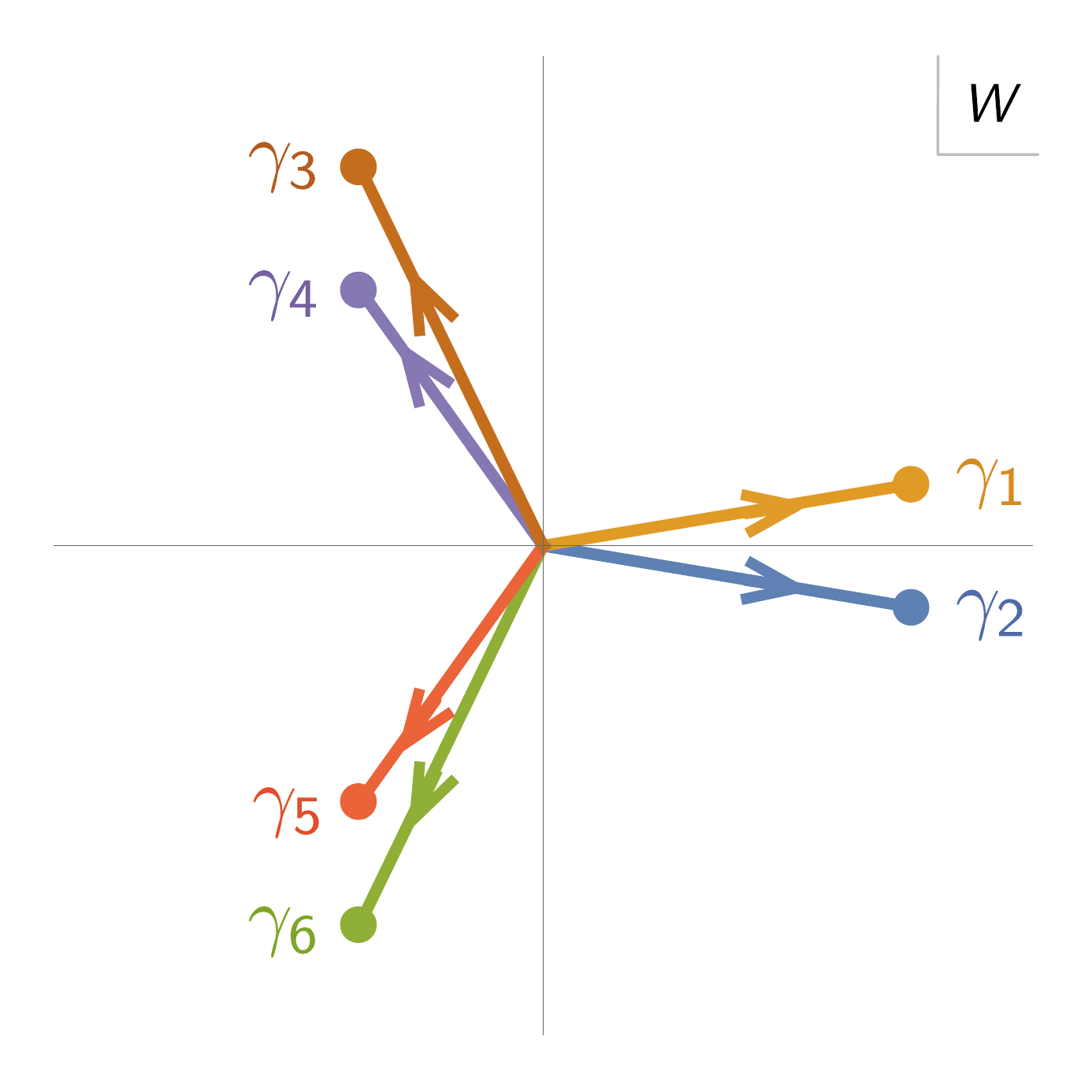}
\caption{Vanishing paths for phase A of $M^{3,2}$.}
	\label{fig:M32-A-a}
\end{figure}

$M^{3,2}$ has various toric phases related by triality. Determining all of them is straightforward, but beyond the scope of this paper. Here we derive one of these phases, which we call phase A. It corresponds to the following choice of Newton polynomial
\begin{align}
P(x,y,z) = x + y + \frac{1}{x y} + \frac{i}{4} \left(z + \frac{1}{z} \right) \,.
\label{M32-ansatz}
\end{align}
The six critical points of $P$ are
\begin{align}
(x^*=y^*,z^*) = (\omega^a , \pm 1) \,,\quad W^* = 3x^* + \frac{i}{2} z^* \,, \quad (a=0,1,2) \,,
\end{align}
with $\omega=-(-1)^{1/3}$. The vanishing paths are shown in \fref{fig:M32-A-a}.

\begin{figure}[h]
	\begin{center}
	\resizebox{\hsize}{!}{
	\includegraphics[height=9cm]{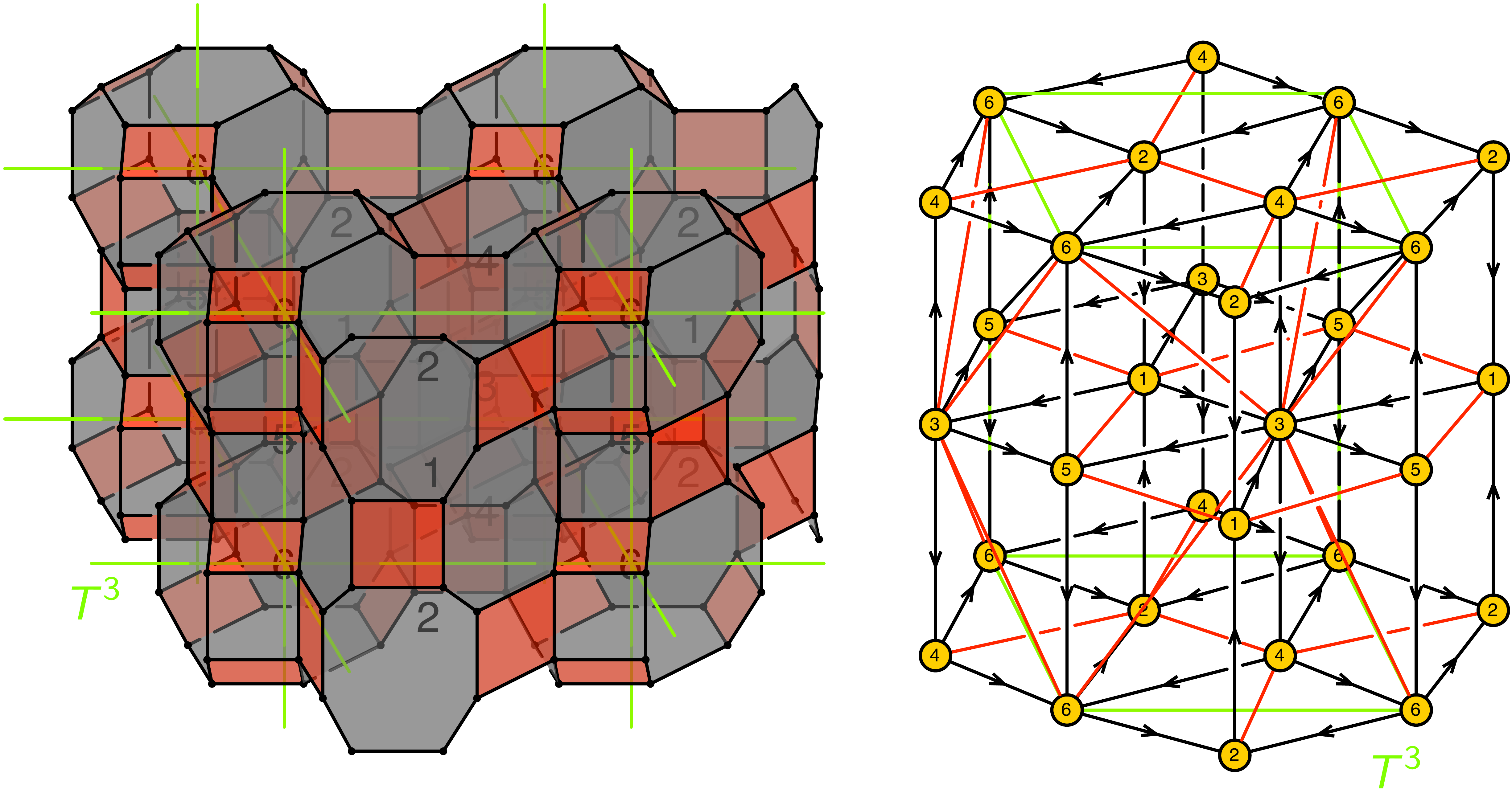}
	}
\caption{Brane brick model and periodic quiver for phase A of $M^{3,2}$. 
} 
	\label{fig:M32-B-quiver}
	\end{center}
\end{figure}

We will now explain how to use the three tomographies to construct the periodic quiver and the brane brick model, which will be presented in \fref{fig:M32-B-quiver}. In order to facilitate comparison with the final result, we will indicate the fields associated to every intersection in the tomographies.

\paragraph{The $z$-tomography.}

\begin{figure}[htbp]
	\centering
	\includegraphics[height=6.5cm]{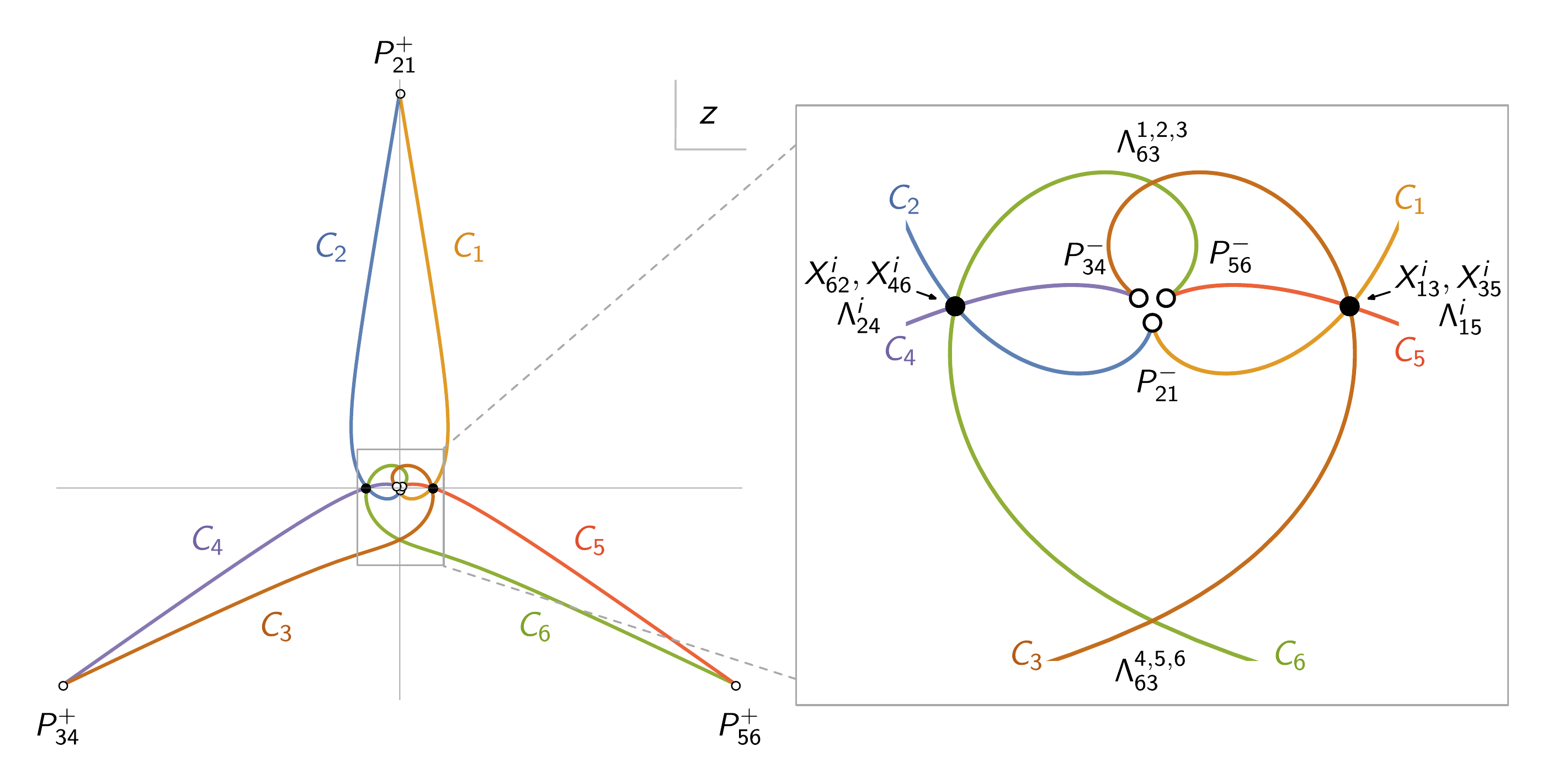}
\caption{The $z$-tomography for phase A of $M^{3,2}$. We indicate the fields associated with each intersection.}
	\label{fig:M32-B-b}
\end{figure}

\fref{fig:M32-B-b} shows the $z$-tomography. We see that all the intersections between nodes $(1,3,5)$, i.e. the matter fields connecting them, are located at $z=1$. Similarly, all intersections between nodes $(2,4,6)$ take place at $z=-1$. This suggests that we should organize the nodes in the periodic quiver into two layers along the $z$ direction. These layers consist of nodes $(1,3,5)$ at $\mathrm{arg}(z)=0$ and nodes $(2,4,6)$ at $\mathrm{arg}(z)=\pi$

Let us consider the intersections in detail, starting from the fields in the two layers along $z$. At $z=1$, $P(x,y,z=1)$ is the same as the Newton polynomial of local $\mathbb{CP}^2$, except that the origin of the $W$-plane is shifted by $i/2$. This implies that $|\langle C_1,C_3 \rangle| = |\langle C_3,C_5 \rangle| = |\langle C_5,C_1 \rangle| = 3$. Similarly, at $z=-1$, we find $|\langle C_2,C_4 \rangle| = |\langle C_4,C_6 \rangle| = |\langle C_6,C_2 \rangle| = 3$.

Let us move to the intersections between pairs of nodes in different layers. The fields connecting the two layers are located on the upper and lower half-planes of the $z$-tomography. From \eref{fig:M32-B-b}, we see that $C_{2k-1}$ and $C_{2k}$ intersect at the two branch points situated at their endpoints. Recalling that branch point intersections always have multiplicity equal to 1, we conclude that $|\langle C_1,C_2\rangle| = |\langle C_3,C_4\rangle| = |\langle C_5,C_6\rangle| = 2$.

Finally, the only other pair of cycles allowed to intersect in \fref{fig:M32-B-b} is $(C_3,C_6)$. 
These intersections occur when $z$ is pure imaginary and $x$ and $y$ are real. A detailed analysis shows that at $\mathrm{arg}(z) = \pi/2$ there are three intersections. The signs for $(x,y)$ at the intersections are $(+,-)$, $(-,+)$, $(-,-)$. Similarly, there are three more intersections at $\mathrm{arg}(z) = -\pi/2$.

Summarizing, the non-vanishing intersection numbers are
\begin{align}
\begin{split}
&|\langle C_1,C_3\rangle| = |\langle C_3,C_5\rangle| = |\langle C_5,C_1\rangle| = 3 \,,
\\  
&|\langle C_2,C_4\rangle| = |\langle C_4,C_6\rangle| = |\langle C_6,C_2\rangle| = 3 \,,
\\
&|\langle C_1,C_2\rangle| = |\langle C_3,C_4\rangle| = |\langle C_5,C_6\rangle| = 2 \,,
\\  
&|\langle C_3,C_6\rangle| = 6 \,.
\end{split}
\end{align}

Let us determine the signs of the intersections, i.e. the types of fields. Let us consider the branch point intersections between $C_{2k-1}$ and $C_{2k}$. The intersections between $C_1$ and $C_2$ can be read off from the $z$-tomography. They are
\begin{align}
(x_0,y_0,z_0) = (1,1,i(6 \pm \sqrt{37})) \,.
\end{align}
The local behavior around these points is given by
\begin{align}
\pm \frac{\sqrt{37}}{2} Z = X^2 + XY + Y^2 \,.
\end{align}
Thus, both fields are chiral and extend from $C_2$ to $C_1$. For the pairs $(C_3, C_4)$ and $(C_5, C_6)$, the local behavior around the intersections becomes
\begin{align}
Z = \pm \kappa_{\pm} (X^2 + XY + Y^2) \,,
\end{align}
where the real part of the constant $\kappa_\pm$ is negative. So, the field type is still chiral, but the orientation is opposite to that of $(C_1,C_2)$. In other words, the fields run from $C_3$ to $C_4$ and from $C_5$ to $C_6$. It is straightforward to repeat the analysis for all intersections. The result is summarized in the periodic quiver shown in \fref{fig:M32-B-quiver}.

\paragraph{The $x$- and $y$-tomographies.}

\begin{figure}[H]
\centering
\includegraphics[height=7.5cm]{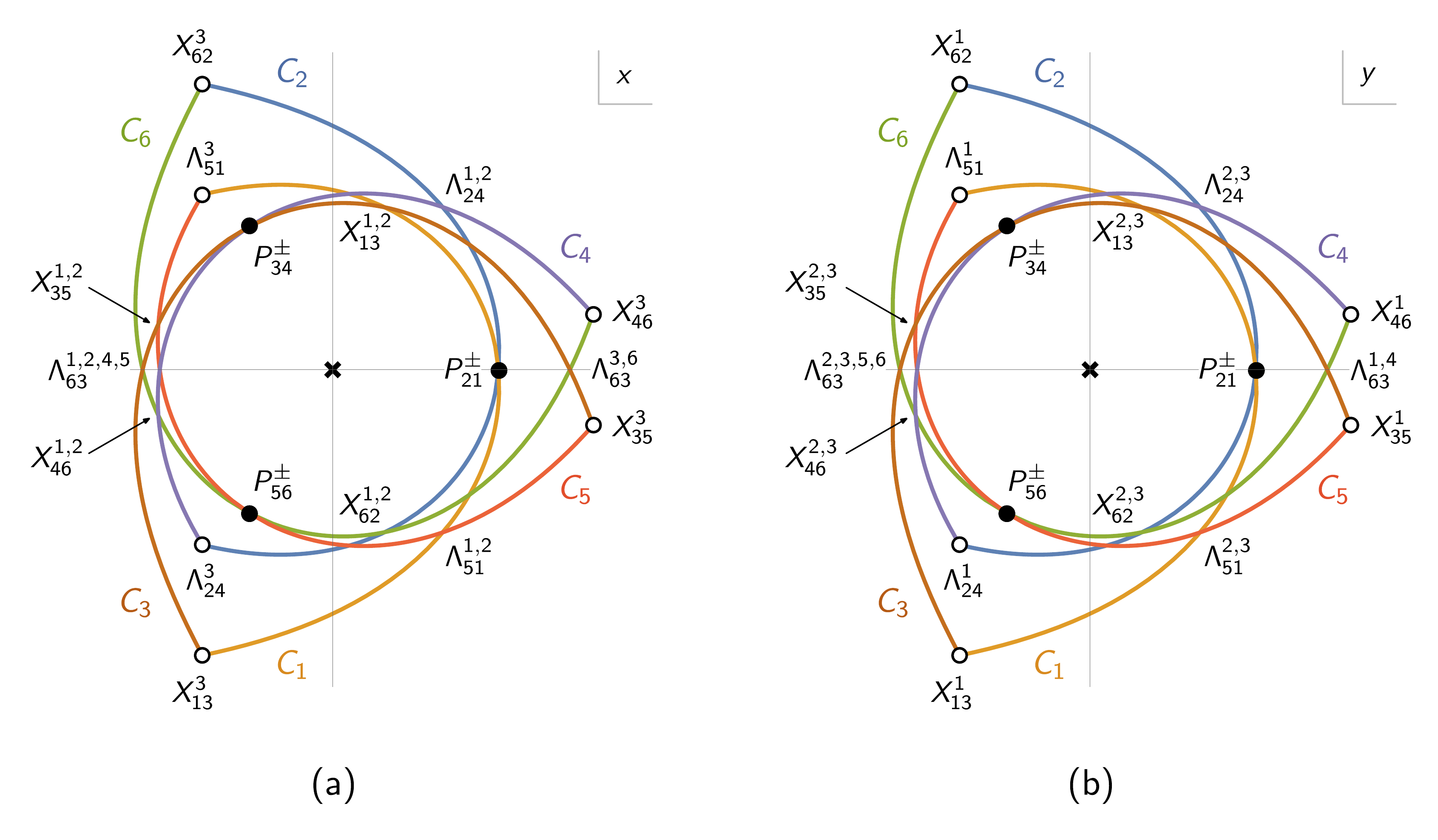} 
\caption{a) $x$- and b) $y$-tomographies for phase A of $M^{3,2}$. We indicate the fields associated with each intersection.}
\label{fig:M32-B-c}
\end{figure}

In order to determine the positions of the fields in the periodic quiver, it is necessary to consider also the $x$- and $y$-tomographies, which are given in \fref{fig:M32-B-c}. Note that, similarly to what occurs for local $\mathbb{CP}^3$, while the projections of the cycles on the $x$- and $y$-planes and the type and multiplicities of every intersection are identical due to the $x\leftrightarrow y$ invariance of \eref{M32-ansatz}, the labels of the corresponding fields differ.

\paragraph{Brane brick model and periodic quiver.}

With the previous analysis we can construct the brane brick model and periodic quiver for phase A of $M^{3,2}$, which are shown in \fref{fig:M32-B-quiver}. From them, we read the $J$- and $E$-terms of the theory:
\beq
\begin{array}{lcccc}
& &  J & &  E \\
\Lambda_{42}^{1} & :\ \ \ & P^{-}_{21} \cdot X^{1}_{13} \cdot P^{+}_{34}  - P^{+}_{21} \cdot X^{1}_{13} \cdot P^{-}_{34}    & \ \ \ \ & X^{3}_{46} \cdot X^{2}_{62}  - X^{2}_{46} \cdot X^{3}_{62} \\
\Lambda_{42}^{2} & :\ \ \ & P^{-}_{21} \cdot X^{2}_{13} \cdot P^{+}_{34}  - P^{+}_{21} \cdot X^{2}_{13} \cdot P^{-}_{34}    & \ \ \ \ & X^{1}_{46} \cdot X^{3}_{62}  - X^{3}_{46} \cdot X^{1}_{62} \\
\Lambda_{42}^{3} & :\ \ \ & P^{-}_{21} \cdot X^{3}_{13} \cdot P^{+}_{34}  - P^{+}_{21} \cdot X^{3}_{13} \cdot P^{-}_{34}    & \ \ \ \ & X^{2}_{46} \cdot X^{1}_{62}  - X^{1}_{46} \cdot X^{2}_{62} \\

\Lambda_{15}^{1} & :\ \ \ & P^{+}_{56} \cdot X^{1}_{62} \cdot P^{-}_{21}  - P^{-}_{56} \cdot X^{1}_{62} \cdot P^{+}_{21}    & \ \ \ \ & X^{3}_{13} \cdot X^{2}_{35}  - X^{2}_{13} \cdot X^{3}_{35} \\
\Lambda_{15}^{2} & :\ \ \ & P^{+}_{56} \cdot X^{2}_{62} \cdot P^{-}_{21}  - P^{-}_{56} \cdot X^{2}_{62} \cdot P^{+}_{21}    & \ \ \ \ & X^{1}_{13} \cdot X^{3}_{35}  - X^{3}_{13} \cdot X^{1}_{35} \\
\Lambda_{15}^{3} & :\ \ \ & P^{+}_{56} \cdot X^{3}_{62} \cdot P^{-}_{21}  - P^{-}_{56} \cdot X^{3}_{62} \cdot P^{+}_{21}    & \ \ \ \ & X^{2}_{13} \cdot X^{1}_{35}  - X^{1}_{13} \cdot X^{2}_{35} \\

\Lambda_{36}^{1} & :\ \ \ & X^{2}_{62} \cdot P^{+}_{21} \cdot X^{3}_{13}  - X^{3}_{62} \cdot P^{+}_{21} \cdot X^{2}_{13}    & \ \ \ \ & P^{-}_{34} \cdot X^{1}_{46}  - X^{1}_{35} \cdot P^{-}_{56} \\
\Lambda_{36}^{2} & :\ \ \ & X^{3}_{62} \cdot P^{+}_{21} \cdot X^{1}_{13}  - X^{1}_{62} \cdot P^{+}_{21} \cdot X^{3}_{13}    & \ \ \ \ & P^{-}_{34} \cdot X^{2}_{46}  - X^{2}_{35} \cdot P^{-}_{56} \\
\Lambda_{36}^{3} & :\ \ \ & X^{1}_{62} \cdot P^{+}_{21} \cdot X^{2}_{13}  - X^{2}_{62} \cdot P^{+}_{21} \cdot X^{1}_{13}    & \ \ \ \ & P^{-}_{34} \cdot X^{3}_{46}  - X^{3}_{35} \cdot P^{-}_{56} \\
\end{array} \nonumber
\eeq

\beq
\begin{array}{lcccc}
& &  J & &  E \\
\Lambda_{36}^{4} & :\ \ \ & X^{2}_{62} \cdot P^{-}_{21} \cdot X^{3}_{13}  - X^{3}_{62} \cdot P^{-}_{21} \cdot X^{2}_{13}    & \ \ \ \ & P^{+}_{34} \cdot X^{1}_{46}  - X^{1}_{35} \cdot P^{+}_{56} \\
\Lambda_{36}^{5} & :\ \ \ & X^{3}_{62} \cdot P^{-}_{21} \cdot X^{1}_{13}  - X^{1}_{62} \cdot P^{-}_{21} \cdot X^{3}_{13}    & \ \ \ \ & P^{+}_{34} \cdot X^{2}_{46}  - X^{2}_{35} \cdot P^{+}_{56} \\
\Lambda_{36}^{6} & :\ \ \ & X^{1}_{62} \cdot P^{-}_{21} \cdot X^{2}_{13}  - X^{2}_{62} \cdot P^{-}_{21} \cdot X^{1}_{13}    & \ \ \ \ & P^{+}_{34} \cdot X^{3}_{46}  - X^{3}_{35} \cdot P^{+}_{56} \\
\end{array}
~.~
\label{es1a2}
\eeq
Explicit computation of the mesonic moduli space of this theory confirms that it indeed corresponds to the geometry in \fref{toric_M32}.

\subsection{$Q^{1,1,1}/\mathbb{Z}_2$ \label{sec:Q111Z2-A}}

\begin{figure}[H]
	\centering
	\includegraphics[width=5cm]{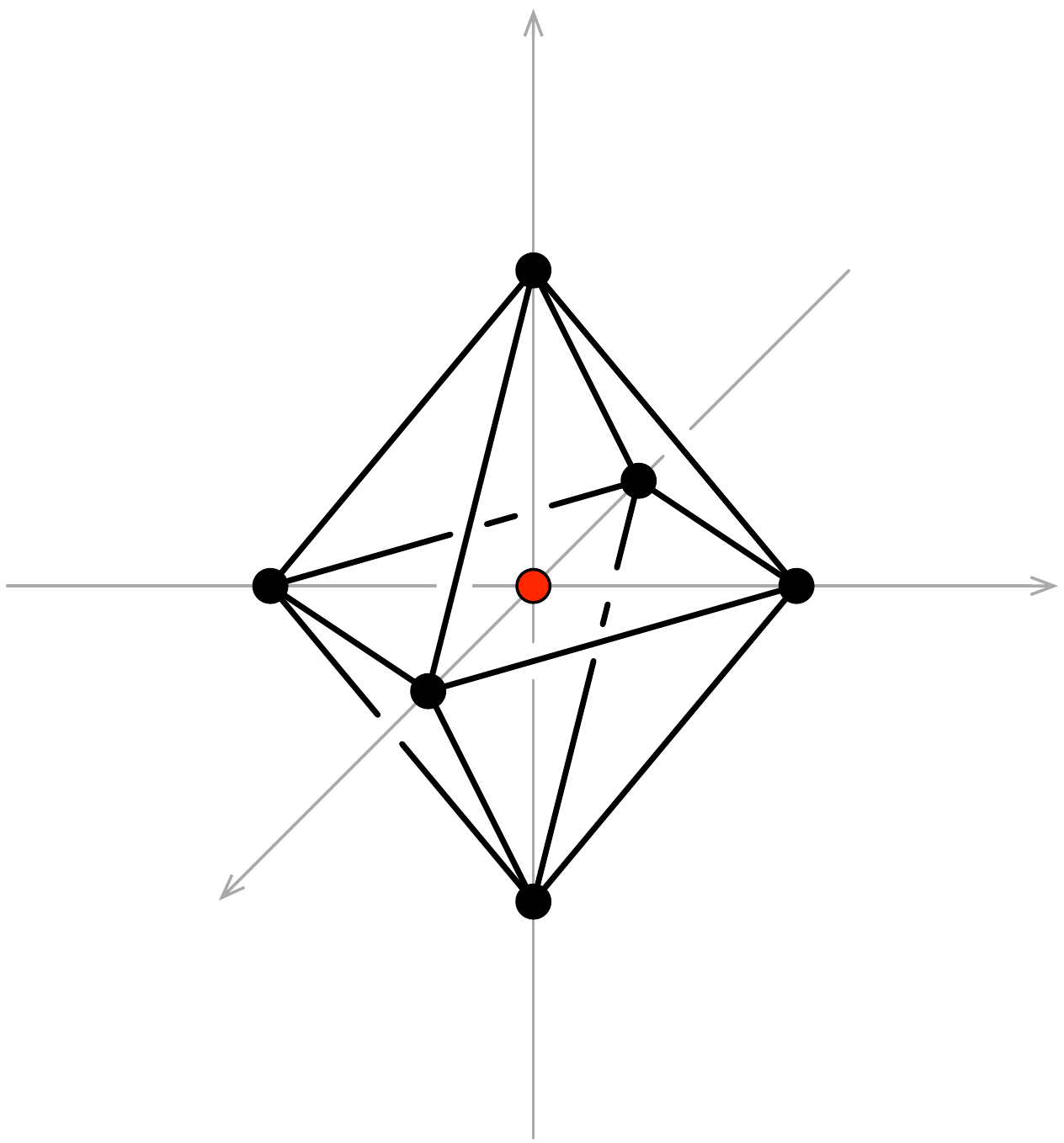}
\caption{Toric diagram for $Q^{1,1,1}/\mathbb{Z}_2$.}
	\label{toricq111z2}
\end{figure}

The toric diagram for $Q^{1,1,1}/\mathbb{Z}_2$ is shown in \fref{toricq111z2}. This geometry gives rise to several toric phases related by triality, whose study was initiated in \cite{Franco:2016nwv}. All of them are captured by appropriate choices of coefficients in the Newton polynomial. Let us consider 
\begin{align}
P(x,y,z) = x + \frac{1}{x} + 2 \left(y + \frac{1}{y} \right) + i \left(z + \frac{1}{z} \right) \,.
\label{P-Q111-ansatz}
\end{align}
This choice turns out to be closely related to the phases of $F_0$ we discussed in section \sref{section_F0}. It gives rise to phase A in the classification of \cite{Franco:2016nwv}.

\begin{figure}[ht]
\begin{center}
\includegraphics[width=8cm]{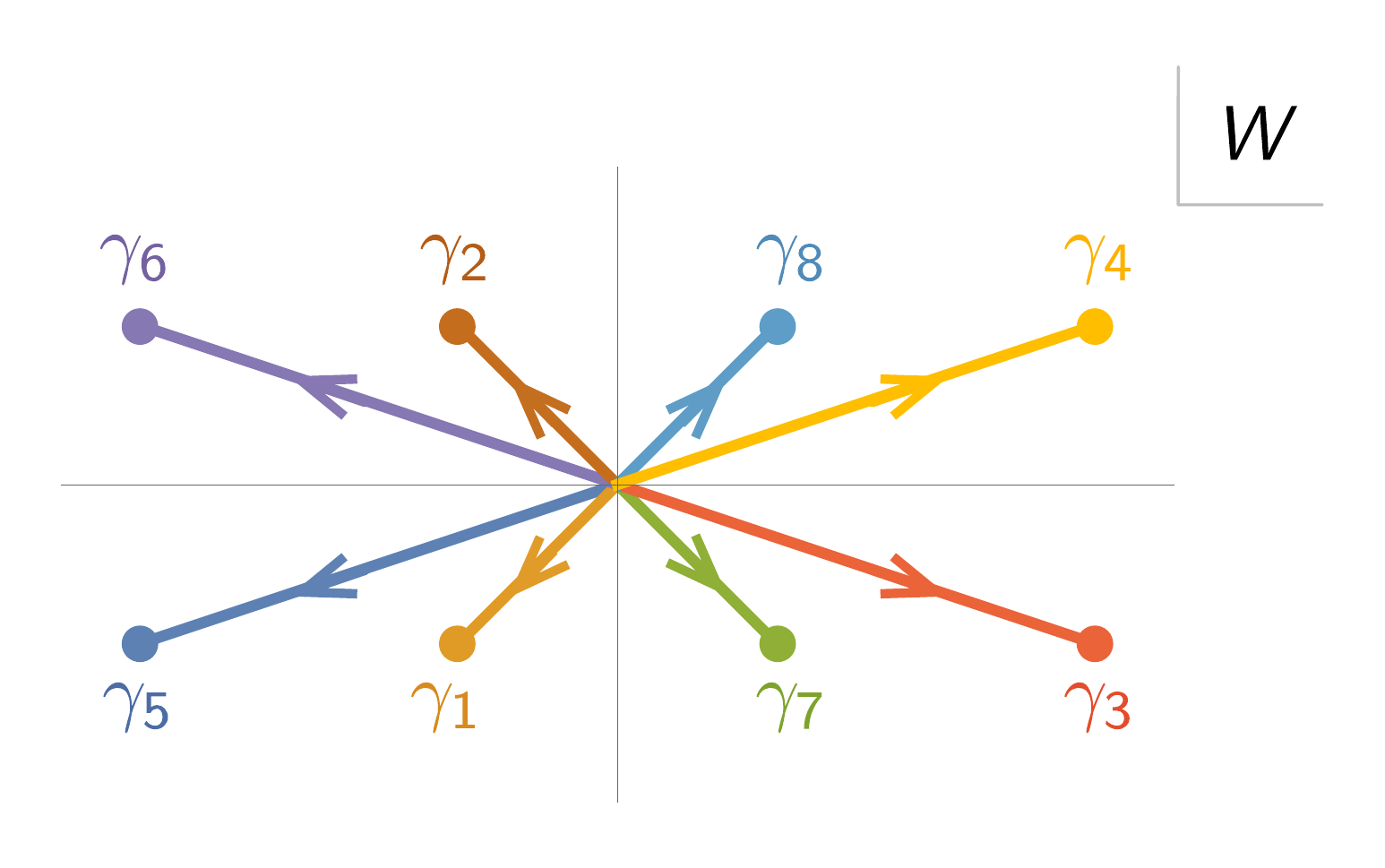}
\caption{Vanishing paths for phase A of $Q^{1,1,1}/\mathbb{Z}_2$.
\label{mirror_Q111Z2_W}}
 \end{center}
 \end{figure}

The eight critical points of $P$ and the corresponding critical values are
\begin{align}
(x^*,y^*,z^*) = (\pm 1, \pm 1, \pm 1) \,,\quad W^* = 2(x^* +2 y^* + i z^*) \,.
\end{align}
The resulting vanishing paths are shown in \fref{mirror_Q111Z2_W}. 

\begin{figure}[H]
\begin{center}
\resizebox{\hsize}{!}{
\includegraphics[trim=0cm 0cm 0cm 0cm,totalheight=10 cm]{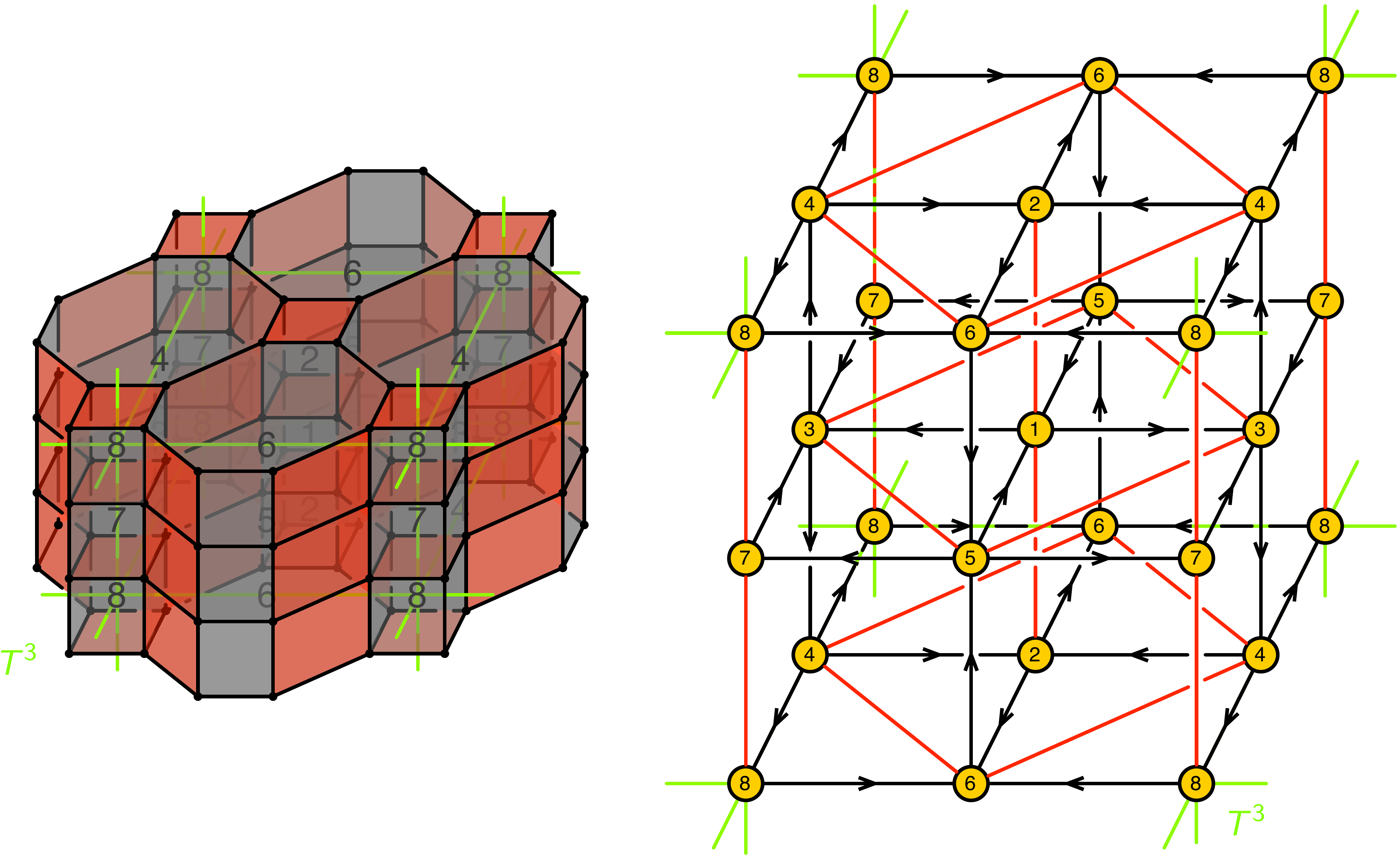}
}  
\vspace{-.3cm}\caption{Brane brick model and periodic quiver for phase A of $Q^{1,1,1}/\mathbb{Z}_2$.
\label{periodic_quiver_Q111Z2_A}}
 \end{center}
 \end{figure}

Below we analyze the three tomographies. Having discussed in detail the computation of intersection numbers in the previous examples, our presentation will be more concise. In particular, we will simply quote the results regarding field types. The final brane brick model and periodic quiver are given in \fref{periodic_quiver_Q111Z2_A}.

\paragraph{The $x$-tomography.}

\begin{figure}[H]
\begin{center}
\includegraphics[width=11cm]{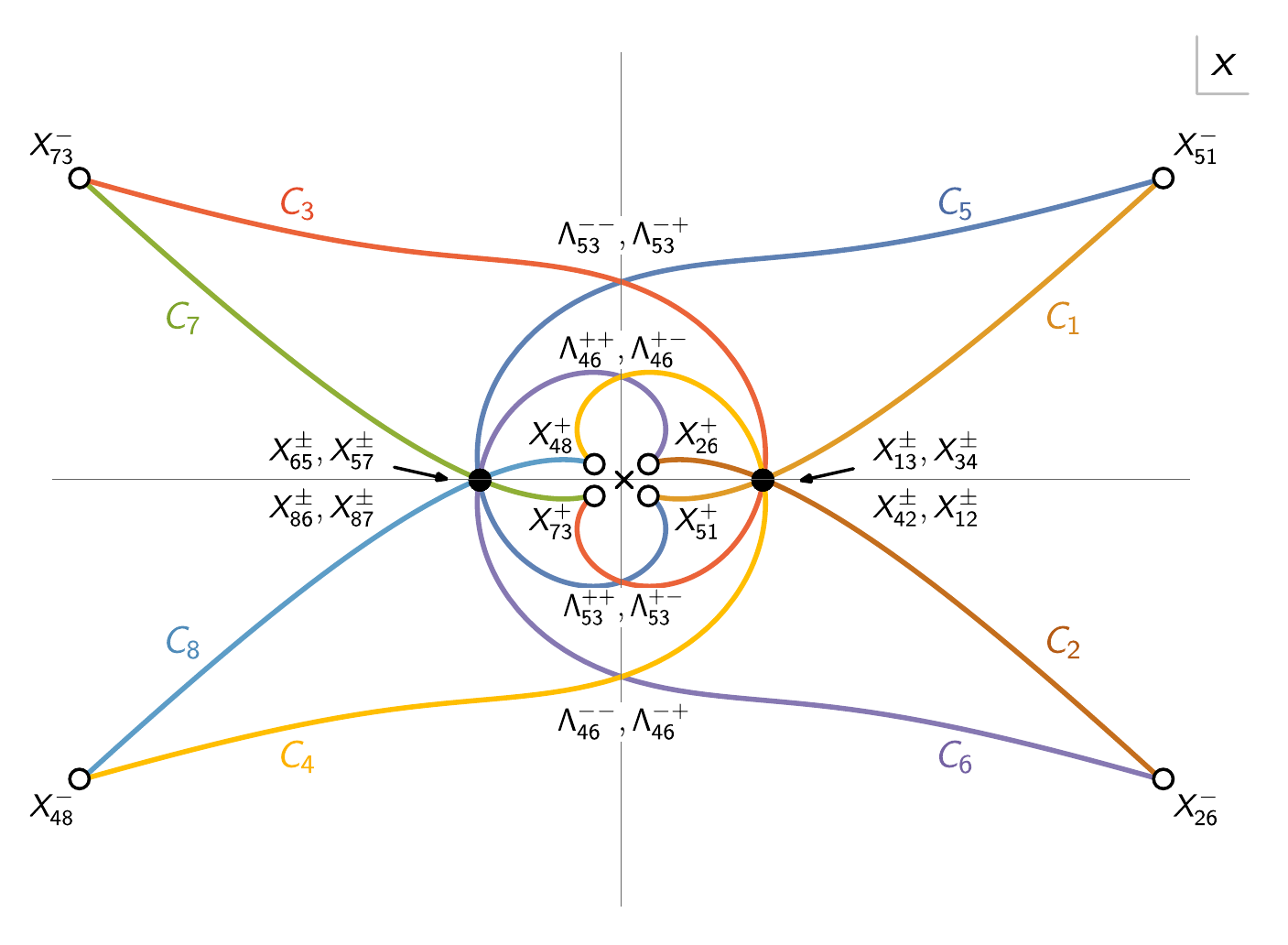}
\caption{The $x$-tomography for phase A of $Q^{1,1,1}/\mathbb{Z}_2$. We indicate the fields associated with each intersection. 
\label{mirror_fields_Q111Z2_x}}
 \end{center}
 \end{figure}

\fref{mirror_fields_Q111Z2_x} shows the $x$-tomography. All the intersections between nodes $(1,2,3,4)$ are at $x=1$, while all intersections between nodes $(5,6,7,8)$ are at $x=-1$. As a result, the nodes in the periodic quiver are arranged into two layers along the $x$ axis: nodes $(1,2,3,4)$ at $\mathrm{arg}(x)=0$ and nodes $(5,6,7,8)$ at $\mathrm{arg}(x)=\pi$. 

Once again, interesting conclusions about some of the intersections can be reached without the need for detailed calculations. Let us consider the two points $x=\pm 1$. $P(x=\pm 1, y,z)$ is isomorphic to the Newton polynomial for phase 1 of $F_0$. Hence, borrowing the $F_0$ result, we conclude that the total intersection number among the four vanishing cycles meeting at each of these points is 8. Moreover, we know that there are two fields between each pair of nodes. In more detail, there are 6 chiral and 2 Fermi fields connecting the nodes on each of these layers. On the first layer we have $X_{13}^\pm$, $X_{34}^\pm$, $X_{42}^\pm$ and $\Lambda_{12}^\pm$, which sit at $x=1$, i.e. $\mathrm{arg}(x)=0$, in \fref{mirror_fields_Q111Z2_x}. On the $\mathrm{arg}(x)=\pi$ layer we have $X_{65}^\pm$, $X_{57}^\pm$, $X_{86}^\pm$ and $\Lambda_{87}^\pm$, coming from the intersections at $x=-1$.

Additional chiral and Fermi fields connect the two layers. The two interlayers correspond to the fields on the upper and lower half-planes in \fref{mirror_fields_Q111Z2_x}, i.e. fields with $0<\mathrm{arg}(x)<\pi$ and $-\pi<\mathrm{arg}(x)<0$, respectively.

\begin{figure}[H]
\begin{center}
\includegraphics[width=10cm]{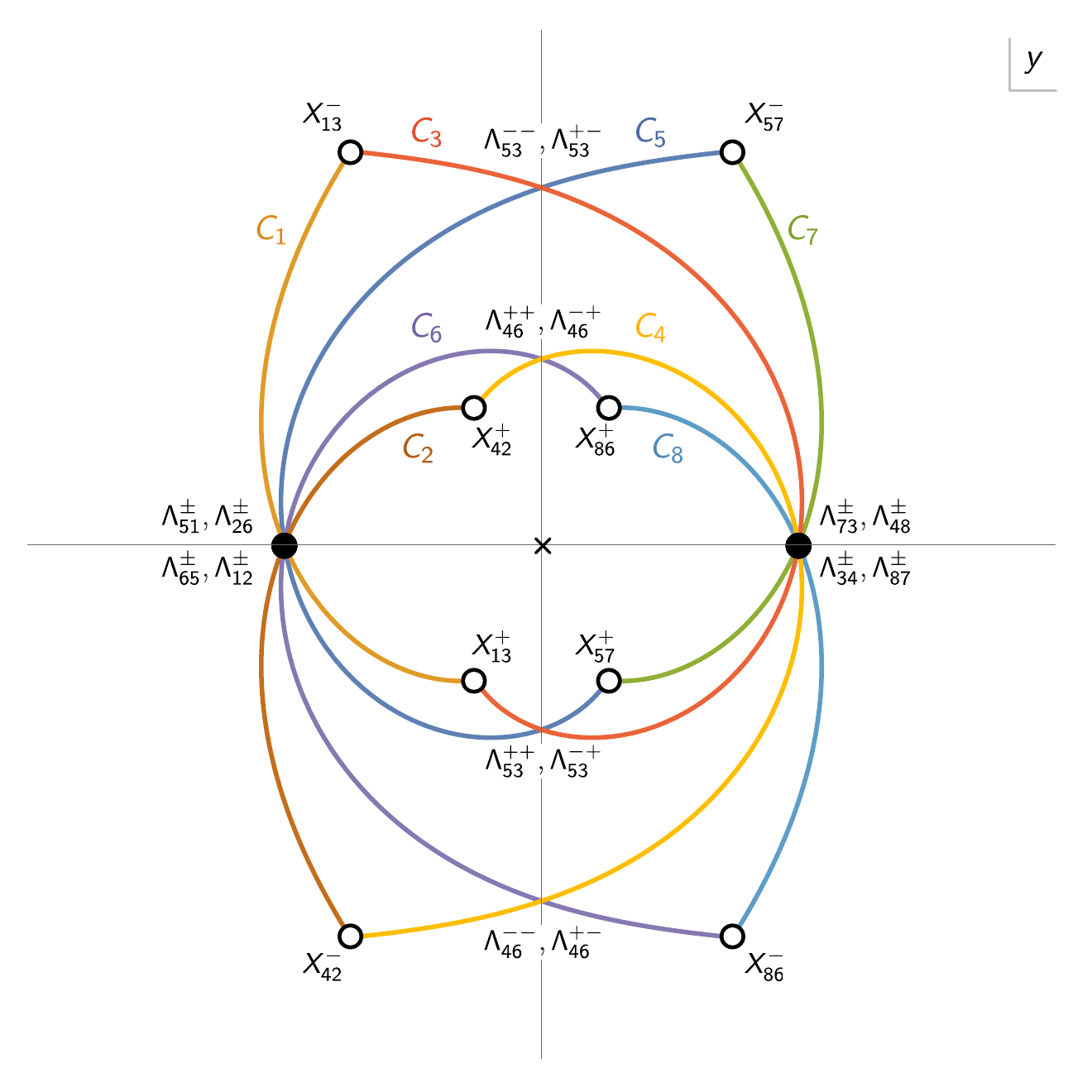}
\caption{The $y$-tomography for phase A of $Q^{1,1,1}/\mathbb{Z}_2$. We indicate the fields associated with each intersection. 
\label{mirror_fields_Q111Z2_y}}
 \end{center}
 \end{figure}

\begin{figure}[H]
\begin{center}
\includegraphics[width=7.3cm]{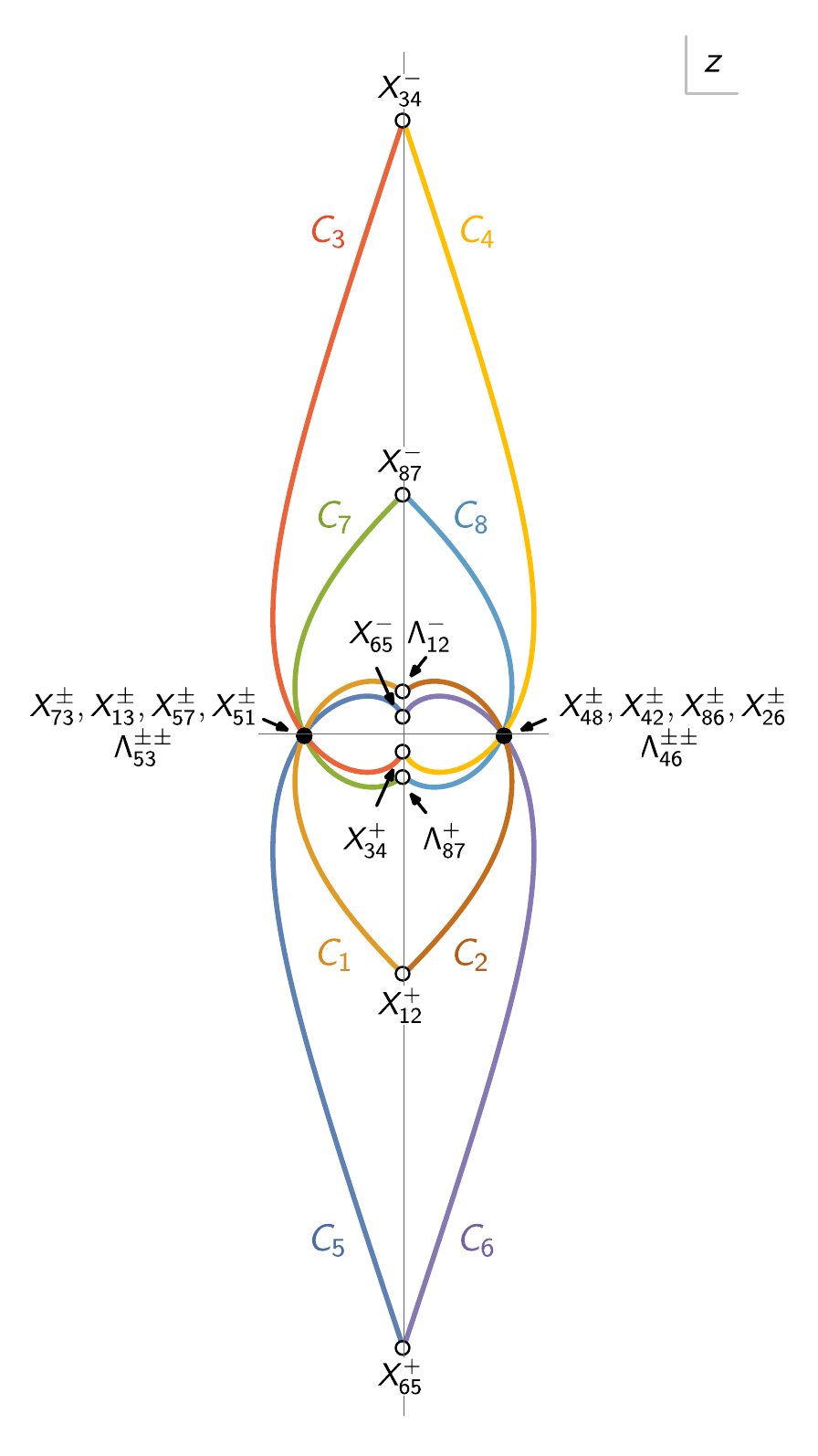}
\caption{The $z$-tomography for phase A of $Q^{1,1,1}/\mathbb{Z}_2$. We indicate the fields associated with each intersection. 
\label{mirror_fields_Q111Z2_z}}
 \end{center}
 \end{figure}

It is also easy to understand why some apparent intersections in \fref{mirror_fields_Q111Z2_x} do not give rise to any field. This is the case for $(C_1,C_4)$ and $(C_2,C_3)$ at $x=1$, and of $(C_5,C_8)$ and $(C_6,C_7)$ at $x=-1$. True intersections between cycles must show up as such when projected onto the three planes $x$, $y$ and $z$. If two cycles do not intersect in some of the tomographies, we conclude there is not actual intersection between them. Looking at Figures \ref{mirror_fields_Q111Z2_y} and \ref{mirror_fields_Q111Z2_z} we see that the pairs of cycles we just mentioned do not intersect.

\paragraph{The $y$-tomography.}

Having discussed the $x$-tomography in detail, we can be more schematic. \fref{mirror_fields_Q111Z2_y} shows the $y$-tomography. Topologically and in terms of the types of fields at each intersection, \fref{mirror_fields_Q111Z2_x} is identical to \fref{mirror_fields_Q111Z2_y}. This implies a symmetry between the $x$ and $y$ directions in the periodic quiver, which is manifest in \fref{periodic_quiver_Q111Z2_A}. The difference in appearance between Figures \ref{mirror_fields_Q111Z2_x} and \ref{mirror_fields_Q111Z2_y} is due to our choice of coefficients in \eref{M32-ansatz}.

Along the $y$ axis, the nodes in the periodic quiver form two layers: nodes $(3,4,7,8)$ at $\mathrm{arg}(y)=0$ and nodes $(1,2,5,6)$ at $\mathrm{arg}(y)=\pi$. The discussion about the fields on each layer and between them, and about intersection numbers is identical to the one for the $x$-tomography.

\paragraph{The $z$-tomography.}

The $z$-tomography is given in \fref{mirror_fields_Q111Z2_z}. Reasoning as before, we conclude that the nodes in the periodic quiver form two layers in the $z$ direction, consisting of nodes $(2,4,6,8)$ at $\mathrm{arg}(z)=0$ and nodes $(1,3,5,7)$ at $\mathrm{arg}(z)=\pi$. 8 chiral and 4 Fermi fields connect the nodes at each of these layers. We can see that there are 12 fields at each of these two intersections by realizing that $P(x, y,z=\pm 1)$ is isomorphic to the Newton polynomial for phase 2 of $F_0$. Furthermore, we can also identify the pairwise intersection numbers between cycles, which are either 2 or 4 as in $F_0$. On the first layer we have $X_{48}^\pm$, $X_{42}^\pm$, $X_{86}^\pm$, $X_{26}^\pm$ and $\Lambda_{46}^{\pm\pm}$, which sit at $z=1$, i.e. $\mathrm{arg}(z)=0$. On the $\mathrm{arg}(z)=\pi$ layer we have $X_{73}^\pm$, $X_{13}^\pm$, $X_{57}^\pm$, $X_{51}^\pm$ and $\Lambda_{53}^{\pm\pm}$, coming from the intersections at $z=-1$. 

The two layers are connected by $X_{34}^-$, $X_{65}^-$, $\Lambda_{12}^-$ and $\Lambda_{87}^-$ at $\mathrm{arg}(z)=\pi/2$, and by $X_{34}^+$, $X_{65}^+$, $\Lambda_{12}^+$ and $\Lambda_{87}^+$ at $\mathrm{arg}(z)=-\pi/2$. These are the vertical fields in \fref{periodic_quiver_Q111Z2_A}.

\paragraph{Brane brick model and periodic quiver.}

The resulting brane brick model and periodic quiver for phase A of $Q^{1,1,1}/\mathbb{Z}_2$ are shown in \fref{periodic_quiver_Q111Z2_A}. They are in full agreement with \cite{Franco:2016nwv}. The $J$- and $E$-terms are:

{\small
\beq
\begin{array}{lcccc}
& &  J & &  E 
\\
   \Lambda_{12}^{-} & :\ \ \ & 
X^{+}_{26} \cdot X^{+}_{65} \cdot X^{-}_{51} - 
X^{-}_{26} \cdot X^{+}_{65} \cdot X^{+}_{51}  
& \ \ \ \ &  
X^{+}_{13} \cdot X^{-}_{34} \cdot X^{-}_{42} - 
X^{-}_{13} \cdot X^{-}_{34} \cdot X^{+}_{42} 
   \\
   \Lambda_{12}^{+} & :\ \ \ & 
X^{-}_{26} \cdot X^{-}_{65} \cdot X^{+}_{51} -
X^{+}_{26} \cdot X^{-}_{65} \cdot X^{-}_{51} 
& \ \ \ \ & 
X^{+}_{13} \cdot X^{+}_{34} \cdot X^{-}_{42}   -
X^{-}_{13}  \cdot X^{+}_{34} \cdot X^{+}_{42} 
\\ 
  \Lambda_{87}^{-} & :\ \ \ &  
 X^{+}_{73} \cdot X^{+}_{34} \cdot X^{-}_{48}   -
  X^{-}_{73} \cdot X^{+}_{34} \cdot X^{+}_{48} 
  & \ \ \ \ &   
  X^{+}_{86} \cdot X^{-}_{65} \cdot X^{-}_{57}  -
   X^{-}_{86} \cdot X^{-}_{65} \cdot X^{+}_{57}    
  \\
   \Lambda_{87}^{+} 
   & :\ \ \ &  
    X^{-}_{73} \cdot X^{-}_{34} \cdot X^{+}_{48}-
    X^{+}_{73} \cdot X^{-}_{34} \cdot X^{-}_{48}    
    & \ \ \ \ &  
     X^{+}_{86}\cdot X^{+}_{65} \cdot X^{-}_{57} - 
    X^{-}_{86} \cdot X^{+}_{65} \cdot X^{+}_{57}    
 \\
   \Lambda_{46}^{--} & :\ \ \ & X^{+}_{65} \cdot X^{+}_{57} \cdot X^{+}_{73} \cdot X^{-}_{34}  - X^{-}_{65} \cdot X^{+}_{51} \cdot X^{+}_{13} \cdot X^{+}_{34} & \ \ \ \ &  
X^{-}_{42} \cdot X^{-}_{26} - X^{-}_{48} \cdot X^{-}_{86}  
\\
    \Lambda_{46}^{++} & :\ \ \ & 
    X^{+}_{65} \cdot X^{-}_{51} \cdot X^{-}_{13} \cdot X^{-}_{34} -
    X^{-}_{65} \cdot X^{-}_{57} \cdot  X^{-}_{73} \cdot X^{+}_{34}
    & \ \ \ \ &  
X^{+}_{42} \cdot X^{+}_{26} - X^{+}_{48} \cdot X^{+}_{86}   
\\
    \Lambda_{46}^{-+} & :\ \ \ & 
    X^{-}_{65}\cdot X^{+}_{51}\cdot X^{-}_{13}\cdot X^{+}_{34}    -
    X^{+}_{65}\cdot X^{-}_{57} \cdot X^{+}_{73}  \cdot X^{-}_{34} 
    & \ \ \ \ &  
X^{+}_{42} \cdot X^{-}_{26} - X^{-}_{48} \cdot X^{+}_{86}     
\\
   \Lambda_{46}^{+-} & :\ \ \ & X^{-}_{65}\cdot X^{+}_{57} \cdot X^{-}_{73}\cdot X^{+}_{34} - X^{+}_{65}\cdot X^{-}_{51}\cdot X^{+}_{13}  \cdot  X^{-}_{34}    & \ \ \ \ &  
X^{-}_{42} \cdot X^{+}_{26} - X^{+}_{48} \cdot X^{-}_{86} 
\\ 
   \Lambda_{53}^{--} & :\ \ \ &  X^{+}_{34}\cdot X^{+}_{42}\cdot X^{+}_{26}\cdot X^{-}_{65}   - X^{-}_{34} \cdot X^{+}_{48}\cdot X^{+}_{86}\cdot X^{+}_{65}      & \ \ \ \ &  
X^{-}_{57}\cdot X^{-}_{73}  - X^{-}_{51} \cdot X^{-}_{13}    
\\
    \Lambda_{53}^{++} & :\ \ \ & X^{+}_{34} \cdot X^{-}_{48} \cdot X^{-}_{86} \cdot X^{-}_{65} 
 - 
X^{-}_{34} \cdot X^{-}_{42} \cdot X^{-}_{26} \cdot X^{+}_{65}   
 & \ \ \ \ &  
X^{+}_{57} \cdot X^{+}_{73}  - X^{+}_{51} \cdot X^{+}_{13}   
\\ 
   \Lambda_{53}^{-+} & :\ \ \ & 
  X^{-}_{34}\cdot X^{+}_{48}\cdot X^{-}_{86} \cdot  X^{+}_{65}  -
 X^{+}_{34} \cdot X^{-}_{42} \cdot X^{+}_{26} \cdot  X^{-}_{65}  
   & \ \ \ \ &  
X^{+}_{57}\cdot X^{-}_{73}  - X^{-}_{51}\cdot X^{+}_{13}     
\\
    \Lambda_{53}^{+-} & :\ \ \ &  X^{-}_{34} \cdot X^{+}_{42}\cdot X^{-}_{26} \cdot X^{+}_{65}  -  X^{+}_{34} \cdot X^{-}_{48}\cdot X^{+}_{86} \cdot X^{-}_{65}   & \ \ \ \ &  
X^{-}_{57} \cdot X^{+}_{73} - X^{+}_{51}  \cdot X^{-}_{13} 
\\ 
\end{array}
\label{es1a1}
\eeq
}

\newpage

\section{Geometric Transitions and Triality}

\label{section_triality}

In this section we explain how triality follows from a simple geometric transition in the mirror.

\subsection{Triality}

\begin{figure}[H]
\centering
\includegraphics[height=6cm]{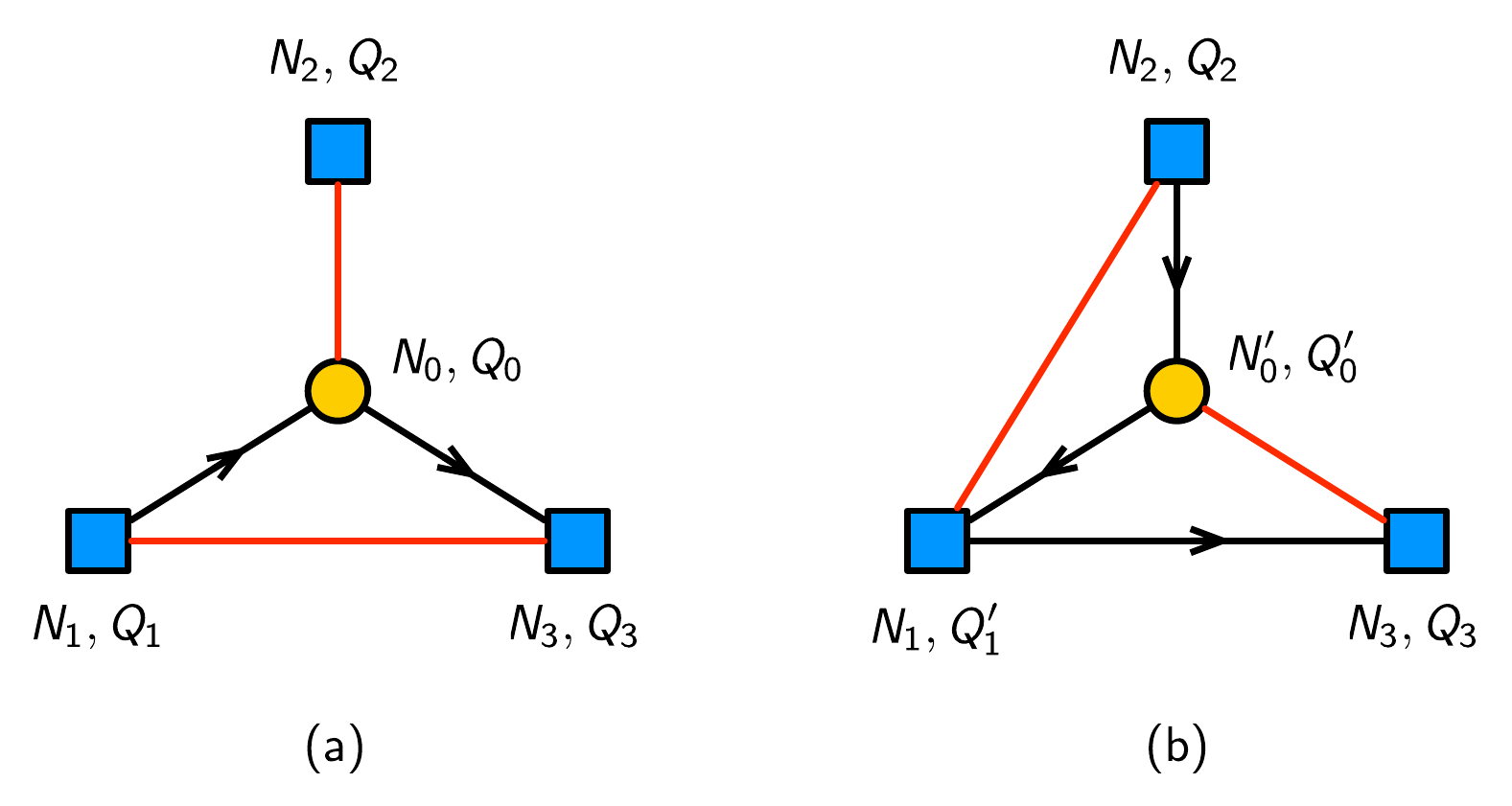}
\label{fig:test1}\label{fig:test2}
\caption{Local quivers for triality. The $Q_i$ are the D-brane charge vectors for the different nodes.}
\label{quiver_triality}
\end{figure}

Let us briefly review the basics of triality. We refer the reader to \cite{Gadde:2013lxa} for further details. Without loss of generality, we can restrict our consideration to the four node quiver shown in \fref{quiver_triality}.a. The yellow node represents the gauge group that undergoes triality, while the blue nodes are flavor groups.\footnote{In the theories on D-branes that we consider, the flavor nodes contain additional matter and are gauged.} The multiplicities of flavors are absorbed into the ranks of the flavor nodes. It is straightforward to extend our discussion to non-trivial multiplicities of the flavor arrows, to multiple flavor nodes of each type and to include fields stretching between flavor nodes. Nodes that are not connected to the dualized one are irrelevant for our analysis. For later use, the $Q_i$ indicate D-brane charge vectors of the different nodes. 

The quiver for the triality dual is shown in \fref{quiver_triality}.b. Anomaly cancellation constraints the rank of the central node in the two theories to be
\beq
N_0 = {N_1 +N_3 - N_2 \over 2} \ \ \ , \ \ \ 
N_0' =  {N_1 +N_2 - N_3 \over 2} \, .
\eeq
This implies that under triality it transforms according to
\beq
N_0'= N_1 -N_0 \, .
\label{triality_transformation_rank}
\eeq
The dual theory contains $J$-and $E$-terms associated to the two triangles in the quiver. Acting with triality three times on the same node we recover the initial theory.

\subsection{Triality in the Mirror}

\begin{figure}[htbp]
\begin{center}
\includegraphics[width=6cm]{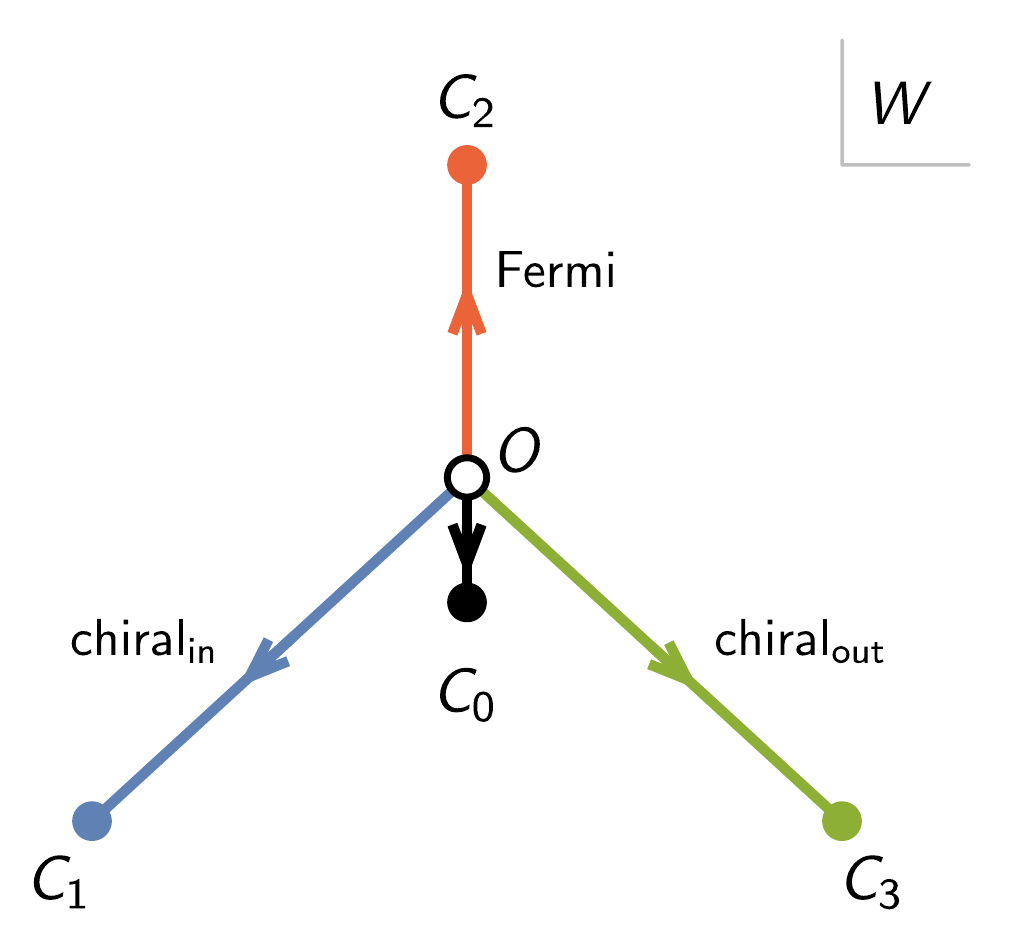}
\caption{Cyclic ordering of vanishing paths on the $W$-plane.
\label{mirror_cyclic_order}}
 \end{center}
 \end{figure}

Let us start by discussing general properties of the vanishing paths on the $W$-plane. For any cycle $C_0$, the corresponding flavor cycles are always distributed as shown in \fref{mirror_cyclic_order}. $C_1$, $C_2$ and $C_3$ may represent collections of cycles. The fields contributed to $C_0$ by the cycles in each collection are of the same type and, starting from $C_0$, they appear in the following cyclic order around the origin: chiral in, Fermi, chiral out.\footnote{This ordering might be clockwise or counterclockwise. It can be reversed simply by conjugating all fields. By convention, all our examples will be ordered in the clockwise direction.} The chiral in and chiral out cycles sit at both sides of $C_0$ due to the symmetry of the theory under conjugation of all fields. In the figure, we have arranged the positions of the critical points to simplify the comparison with \fref{quiver_triality}.

All the explicit examples considered in this paper satisfy the ordering of vanishing paths on the $W$-plane, with the exception of phase S of $Q^{1,1,1}/\mathbb{Z}_2$, which is presented in appendix \ref{section_Q111_phases}. An apparent violation of the ordering rule should be regarded as an indication that connecting critical points to the origin by straight segments to form vanishing paths is an invalid approximation. In such cases, the correct geometry is only captured by curved vanishing paths. Furthermore, as we explain below, this property is preserved by triality.  

The cyclic ordering of vanishing paths has a practical application. Combined with anomaly cancellation, it provides a simple method for determining the type of field that is associated to every intersection. This approach bypasses the computation of the signs of intersections discussed in section \sref{sec:field-type}. 

\begin{figure}[htbp]
\begin{center}
\includegraphics[width=13cm]{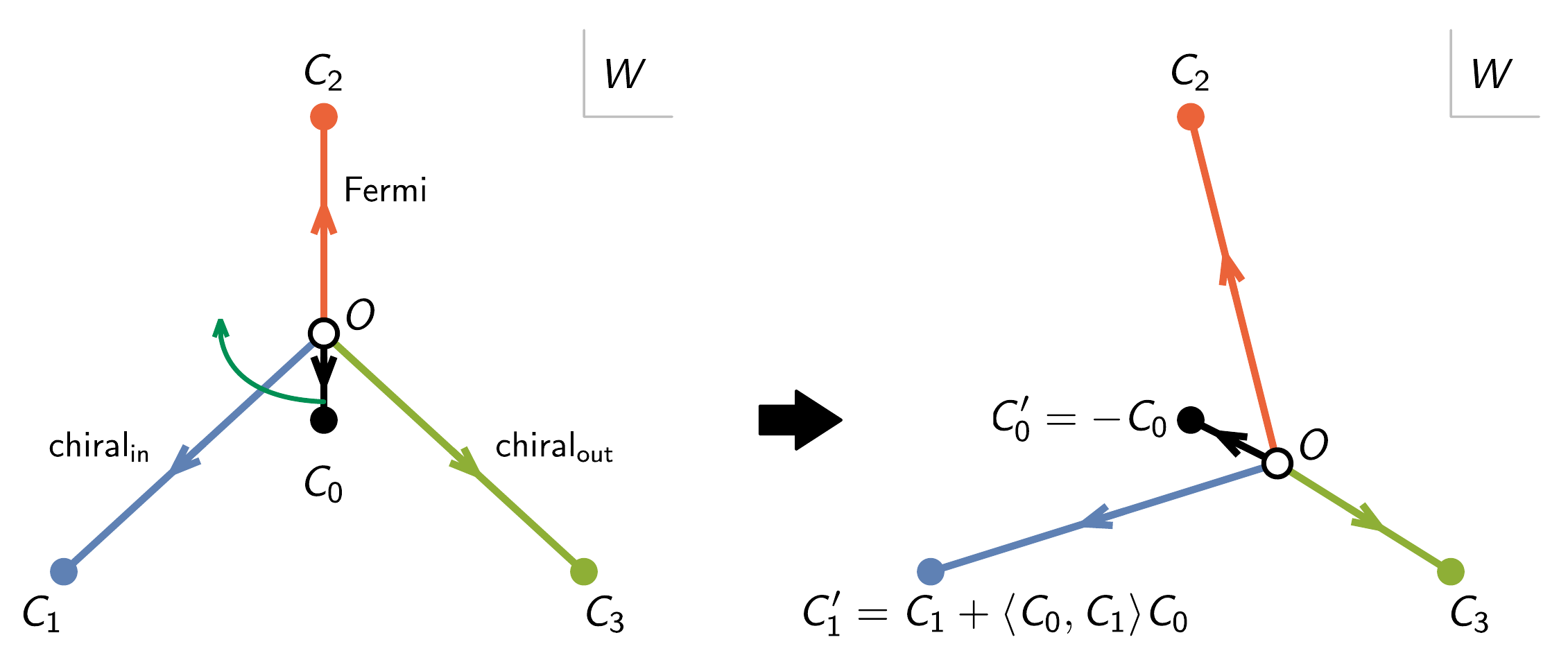}
\caption{Triality on the $W$-plane.
\label{mirror_triality}}
 \end{center}
 \end{figure}

Triality has a simple implementation in the mirror geometry, extending a similar case studied in \cite{Cachazo:2001sg}. It corresponds to shrinking $C_0$ to zero size and regrowing it with the opposite orientation between $C_1$ and $C_2$ on the $W$-plane, namely in the chiral in-Fermi wedge. This process is illustrated in \fref{mirror_triality}, where we have fixed the critical points and moved around the origin. Since the different types of fields divide the $W$-plane into three wedges, this implementation of triality makes it manifest that it is a duality of order 3. Inverse triality corresponds to moving $C_0$ to the wedge between $C_3$ and $C_2$. Alternatively, it can be obtained by acting with triality twice.

\begin{figure}[htbp]
\begin{center}
\includegraphics[width=6cm]{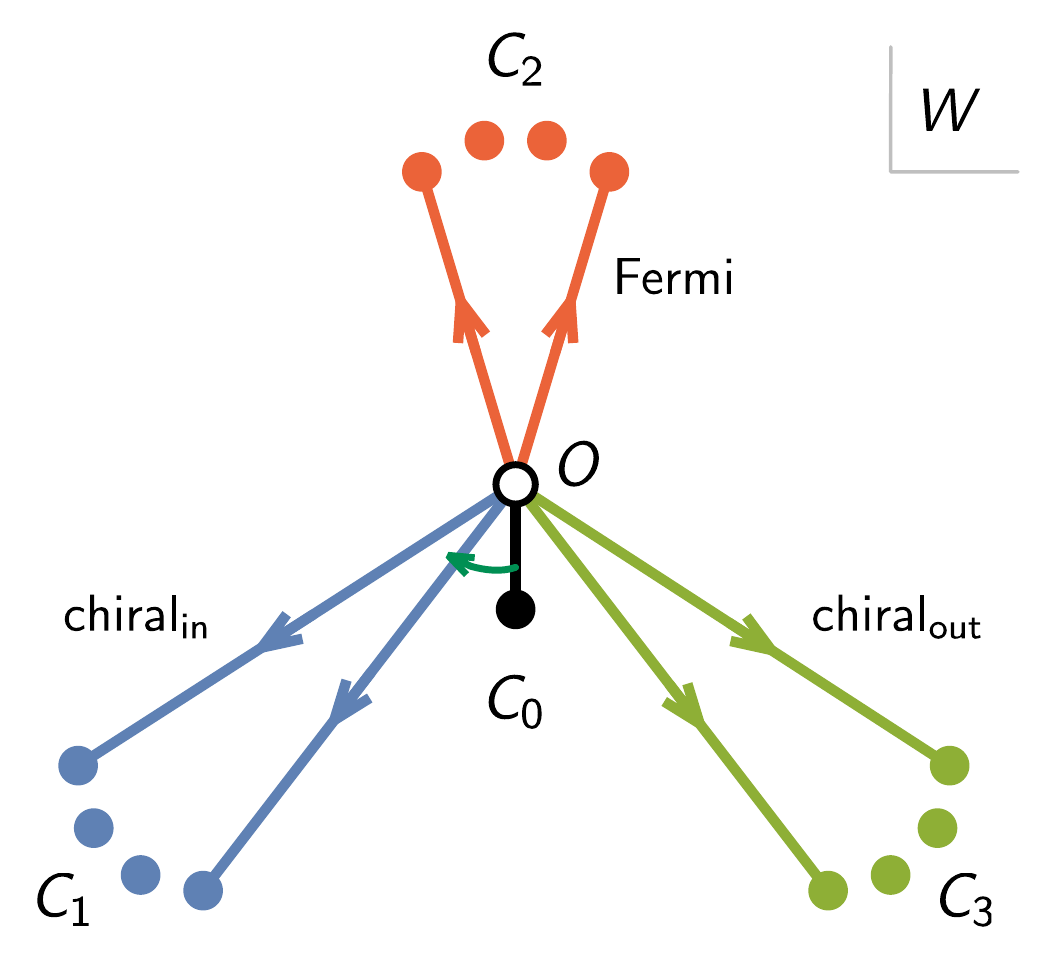}
\caption{A partial motion of $C_0$ over the cycles contributing incoming chiral fields does not lead to a supersymmetric configuration. 
\label{mirror_partial_transition}}
 \end{center}
 \end{figure}

Naively, it appears that the string construction allows more general transformations than triality. In particular, it is possible to shrink $C_0$ and regrow it in any of the wedges defined by the other vanishing paths. It is sufficient to consider the case in which there are multiple cycles contributing incoming chirals and $C_0$ moves only over a subset of them, as shown in \fref{mirror_partial_transition}. Other configurations reduce to this one after a number of trialities. This configuration is analogous to the one obtained by starting from the one engineering $4d$ SQCD \cite{Elitzur:1997fh,Giveon:1998sr} and partially moving flavor branes over color branes, instead of moving all of them and producing the Seiberg dual. In analogy with the $4d$ counterpart, we expect such general transformations to break SUSY.

Below we analyze the triality transition in terms of D-brane charges associated to the nodes in the quiver.

\paragraph{Intersection numbers.}

Let us first discuss the intersection numbers between branes, $\langle C_i,C_j \rangle=\langle Q_i,Q_j \rangle$. The intersection matrix in a CY$_4$ is symmetric, i.e. $\langle Q_i,Q_j\rangle = \langle Q_j,Q_i\rangle$. For the initial theory in \fref{quiver_triality}, we have
\beq
\begin{array}{ccl}
\langle Q_0,Q_1\rangle & = & 1 \\[.15cm]
\langle Q_0,Q_2\rangle & = & - 1 \\[.15cm]
\langle Q_0,Q_3\rangle & = & 1 
\end{array}
\eeq
Positive and negative intersection numbers distinguish between chiral and Fermi fields, respectively. Our assumption of no fields between flavor nodes implies $\langle Q_1,Q_3\rangle=\langle Q_1,Q_2\rangle=\langle Q_2,Q_3\rangle=0$. 

Finally, it is important to take into account that 4-cycles in a CY$_4$ have non-vanishing self-intersections \cite{Cvetic:2000db}. In particular,
\beq
\langle Q_0,Q_0\rangle = -2 \,.
\eeq
Such a contribution is not present in the brane realization of Seiberg duality and is crucial for several features of triality. The $SU(N_0)^2$ gauge anomaly can be compactly written as
\beq
A_{SU(N_0)^2}=\sum_{i=0}^3 \langle Q_0,Q_i \rangle N_i \, .
\eeq

\paragraph{Transformation of the brane charges.}

Triality corresponds to shrinking the cycle $C_0$ to zero size and reemerging on the $W$-plane on the wedge past $C_1$. Then, the brane charges transform as follows:
\beq
\begin{array}{ccl}
Q_0' & = & - Q_0 \\[.15cm]
Q_1'  & = & Q_1 +  \langle Q_0,Q_1\rangle Q_0 = Q_1 + Q_0 \\[.15cm]
Q_2' & = & Q_2 \\[.15cm]
Q_3' & = & Q_3 
\end{array}
\label{change charges}
\eeq
This can be understood by considering a trajectory that keeps $C_0$ at finite volume and moving it over $C_1$. The transformation is analogous to the one that implements Seiberg duality in $4d$ $\mathcal{N}=1$ theories, see e.g. \cite{Cachazo:2001sg,Feng:2002kk}. The minus sign for $Q_0$ accounts for the reversal of the dualized cycle. $Q_1$ picks a contribution proportional to $Q_0$ and its intersection number with it.\footnote{$C_0$ might also have to pass over cycles that have vanishing intersections with it. Such cycles do not affect our discussion.} $Q_2$ and $Q_3$ do not change, since they do not participate in the brane crossing process.

\subsection{Transformation of the Gauge Theory}

We now explain how the transformation of the brane configuration outlined in the previous section accounts for the triality transformation of the gauge theory.

\paragraph{Rank of the gauge group.}

The transformation of the rank of the gauge group follows from conservation of the total brane charge. Initially, we have
\beq
Q_T = \sum_{i=0}^4 N_i \, Q_i \, .
\eeq
Since the ranks of the flavor nodes do not change, after the transition we have
\beq
\begin{array}{ccl}
Q_T' & = & N_0' \, Q_0' + N_1 \, Q_1' + N_3 \, Q_3' + N_2 \, Q_2' \\[.15cm]
& = & -N_0' \, Q_0 + N_1 (Q_1 + Q_0) + N_2 \, Q_2+ N_3 \, Q_3  \\[.15cm]
& = & \left[ -N_0'+ (N_1  -N_0)\right] Q_0 + Q_T \, .
\end{array}
\eeq
Conservation of brane charge implies that $Q_T'=Q_T$, so we conclude that
\beq
N_0' = N_1 -N_0 \, ,
\eeq
which is the correct transformation under triality, given in \eref{triality_transformation_rank}.

\paragraph{Dual flavors.}

Let us now check that the transformation of charges also gives rise to the appropriate transformation of the flavors, i.e. of the fields charged under the dualized gauge group.

Let us first consider the fields between $C_{0}'$ and $C_1'$. The intersection between these two cycles is
\beq
\begin{array}{ccl}
\langle Q_0',Q_1' \rangle & = & \langle -Q_0, Q_1 + Q_0\rangle = -\langle Q_0, Q_1 \rangle - \langle Q_0, Q_0 \rangle \\[.15 cm]
& = & -1 + 2  = 1 \, .
\end{array}
\eeq 
This implies that the multiplicity of lines between the two nodes remains the same. Furthermore, since the intersection number is positive, we conclude that it corresponds to a chiral field.\footnote{Identifying the orientation reversal of the chiral fields requires additional information beyond the intersection number.} This is in full agreement with \fref{quiver_triality}.b. Notice that the self-intersection of $Q_0$ is crucial for producing the correct result.

We also have 
\be
\langle Q_0',Q_2' \rangle = \langle - Q_0,Q_2 \rangle = 1 \, .
\eeq
The multiplicity of lines between the two nodes does not change. The sign of the intersection number however becomes positive, implying that these are chiral fields.  This is in agreement with \fref{quiver_triality}.b.

Similarly,
\beq
\langle Q_0',Q_3' \rangle = \langle - Q_0,Q_3 \rangle = - 1 \, ,
\eeq
implying that we have the same number of lines, but the fields connecting this pair of nodes are now Fermi fields. Once again, this matches \fref{quiver_triality}.b.

\paragraph{Mesons.}

Finally, let us verify that the mesons are appropriately generated, by considering the intersections between the flavor nodes. Between $C_1'$ and $C_2'$, we have
\beq
\langle Q_1',Q_2' \rangle = \langle Q_1 + Q_0 ,Q_2 \rangle = \langle Q_{0}, Q_2 \rangle = -1 .
\eeq
We thus obtain the correct multiplicity and the fact that these mesons are Fermi fields.

Between $C_1'$ and $C_3'$, we get
\beq
\langle Q_1',Q_3' \rangle = \langle Q_1 + Q_0 ,Q_3 \rangle =  \langle Q_{0}, Q_3 \rangle = 1 .
\eeq
The right multiplicity and the fact that these mesons are chiral fields are correctly generated.

Finally, since $Q_2$ and $Q_3$ do not change, their intersection number remains zero and we conclude that no mesons connecting these two nodes are created.

\paragraph{Periodicity.}

The fact that triality is an order 3 duality also follows from the brane charges. If we perform the transformation \eqref{change charges} three times, we obtain
\begin{align}
Q_i''' = Q_i + \langle Q_0,Q_i \rangle Q_0 \,,
\end{align}
for all brane charges, $i=0,\ldots,3$.\footnote{This formula applies not only to the flavor nodes but also to $Q_0$, which becomes $-Q_0$ after three triality transformations.} We recognize this as the Picard-Lefschetz formula. 

The intersection numbers return to their original values:
\begin{align}
\begin{split}
\langle Q'''_i , Q'''_j \rangle &= 
\langle Q_i + \langle Q_0,Q_i  \rangle Q_0 
, Q_j + \langle Q_0,Q_j  \rangle Q_0 
\rangle 
\\[.15cm]
&= \langle Q_i , Q_j \rangle + (1+1-2) \langle Q_0 , Q_i \rangle \langle Q_0 , Q_j \rangle = \langle Q_i , Q_j \rangle\,.
\end{split}
\end{align}
Note that the self-intersection, $\langle Q_0 , Q_0 \rangle = -2$, plays a crucial role.

\subsection{Triality and Tomography}

\label{section_detailed_triality_Q111/Z2}

\begin{figure}[H]
\begin{center}
\includegraphics[trim={0 2cm 0 0},width=15.5cm]{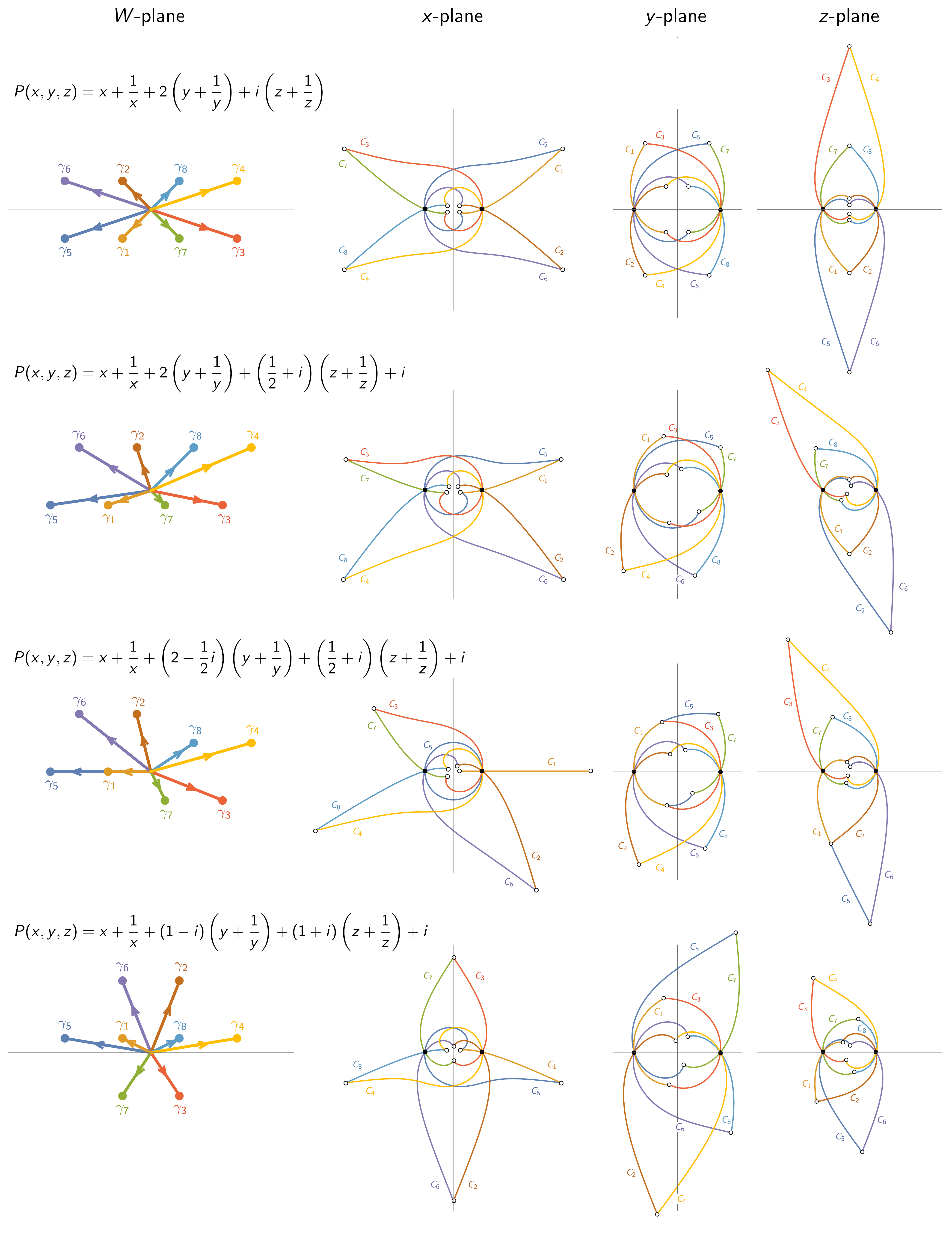}
\caption{Continuous transition between phases A and B corresponding to triality on node 1. We show the variation in $P(x,y,z)$ and the transformation on the $W$-plane and the $x$-, $y$- and $z$-tomographies. In order to facilitate the combination of figures, the different planes have relative rescalings. 
\label{evq111z2flowAB}}
 \end{center}
 \end{figure}

Tomography beautifully captures the continuous transition between two toric phases connected by triality. For illustration, let us consider phase A of $Q^{1,1,1}/\mathbb{Z}_2$, as shown in \fref{periodic_quiver_Q111Z2_A}, and act with triality on node 1, obtaining phase B. The details of phase B are presented in appendix \ref{section_Q111_phases}. In phase A, node 1 is such that it has: two incoming chiral arrows from node 5, two Fermi lines going to node 2 and two outgoing chiral arrows going to node 3. Following the general prescription, triality corresponds to moving $C_1$ over $C_5$ on the $W$-plane. \fref{evq111z2flowAB} shows the continuous deformation connecting the phases A and B on the $W$-plane and the three tomographies.

\section{M-Theory Lift: M5-Branes on 4-Manifolds}

\label{section_4-manifolds}

An alternative approach for engineering $2d$ $(0,2)$ theories is in terms of M5-branes wrapping 4-manifolds \cite{Gauntlett:2000ng,Gauntlett:2001jj,Benini:2013cda,Gadde:2013sca}. This framework connects the geometry and topology of the 4-manifolds to properties of the field theories. Brane brick models provide a direct link between the $2d$ $(0,2)$ theories and such setups. The M5-brane configuration is simply the M-theory lift of the Type IIA brane brick model, along the lines of \cite{Witten:1997sc}. 

Table \ref{M-theory_lift} shows how the building blocks of brane brick models are individually lifted. 
Before the lift, the D4-branes and the NS5-brane wrap supersymmetric cycles in $\mathbb{R}^3_{3,5,7}\times T^3_{2,4,6}$. It is natural to regard this 6-dimensional space as a (flat) CY 3-fold. The D4-branes wrap $T^3_{2,4,6}$ which is a special Lagrangian 3-cycle, while the NS5-brane wraps the holomorphic 4-cycle $\Sigma$. 

The D4-branes lift to M5-branes wrapped over the 4-torus $T^4_{2,4,6,10}$ consisting of the original brane brick model 3-torus together with the M-theory circle. The NS5-brane becomes an M5-brane wrapping the 4-cycle $\Sigma$. As in the M-theory lift of \cite{Witten:1997sc}, the two types of 4-cycles merge into a single one. This agrees with the fact that M5-branes wrapping a coassociative 4-cycle in a $G_2$ holonomy manifold gives rise to a 2d $(0,2)$ field theory \cite{Gauntlett:2000ng,Gauntlett:2001jj,Benini:2013cda,Gadde:2013sca}. In the current set-up, the unified M5-brane wraps a coassociative 4-cycle $M_4$ in $\mathbb{R}^3_{3,5,7}\times T^4_{2,4,6,10}$ regarded as a (flat) $G_2$ holonomy manifold.  

\begin{table}[ht!!]
\centering
\begin{tabular}{l|cccccccccc|c}
\; & 0 & 1 & 2 & 3 & 4 & 5 & 6 & 7 & 8 & 9 & 10 \\
\hline
$\text{M5}_\text{D4}$ & $\times$ & $\times$ & $\times$ & $\cdot$ & $\times$ & $\cdot$ & $\times$ & $\cdot$ & $\cdot$ & $\cdot$ & $\times$ \\
$\text{M5}_\text{NS5}$ & $\times$ & $\times$ & \multicolumn{6}{c}{----------- \ $\Sigma$ \ ------------} & $\cdot$ & $\cdot$ & $\cdot$ \\
\end{tabular}
\caption{M-theory lift of the building blocks of a brane brick model.}
\label{M-theory_lift}
\end{table}

As part of the information that can be gleaned from $M_4$, in \cite{Gadde:2013sca} it was suggested that the Betti number $b_2^-$ is the number of Fermi multiplets. It is not clear whether we can directly apply this relation to the field content of the brane brick models. The M-theory lift \cite{Witten:1997sc} tends to probe the IR dynamics of a gauge theory, whereas our prescription for constructing the brane brick model specifies the UV field content of the gauge theory. Given the transparent relation between brane brick models and gauge theories, it is natural to conjecture that, for the class of theories under consideration, they provide the analogue of a simplicial decomposition of $M_4$. It would be interesting to investigate whether this is indeed the case and what new lessons can be learnt from this line of thought.

\section{Conclusions}

\label{section_conclusions}

We applied mirror symmetry to the study of the $2d$ $(0,2)$ gauge theories on D1-branes probing toric CY 4-folds. The mirror configuration consists of D5-branes wrapping $S^4$'s. These $S^4$'s are in one-to-one correspondence with $S^2$'s on the holomorphic surface $\Sigma$, given by $P(x,y,z)=0$. The gauge theory is determined by how the $S^2$'s intersect on $\Sigma$. A significant development introduced by our work is the identification of the type of matter fields, Fermi or chiral, based on the sign of the intersections. We exploited the concept of tomography to get a detailed understanding of the geometry of the D5-branes. Combined with the coamoeba projection, tomography provides a systematic approach for constructing brane brick models starting from geometry.

We also explained how $2d$ $(0,2)$ triality is realized in terms of geometric transitions in the mirror geometry. Our analysis applies to generic trialities, generalizing the earlier work in \cite{Franco:2016nwv}, which was restricted to toric phases. Perhaps one of the most remarkable insights of mirror symmetry in this context is that it provides a geometric unification of field theory dualities in different dimensions. Mirror symmetry naturally explains why $(10-2n)$-dimensional field theories, which are associated with CY $n$-folds, exhibit duality symmetries of order $n-1$. Extrapolating these ideas to $D(-1)$-branes on CY 5-folds leads us to conjecture quadrality for $\mathcal{N}=1$ matrix models \cite{quadrality}.

Finally, we discussed how the M-theory lift of brane brick models connects with the classification of $2d$ $(0,2)$ theories in terms of 4-manifolds.

\acknowledgments

We would like to thank S. Gukov, M. Romo and P. Putrov for useful and enjoyable discussions. We are also grateful to D. Ghim for collaboration on related topics. We gratefully acknowledge support from the Simons Center for Geometry and Physics, Stony Brook University, where some of the research for this paper was performed during the 2016 Simons Summer Workshop. The work of S. F. is supported by the U.S. National Science Foundation grant PHY-1518967 and by a PSC-CUNY award. The work of S. L. was supported by Samsung Science and Technology Foundation under Project Number SSTF-BA1402-08. The work of S. L. was also performed in part at the Institute for Advanced Study supported by the IBM Einstein Fellowship of the Institute for Advanced Study, and at the Aspen Center for Physics supported by National Science Foundation grant PHY-1066293. The work of C.V. is supported in part by NSF grant PHY-1067976.


\newpage

\appendix

\section{Phases of $Q^{1,1,1}/\mathbb{Z}_2$}

\label{section_Q111_phases}

As mentioned earlier, $Q^{1,1,1}/\mathbb{Z}_{2}$ has several toric phases related by triality. Many of them were found and studied in \cite{Franco:2016nwv}. They are generated by different choices of coefficients in the Newton polynomial. Section \sref{sec:Q111Z2-A} discussed phase A in detail. In this appendix we review the mirror description of two additional phases, B and S, which are mentioned in the main body of the paper. We first present the periodic quivers and brane brick models, which for these theories were found in \cite{Franco:2016nwv}, and then briefly discuss how they are constructed using mirror symmetry.

\subsection*{Phase B}

\begin{figure}[ht]
\begin{center}
\resizebox{\hsize}{!}{
\includegraphics[trim=0cm 0cm 0cm 0cm,totalheight=10 cm]{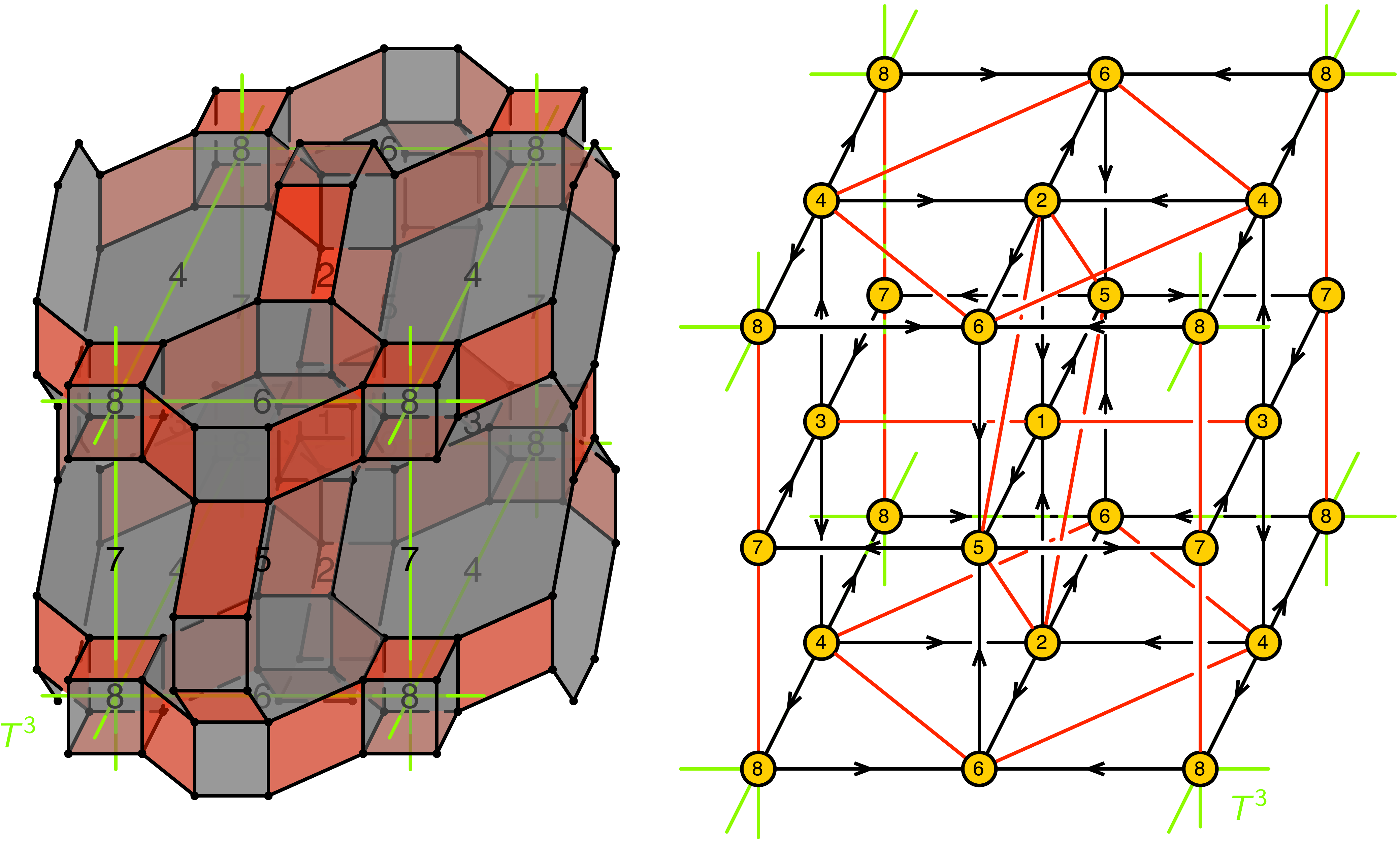}
}  
\vspace{-.3cm}\caption{Brane brick model and periodic quiver for phase B of $Q^{1,1,1}/\mathbb{Z}_2$. 
\label{periodic_quiver_Q111Z2_B}} 
 \end{center}
 \end{figure}

Phase B is obtained by starting from phase A, as given in \fref{periodic_quiver_Q111Z2_A}, and acting with a triality transformation on node 1. \fref{periodic_quiver_Q111Z2_B} shows the brane brick model and periodic quiver for this theory. The $J$- and $E$-terms are:
{\small
\beq
\begin{array}{lcccc}
& &  J  & &  E  
\\  
   \Lambda_{31}^{+} 
 & :\ \ \ &  
  X^{+}_{15} \cdot X^{-}_{57} \cdot X^{-}_{73} -
    X^{-}_{15} \cdot X^{-}_{57} \cdot X^{+}_{73}    
  & \ \ \ \ &   
   X^{+}_{34} \cdot X^{+}_{42} \cdot X^{-}_{21} - 
       X^{-}_{34} \cdot X^{+}_{42} \cdot X^{+}_{21} 
  \\
 \Lambda_{31}^{-} & :\ \ \ &  
 X^{-}_{15} \cdot X^{+}_{57} \cdot X^{+}_{73} -
 X^{+}_{15} \cdot X^{+}_{57} \cdot X^{-}_{73}    
 & \ \ \ \ &   
   X^{+}_{34} \cdot X^{-}_{42} \cdot X^{-}_{21} -
   X^{-}_{34} \cdot X^{-}_{42} \cdot X^{+}_{21}
   \\  
  \Lambda_{78}^{+} & :\ \ \ &  
  X^{+}_{86} \cdot X^{-}_{65} \cdot X^{-}_{57}  -
  X^{-}_{86} \cdot X^{-}_{65} \cdot X^{+}_{57}   
  & \ \ \ \ & 
  X^{+}_{73} \cdot X^{+}_{34} \cdot X^{-}_{48} -    
  X^{-}_{73} \cdot X^{+}_{34} \cdot X^{+}_{48}  
  \\
   \Lambda_{78}^{-} 
   & :\ \ \ &  
   X^{-}_{86} \cdot X^{+}_{65} \cdot X^{+}_{57} -
    X^{+}_{86} \cdot X^{+}_{65} \cdot X^{-}_{57}    
    & \ \ \ \ &  
X^{+}_{73} \cdot X^{-}_{34} \cdot X^{-}_{48}  - 
X^{-}_{73} \cdot X^{-}_{34} \cdot X^{+}_{48}  
   \\
   \Lambda_{46}^{++} & :\ \ \ &  
   X^{+}_{65} \cdot X^{-}_{57} \cdot X^{-}_{73} \cdot X^{-}_{34}  -
   X^{-}_{65} \cdot X^{-}_{57} \cdot X^{-}_{73} \cdot X^{+}_{34}     
   & \ \ \ \ &   
X_{42}^{+} \cdot X_{26}^{+} - X_{48}^{+} \cdot X_{86}^{+}   
\\
         \Lambda_{46}^{--} & :\ \ \ &  X^{+}_{65} \cdot X^{+}_{57} \cdot X^{+}_{73} \cdot X^{-}_{34} - X^{-}_{65} \cdot X^{+}_{57} \cdot X^{+}_{73} \cdot X^{+}_{34}   
& \ \ \ \ &   
X_{42}^{-} \cdot X_{26}^{-} - X_{48}^{-} \cdot X_{86}^{-}   
\\
   \Lambda_{46}^{+-} & :\ \ \ &  
X^{-}_{65} \cdot X^{-}_{57} \cdot X^{+}_{73} \cdot X^{+}_{34}   -
   X^{+}_{65} \cdot X^{-}_{57} \cdot X^{+}_{73} \cdot X^{-}_{34}    
& \ \ \ \ &   
X_{42}^{+} \cdot X_{26}^{-} - X_{48}^{-} \cdot X_{86}^{+}      
\\
   \Lambda_{46}^{-+} & :\ \ \ &  X^{-}_{65} \cdot X^{+}_{57} \cdot X^{-}_{73} \cdot X^{+}_{34} - X^{+}_{65} \cdot X^{+}_{57} \cdot X^{-}_{73} \cdot X^{-}_{34}   & \ \ \ \ &   
X_{42}^{-} \cdot X_{26}^{+} - X_{48}^{+} \cdot X_{86}^{-}    
\\
\Lambda_{25}^{++} & :\ \ \ &  
X^{+}_{57} \cdot X^{-}_{73} \cdot X^{-}_{34} \cdot X^{-}_{42}   -
X^{-}_{57} \cdot X^{-}_{73} \cdot X^{-}_{34} \cdot X^{+}_{42} 
& \ \ \ \ &   X^{+}_{26} \cdot X^{+}_{65} - X^{+}_{21} \cdot X^{+}_{15}   \\
\Lambda_{25}^{--} & :\ \ \ &  X^{+}_{57} \cdot X^{+}_{73} \cdot X^{+}_{34} \cdot X^{-}_{42} - X^{-}_{57} \cdot X^{+}_{73} \cdot X^{+}_{34} \cdot X^{+}_{42}   & \ \ \ \ &   X^{-}_{26} \cdot X^{-}_{65} - X^{-}_{21} \cdot X^{-}_{15}   \\
   \Lambda_{25}^{+-} 
   & :\ \ \ &  
   X^{-}_{57} \cdot X^{-}_{73} \cdot X^{+}_{34} \cdot X^{+}_{42}   -
   X^{+}_{57} \cdot X^{-}_{73} \cdot X^{+}_{34} \cdot X^{-}_{42}    & \ \ \ \ &   
   X^{+}_{26} \cdot X^{-}_{65} - X^{-}_{21} \cdot X^{+}_{15}   
   \\
   \Lambda_{25}^{-+} & :\ \ \ &  X^{-}_{57} \cdot X^{+}_{73} \cdot X^{-}_{34} \cdot X^{+}_{42} - X^{+}_{57} \cdot X^{+}_{73} \cdot X^{-}_{34} \cdot X^{-}_{42}   & \ \ \ \ & X^{-}_{26} \cdot X^{+}_{65} - X^{+}_{21} \cdot X^{-}_{15}   \\
\end{array}
\label{q111z2b-je}
\eeq
}

\begin{figure}[H]
\begin{center}
\includegraphics[width=10cm]{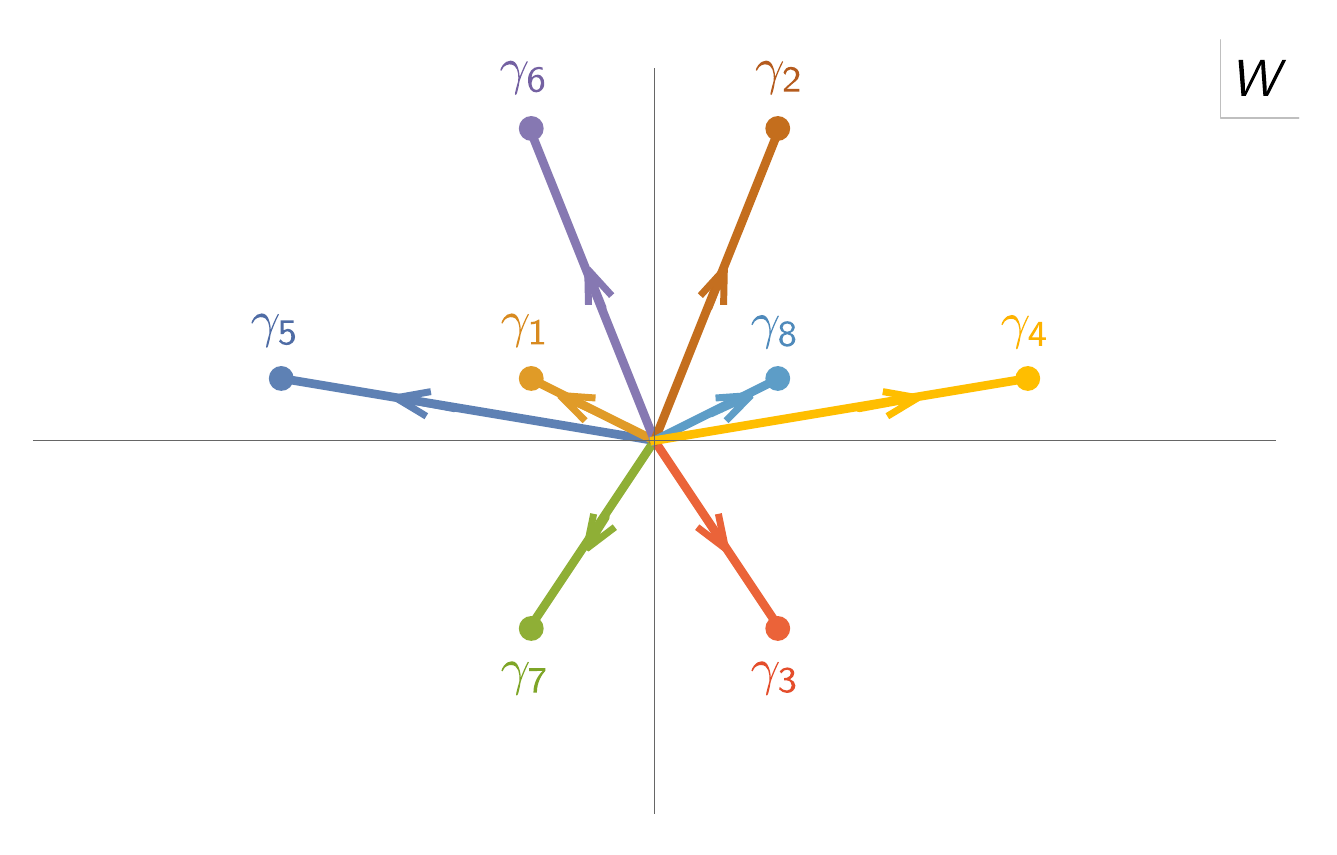}
\caption{Vanishing paths for phase B of $Q^{1,1,1}/\mathbb{Z}_2$.
\label{vcq111z2Bew}}
 \end{center}
 \end{figure}

We can obtain this theory by picking the Newton polynomial as follows
\beal{es200B1}
P(x,y,z) =  \left(x + \frac{1}{x}\right) + (1-i) \left(y + \frac{1}{y}\right) + (1+i) \left(z + \frac{1}{z}\right) + i ~.~
\eea
The eight critical points are $(x^*,y^*,z^*) = (\pm 1, \pm 1, \pm 1) $ and the critical values are
\beal{es200B3}
W^* = 
\pm 2+i~,~
\pm 2-3i~,~
\pm 2+5i~,~
\pm 6+i~,~
\eea
\fref{vcq111z2Bew} shows the vanishing paths.

Let us study how the tomographies reconstruct the periodic quiver.

\paragraph{The $x$-tomography.}

\begin{figure}[H]
\begin{center}
\includegraphics[width=8cm]{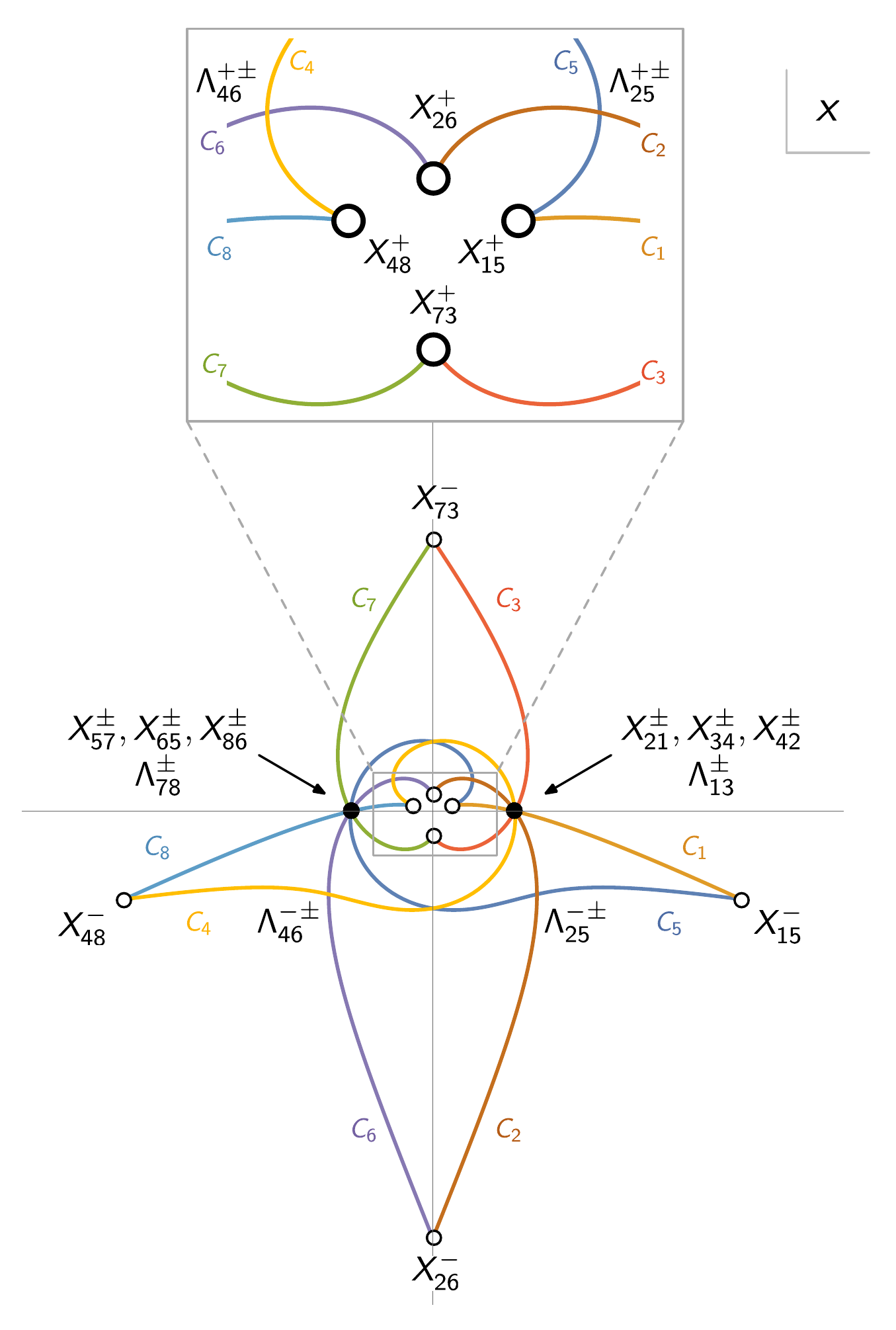}
\caption{The $x$-tomography for phase B of $Q^{1,1,1}/\mathbb{Z}_2$ on the $x$-plane. We indicate the fields associated with each intersection.
\label{mirror_fields_Q111Z2_B_x}}
 \end{center}
 \end{figure}

\fref{mirror_fields_Q111Z2_B_x} shows the $x$-tomography. Nodes in the periodic quiver are arranged in two layers along the $x$ direction, consisting of $(1,2,3,4)$ at $\mathrm{arg}(x)=0$ and $(5,6,7,8)$ at $\mathrm{arg}(x)=\pi$. These layers correspond to the points $x=\pm 1$. At these points, the Newton polynomial becomes equivalent to the one for phase 1 of $F_0$, implying that each of these layers contains 8 fields and that the pairwise intersections are double. More specifically, there are 6 chirals and 2 Fermis on each layer. The fields between the layers correspond to the intersections with $-\pi<\mathrm{arg}(x)<0$ and $0<\mathrm{arg}(x)<\pi$. All bulk intersections in these regions are double. 

Two cycles intersect only if they meet in all three tomographies. For example, while $C_4$ and $C_5$ seem to intersect in the $x$-tomography, they are clearly separated in the $y$- and $z$-tomographies, hence $\langle C_4,C_5\rangle=0$.

\paragraph{The $y$-tomography.}

\begin{figure}[H]
\begin{center}
\includegraphics[width=8cm]{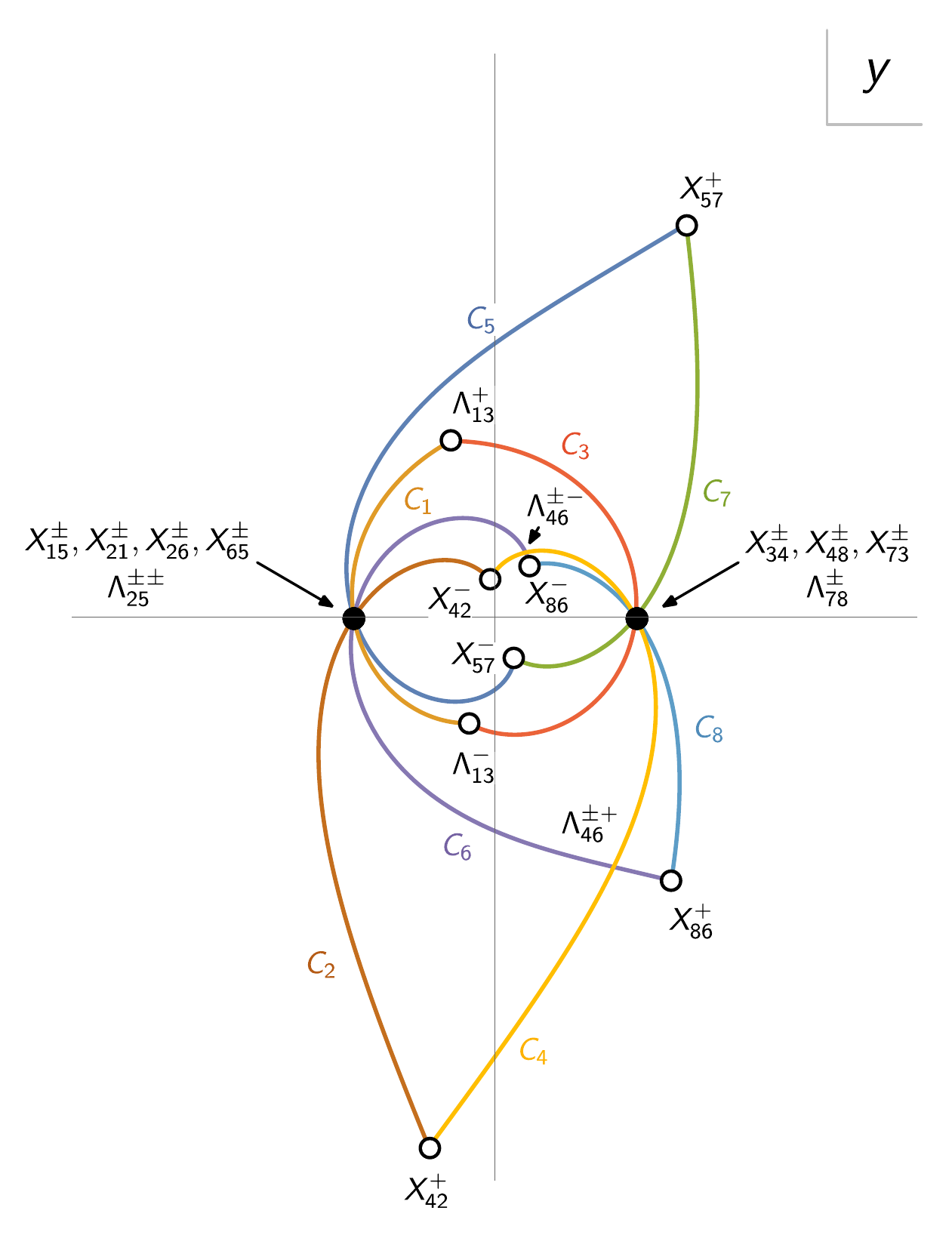}
\caption{The $y$-tomography for phase B of $Q^{1,1,1}/\mathbb{Z}_2$. We indicate the fields associated with each intersection.
\label{mirror_fields_Q111Z2_B_y}}
 \end{center}
 \end{figure}

\fref{mirror_fields_Q111Z2_B_y} gives the $y$-tomography. The nodes in the periodic quiver form two layers in the $y$ direction, consisting of $(3,4,7,8)$ at $\mathrm{arg}(y)=0$ and $(1,2,5,6)$ at $\mathrm{arg}(y)=\pi$. The $(3,4,7,8)$ layer sits at $y=1$, where the Newton polynomial reduces to the one for phase 1 of $F_0$. We thus conclude that this layer contains 8 fields. Detailed analysis reveals that these are 6 chiral and 2 Fermi fields. At the $(1,2,5,6)$ layer, which is located at $y=-1$, the Newton polynomial becomes instead that for phase 2 of $F_0$. This implies that there are 12 fields on this layer. 

\paragraph{The $z$-tomography.}

\begin{figure}[H]
\begin{center}
\includegraphics[width=8cm]{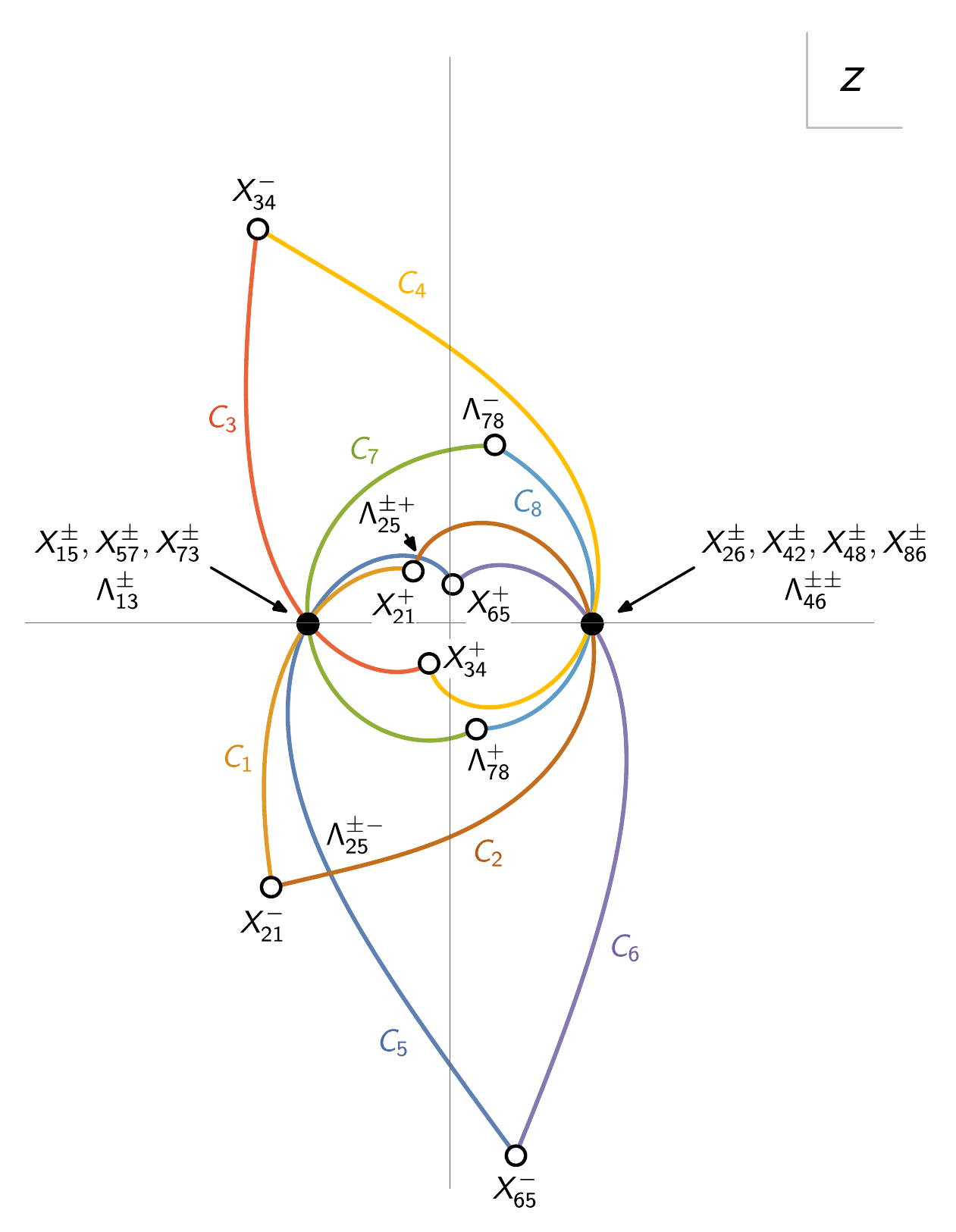}
\caption{The $z$-tomography for  phase B of $Q^{1,1,1}/\mathbb{Z}_2$. We indicate the fields associated with each intersection.
\label{mirror_fields_Q111Z2_B_z}}
 \end{center}
 \end{figure}

The $z$-tomography is shown in \fref{mirror_fields_Q111Z2_B_z}. Up to relabeling of cycles and reflection the configuration is identical to the one on the $y$-plane, so the previous analysis extends with minor modifications.

\subsection*{Phase S}

\begin{figure}[H]
\begin{center}
\resizebox{\hsize}{!}{
\includegraphics[trim=0cm 0cm 0cm 0cm,totalheight=10 cm]{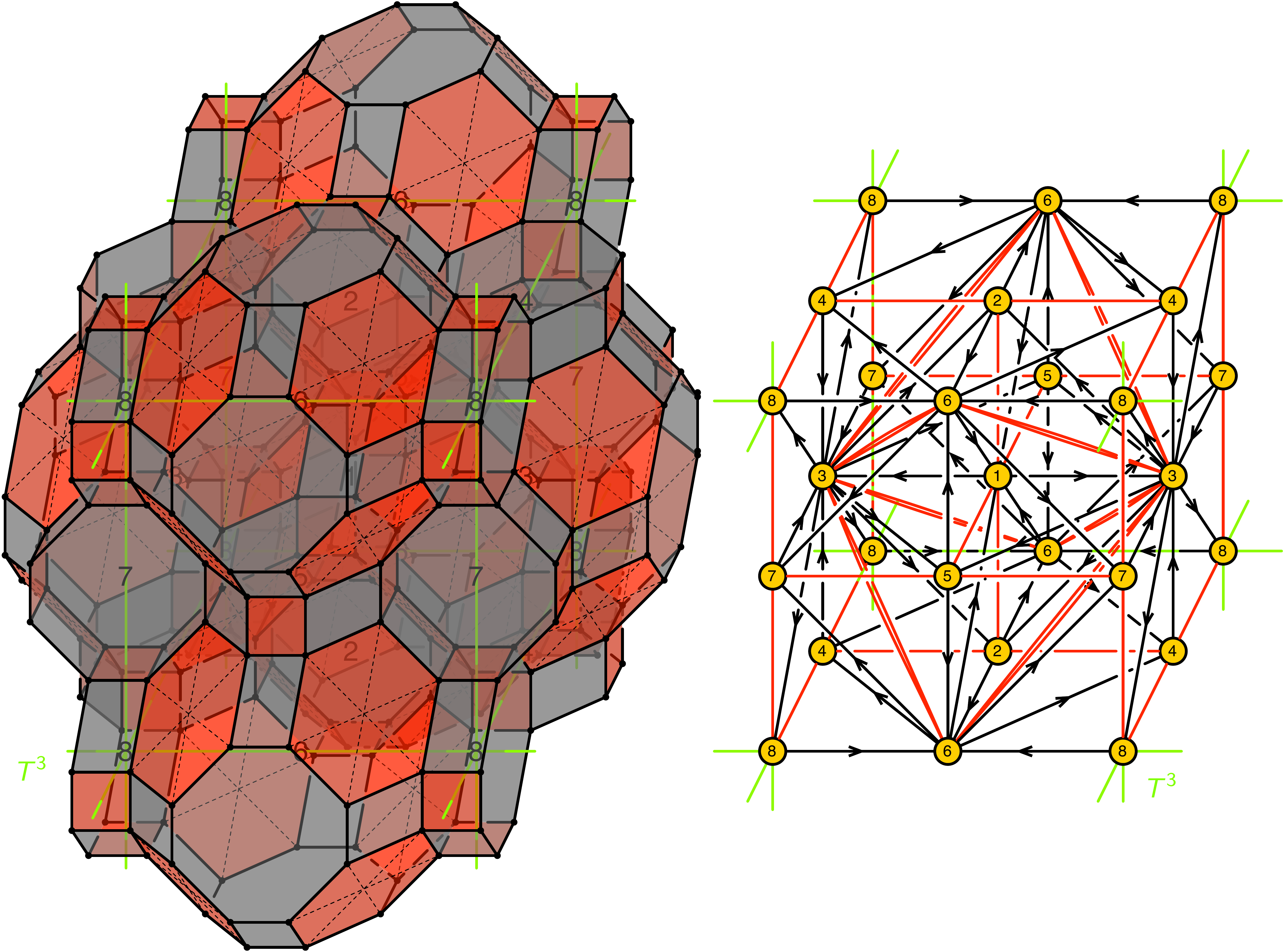}
}  
\vspace{-.3cm}\caption{Brane brick model and periodic quiver for phase S of $Q^{1,1,1}/\mathbb{Z}_2$. 
\label{periodic_quiver_Q111Z2_S}}
 \end{center}
 \end{figure}

Starting from phase A, as given in \fref{periodic_quiver_Q111Z2_A}, and performing consecutive triality transformations on nodes 4 and 5, we obtain a new phase that is described by the brane brick model and periodic quiver shown in \fref{periodic_quiver_Q111Z2_S}. We denote this phase $S$, for {\it symmetric}, since it has a manifest octahedral symmetry. The $J$- and $E$-terms are:

\beq
\begin{array}{rcrc}
& J & & E 
\\
\Lambda^{+00}_{84} : \ \ \  &  
X^{00-}_{43} \cdot Y^{-0+}_{38} - X^{00+}_{43} \cdot Y^{-0-}_{38}     & \ \ \ \ & 
X^{0-0}_{86} \cdot Y^{++0}_{64}   - X^{0+0}_{86}\cdot Y^{+-0}_{64}     
\\ 
\Lambda^{-00}_{84}  : \ \ \  &  
X^{00-}_{43}\cdot Y^{+0+}_{38}  - X^{00+}_{43}\cdot Y^{+0-}_{38}    & \ \ \ \ & 
X^{0+0}_{86}\cdot Y^{--0}_{64}   - X^{0-0}_{86} \cdot Y^{-+0}_{64}   
\\
\Lambda^{+00}_{15}  : \ \ \  &  
X^{00-}_{56} \cdot Y^{-0+}_{61} - X^{00+}_{56}\cdot Y^{-0-}_{61}     & \ \ \ \ & 
X^{0-0}_{13} \cdot Y^{++0}_{35}   - X^{0+0}_{13}\cdot Y^{+-0}_{35}      
\\ 
\Lambda^{-00}_{15}  : \ \ \  &  
X^{00-}_{56}\cdot Y^{+0+}_{61}  - X^{00+}_{56}\cdot Y^{+0-}_{61}    & \ \ \ \ & 
X^{0+0}_{13}\cdot Y^{--0}_{35}    - X^{0-0}_{13} \cdot Y^{-+0}_{35}    
\\
\Lambda^{0+0}_{57}  : \ \ \  &  
X^{-00}_{73} \cdot Y^{+-0}_{35}  - X^{+00}_{73} \cdot Y^{--0}_{35}   & \ \ \ \ & 
X^{00-}_{56} \cdot Y^{0++}_{67}   - X^{00+}_{56}\cdot Y^{0+-}_{67}    
\\ 
\Lambda^{0-0}_{57} : \ \ \  &  
X^{-00}_{73}  \cdot Y^{++0}_{35}- X^{+00}_{73} \cdot Y^{-+0}_{35}   & \ \ \ \ & 
X^{00+}_{56}\cdot Y^{0--}_{67}   - X^{00-}_{56}  \cdot Y^{0-+}_{67}   
\\
\Lambda^{0+0}_{42}  : \ \ \  &  
X^{-00}_{26} \cdot Y^{+-0}_{64}  - X^{+00}_{26}  \cdot Y^{--0}_{64}   & \ \ \ \ & 
X^{00-}_{43}\cdot Y^{0++}_{32}   - X^{00+}_{43}\cdot Y^{0+-}_{32}     
\\ 
\Lambda^{0-0}_{42}  : \ \ \  &  
X^{-00}_{26}   \cdot Y^{++0}_{64}- X^{+00}_{26}  \cdot Y^{-+0}_{64}   & \ \ \ \ & 
X^{00+}_{43}\cdot Y^{0--}_{32}    - X^{00-}_{43}  \cdot Y^{0-+}_{32}    
\\
\Lambda^{00+}_{21}  : \ \ \  &  
X^{0-0}_{13} \cdot Y^{0+-}_{32}   - X^{0+0}_{13} \cdot Y^{0--}_{32}    & \ \ \ \ & 
X^{-00}_{26} \cdot Y^{+0+}_{61}  - X^{+00}_{26} \cdot Y^{-0+}_{61}    
\\ 
\Lambda^{00-}_{21}  : \ \ \  &  
X^{0-0}_{13}\cdot  Y^{0++}_{32} - X^{0+0}_{13}\cdot  Y^{0-+}_{32}  & \ \ \ \ & 
X^{+00}_{26} \cdot Y^{-0-}_{61}  - X^{-00}_{26}\cdot Y^{+0-}_{61}    
\\
\Lambda^{00+}_{78}  : \ \ \  &  
X^{0-0}_{86}\cdot Y^{0+-}_{67}   - X^{0+0}_{86} \cdot Y^{0--}_{67}    & \ \ \ \ & 
X^{-00}_{73}\cdot Y^{+0+}_{38}  - X^{+00}_{73} \cdot Y^{-0+}_{38}    
\\ 
\Lambda^{00-}_{78} : \ \ \  &  
X^{0-0}_{86}\cdot  Y^{0++}_{67} - X^{0+0}_{86}\cdot  Y^{0-+}_{67}  & \ \ \ \ & 
X^{+00}_{73} \cdot Y^{-0-}_{38}  - X^{-00}_{73}\cdot Y^{+0-}_{38}    
\\[.3cm]
\Psi^{+--}_{36} : \ \ \  &  
Y^{-+0}_{64}\cdot X^{00+}_{43}   - Y^{0++}_{67} \cdot X^{-00}_{73}    & \ \ \ \ & 
Y^{0--}_{32} \cdot X^{+00}_{26}  - Y^{+0-}_{38} \cdot X^{0-0}_{86}   
\\ 
\Psi^{-++}_{36} : \ \ \  &  
Y^{0--}_{67}\cdot X^{+00}_{73}   - Y^{+-0}_{64}\cdot X^{00-}_{43}     & \ \ \ \ & 
Y^{0++}_{32}\cdot X^{-00}_{26}  - Y^{-0+}_{38}\cdot X^{0+0}_{86}    
\\
\Psi^{-+-}_{36} : \ \ \  &  
Y^{0-+}_{67} \cdot X^{+00}_{73}  - Y^{+0+}_{61} \cdot X^{0-0}_{13}    & \ \ \ \ & 
Y^{-0-}_{38} \cdot X^{0+0}_{86} - Y^{-+0}_{35} \cdot X^{00-}_{56}  
\\ 
\Psi^{+-+}_{36} : \ \ \  &  
Y^{-0-}_{61} \cdot X^{0+0}_{13}   - Y^{0+-}_{67}\cdot X^{-00}_{73}   & \ \ \ \ & 
Y^{+0+}_{38} \cdot X^{0-0}_{86}   - Y^{+-0}_{35} \cdot X^{00+}_{56} 
\\
\Psi^{--+}_{36} : \ \ \  &  
Y^{+0-}_{61}  \cdot X^{0+0}_{13} - Y^{++0}_{64}\cdot X^{00-}_{43}    & \ \ \ \ & 
Y^{--0}_{35}\cdot X^{00+}_{56}   - Y^{0-+}_{32} \cdot X^{-00}_{26}   
\\ 
\Psi^{++-}_{36} : \ \ \  &  
Y^{--0}_{64} \cdot X^{00+}_{43}  - Y^{-0+}_{61} \cdot X^{0-0}_{13}    & \ \ \ \ & 
Y^{++0}_{35} \cdot X^{00-}_{56} - Y^{0+-}_{32}  \cdot X^{+00}_{26}  
\\
\Psi^{+++}_{36} : \ \ \  &  
Y^{--0}_{64}\cdot X^{00-}_{43}  - Y^{0--}_{67}\cdot X^{-00}_{73}    & \ \ \ \ & 
Y^{++0}_{35}\cdot X^{00+}_{56}  - Y^{+0+}_{38} \cdot X^{0+0}_{86}   
\\ 
\Psi^{---}_{36} : \ \ \  &  
Y^{0++}_{67} \cdot X^{+00}_{73} - Y^{++0}_{64} \cdot X^{00+}_{43}     & \ \ \ \ & 
Y^{--0}_{35} \cdot X^{00-}_{56}  - Y^{-0-}_{38} \cdot X^{0-0}_{86} \\[.3cm]
\Psi^{+--}_{63} : \ \ \  &  
Y^{-+0}_{35}\cdot X^{00+}_{56}   - Y^{0++}_{32} \cdot X^{-00}_{26}    & \ \ \ \ & 
Y^{0--}_{67} \cdot X^{+00}_{73}  - Y^{+0-}_{61} \cdot X^{0-0}_{13}   
\\ 
\Psi^{-++}_{63} : \ \ \  &  
Y^{0--}_{32}\cdot X^{+00}_{26}   - Y^{+-0}_{35}\cdot X^{00-}_{56}     & \ \ \ \ & 
Y^{0++}_{67}\cdot X^{-00}_{73}  - Y^{-0+}_{61}\cdot X^{0+0}_{13}    
\\
\Psi^{-+-}_{63} : \ \ \  &  
Y^{0-+}_{32} \cdot X^{+00}_{26}  - Y^{+0+}_{38} \cdot X^{0-0}_{86}    & \ \ \ \ & 
Y^{-0-}_{61} \cdot X^{0+0}_{13} - Y^{-+0}_{64} \cdot X^{00-}_{43}  
\\ 
\Psi^{+-+}_{63} : \ \ \  &  
Y^{-0-}_{38} \cdot X^{0+0}_{86}   - Y^{0+-}_{32} \cdot X^{-00}_{26}   & \ \ \ \ & 
Y^{+0+}_{61} \cdot X^{0-0}_{13}   - Y^{+-0}_{64} \cdot X^{00+}_{43} 
\\
\Psi^{--+}_{63} : \ \ \  &  
Y^{+0-}_{38}  \cdot X^{0+0}_{86} - Y^{++0}_{35}\cdot X^{00-}_{56}    & \ \ \ \ & 
Y^{--0}_{64}\cdot X^{00+}_{43}   - Y^{0-+}_{67} \cdot X^{-00}_{73}   
\\ 
\Psi^{++-}_{63} : \ \ \  &  
Y^{--0}_{35} \cdot X^{00+}_{56}  - Y^{-0+}_{38} \cdot X^{0-0}_{86}    & \ \ \ \ & 
Y^{++0}_{64} \cdot X^{00-}_{43} - Y^{0+-}_{67}  \cdot X^{+00}_{73}  
\\
\Psi^{+++}_{63} : \ \ \  &  
Y^{--0}_{35}\cdot X^{00-}_{56}  - Y^{0--}_{32}\cdot X^{-00}_{26}    & \ \ \ \ & 
Y^{++0}_{64}\cdot X^{00+}_{43}  - Y^{+0+}_{61} \cdot X^{0+0}_{13}   
\\ 
\Psi^{---}_{63} : \ \ \  &  
Y^{0++}_{32} \cdot X^{+00}_{26} - Y^{++0}_{35} \cdot X^{00+}_{56}     & \ \ \ \ & 
Y^{--0}_{64} \cdot X^{00-}_{43}  - Y^{-0-}_{61} \cdot X^{0-0}_{13} 
\end{array}
\label{q111/Z2_S_JE}
\eeq
Subindices specify gauge quantum numbers of fields and superindices indicate their orientations along the $(x,y,z)$ directions of $T^3$.

\begin{figure}[H]
\begin{center}
\resizebox{0.7\hsize}{!}{
\includegraphics[trim=0cm 0cm 0cm 0cm,totalheight=10 cm]{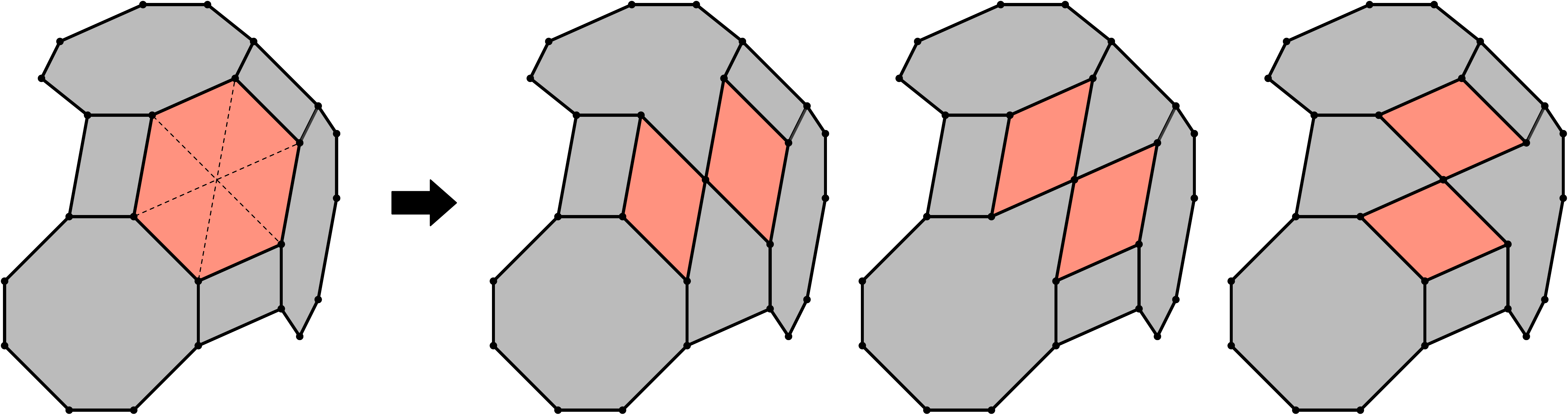}
}  
\vspace{-.3cm}\caption{The hexagonal Fermi faces in \fref{periodic_quiver_Q111Z2_S} represent pairs of regular 4-sided Fermi faces. The figures on the right correspond to the three possible local choices of $J$- and $E$-terms.
\label{fhexagonfermi}}
 \end{center}
 \end{figure}

The red hexagons in the brane brick model of \fref{periodic_quiver_Q111Z2_S} represent the pairs of coincident Fermi fields in the periodic quiver. Any choice of $J$- and $E$-terms leads to a spontaneous breaking of the octahedral symmetry \cite{Franco:2016nwv}. The resulting chiral ring is, however, invariant under the full octahedral symmetry. The hexagonal Fermi faces should be regarded as representing pairs of regular 4-sided Fermi faces, as shown in \fref{fhexagonfermi}. 
 
A possible choice of coefficients in the Newton polynomial leading to this phase is
\beal{es200a1}
P(x,y,z) =  \left(x + \frac{1}{x}\right) + 2 \left(y + \frac{1}{y}\right) -\frac{1}{2} \left(z + \frac{1}{z}\right) + e ~,~
\eea
where we left the parameter $e$ undetermined for later convenience. 
The eight critical points are $(x^*,y^*,z^*) = (\pm 1, \pm 1, \pm 1) $ and the critical values are
\beal{es200a3}
W^* = 
\pm 7 + e~,~
\pm 5 + e~,~
\pm 3 + e~,~
\pm 1 + e\,.~
\eea
It is easy to see that one can continuously connect the Newton polynomial of phase A \eqref{P-Q111-ansatz} to that of phase S \eqref{es200a1} by tuning the coefficients of $(z+1/z)$ and the constant $e$ while keeping the coefficients of $(x+1/x)$ and $(y+1/y)$ fixed. 

\begin{figure}[H]
\begin{center}
\includegraphics[width=9cm]{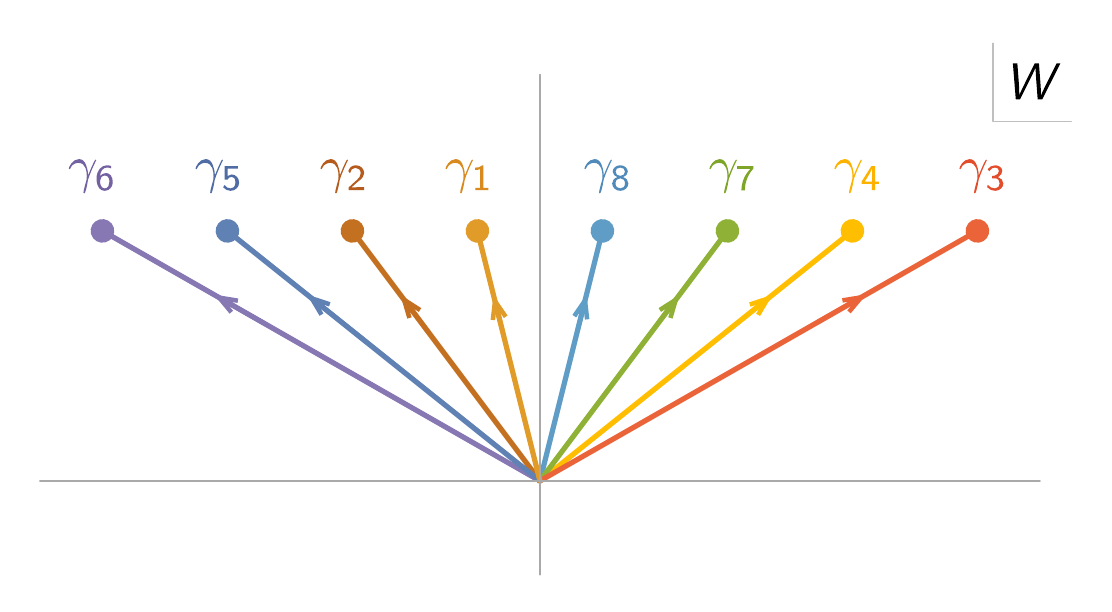}
\caption{Naive straight vanishing paths for phase S of $Q^{1,1,1}/\mathbb{Z}_2$. 
\label{evq111z2ew}}
 \end{center}
 \end{figure}

\begin{figure}[H]
\begin{center}
\includegraphics[width=15cm]{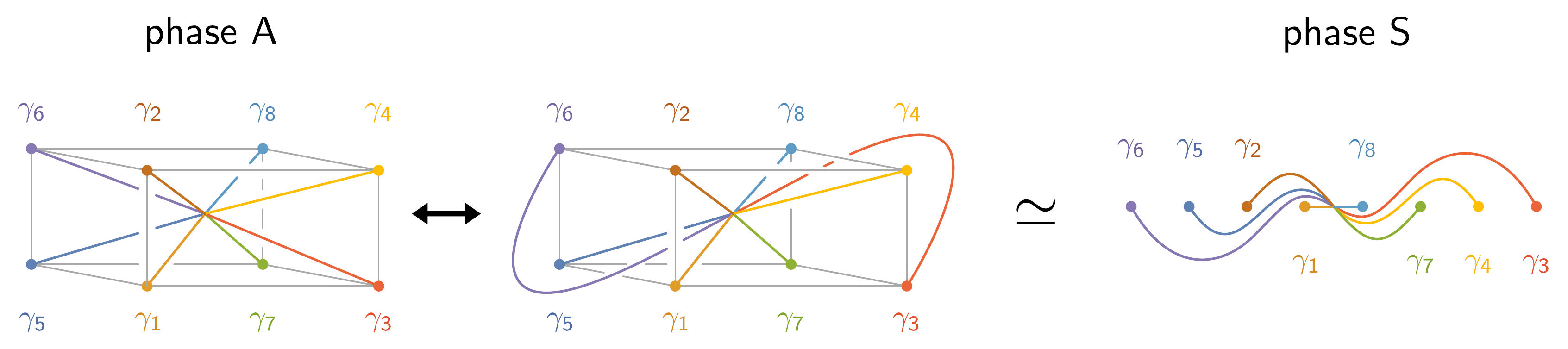}
\caption{Curved vanishing paths for phase S of $Q^{1,1,1}/\mathbb{Z}_2$ 
suggested by triality from phase A and the cyclic ordering condition.}
\label{W-triality-AS}
 \end{center}
 \end{figure}

Naively, one would use a straight segment on the $W$-plane as the vanishing path for each critical point. If we set $e=0$, many straight paths would simultaneously overlap. If instead we set $e=4i$, we would obtain the paths shown in \fref{evq111z2ew}. However, the rules for the triality transformation and cyclic ordering of vanishing paths explained in section \sref{section_triality} imply we should consider different, {\em curved}, paths as shown in \fref{W-triality-AS}. The curved paths may initially look contrived. However, if we reconsider the geometric origin of the vanishing paths, we realize that there is no a priori reason for them to be straight. In simple examples, such as local $\mathbb{CP}^3$, the symmetries of the toric diagram guarantee that the vanishing paths are straight. In general, however, the precise shapes of vanishing paths are determined by the fact that the 4-cycles $\mathcal{C}_i$ are calibrated by the real part of the holomorphic 4-form. This condition turns into a non-linear partial differential equation that is in general not solvable by elementary means. Fortunately, the intersection numbers that determine the field types are topological and insensitive to small deformations of the paths.

For phases A and B of $Q^{1,1,1}/\mathbb{Z}_2$, we are confident that the straight vanishing paths that we considered are homotopic to their true shapes. In contrast, for phase S, triality strongly suggests that we cannot approximate the true configuration by straight paths. It would be interesting to confirm the curved configuration in \fref{W-triality-AS} from an argument independent from triality.

Having discussed the main novel feature of phase S, we omit its tomographies for two practical reasons. One is simply that their determination is more time consuming for curved paths. In addition, the intersections corresponding to degenerate Fermi lines in the periodic quiver shown in \fref{periodic_quiver_Q111Z2_S} are more difficult to resolve than well separated intersections.

\section{Sign of Intersection and Field Type \label{appendix:angles}} 

We can simulate the local geometry of intersections of special Lagrangian $n$-cycles in a CY $n$-fold by branes intersecting at $SU(n)$ angles in $\mathbb{C}^n$. The discussion below is a straightforward generalization of  \cite{Berkooz:1996km}.  

Consider two D5-branes sharing the $x^0$ and $x^9$ directions and each occupying different real 4-planes in the transverse $\mathbb{C}^4$. Let $z_i = x^i + i x^{i+4} \equiv x^i + i y^{i} $ be complex coordinates of $\mathbb{C}^4$. Assume that one of the D5-branes, call it brane A, is extended along the $x^i$-directions. The other D5-brane, call it brane B, is rotated with respect to brane A by an $SU(4)$ rotation of the form $U(\theta) = \mathrm{diag}(e^{i\theta_1}, e^{i\theta_2},e^{i\theta_3},e^{i\theta_4})$ with $\theta_1+\theta_2+\theta_3+\theta_4 = 0$ (mod $2\pi$).

On each complex plane, the complex bosonic field associated to an open string
stretching from brane A to brane B satisfies the boundary conditions
\begin{align}
\begin{split}
&\mathrm{Im}(Z)|_{\sigma = 0 } = 0 \,, \quad 
\mathrm{Re}(\partial_\sigma Z) |_{\sigma = 0 } = 0 \,, 
\\
&\mathrm{Im}(e^{-i\theta} Z)|_{\sigma = \pi } = 0 \,, \quad 
\mathrm{Re}(e^{-i\theta} \partial_\sigma Z) |_{\sigma = \pi } = 0 \,.
\end{split}
\end{align}
The boundary condition shift the mode expansion by $\alpha = \theta/\pi$, and we obtain 
\begin{align}
\begin{split}
Z(\tau,\sigma) 
=\; &
(a\, e^{-i|\alpha|\tau} + a^\dagger e^{+i|\alpha|\tau}) e^{i\alpha \sigma} 
\\
& +\sum_{n\in\mathbb{Z}\backslash \{0\}} \left( \tilde{z}_{n-\alpha} e^{-i(n-\alpha)(\tau+\sigma)} + z_{n+\alpha} e^{-i(n+\alpha)(\tau-\sigma)} \right) \,.
\end{split}
\end{align}
The mode operators satisfy 
\begin{align}
[ a, a^\dagger] = \frac{1}{|\alpha|} 
\,,
\quad
[ \tilde{z}_{r}, z_s ] = \frac{1}{r} \delta_{r+s,0} 
\,,
\quad 
(\tilde{z}_{r})^\dagger = z_{-r}\,. 
\end{align}
The mode expansion of the R-sector fermion is similar to that of the boson. Explicitly, 
\begin{align}
\begin{split}
\Psi(\tau,\sigma) 
=\; &
(\gamma\, e^{-i|\alpha|\tau} + \gamma^\dagger e^{+i|\alpha|\tau}) e^{i\alpha \sigma} 
\\
& +\sum_{n\in\mathbb{Z}\backslash \{0\}} \left( \tilde{\psi}_{n-\alpha} e^{-i(n-\alpha)(\tau+\sigma)} + \psi_{n+\alpha} e^{-i(n+\alpha)(\tau-\sigma)} \right) \,,
\end{split}
\end{align}
with 
\begin{align}
\{ \gamma, \gamma^\dagger \} = 1
\,,
\quad
\{ \tilde{\psi}_{r}, \psi_s \} =  \delta_{r+s,0} 
\,,
\quad 
(\tilde{\psi}_{r})^\dagger = \psi_{-r}\,. 
\end{align}
In the NS-sector, the mode expansion is given by  
\begin{align}
\Psi(\tau,\sigma) 
=\sum_{n\in\mathbb{Z}+1/2} \left( \tilde{\psi}_{n-\alpha} e^{-i(n-\alpha)(\tau+\sigma)} + \psi_{n+\alpha} e^{-i(n+\alpha)(\tau-\sigma)} \right) \,,
\end{align}
with 
\begin{align}
\{ \tilde{\psi}_{r}, \psi_s \} =  \delta_{r+s,0} 
\,,
\quad 
(\tilde{\psi}_{r})^\dagger = \psi_{-r}\,. 
\end{align}

Recall that the contribution to the worldsheet vacuum energy of a complex boson/NS-fermion field, whose modes are shifted by $\alpha$, is 
\begin{align}
\begin{split}
\epsilon_B(\alpha) &= -\frac{1}{12} + \frac{1}{2} |\alpha|( 1-|\alpha| ) \,,
\\
\epsilon_F(\alpha) &= +\frac{1}{12} - \frac{1}{2} \left( \frac{1}{4}-\alpha^2 \right) \,.
\end{split}
\end{align}
The total vacuum energy is
\begin{align}
E_0 = \frac{1}{2}(-1 + |\alpha_1| + |\alpha_2| + |\alpha_3| + |\alpha_4|) \,.
\end{align}
Without loss of generality, we can consider two distinct cases.

\subsubsection*{\underline{$\alpha_1 > \alpha_2 > \alpha_3 > 0 > \alpha_4$}}

Using the fact that $|\alpha_4| = - \alpha_4 = \alpha_1 + \alpha_2 + \alpha_3$, we can rewrite the vacuum energy as 
\begin{align}
E_0 = - \frac{1}{2} + \alpha_1 + \alpha_2 + \alpha_3  \,.
\end{align}
For now, let us assume that $\alpha_1 + \alpha_2 + \alpha_3 < 1/2$, so that the vacuum remains tachyonic (we will relax this restriction shortly). The first excited states, $\tilde{\psi}^i_{-1/2-\alpha_i}|0\rangle$ and $\psi^i_{-1/2+\alpha_i}|0\rangle$, have mass spectrum
\begin{align}
\begin{array}{|c|c|c|c|c|}
\hline
i & 1 & 2 & 3 & 4
\\
\hline
E(\tilde{\psi}^i_{-1/2-\alpha_i}|0\rangle) & 2 \alpha_1 + \alpha_2 + \alpha_3 &  \alpha_1 + 2\alpha_2 + \alpha_3 & \alpha_1 + \alpha_2 + 2\alpha_3 & 0
\\
\hline
E(\psi^i_{-1/2+\alpha_i}|0\rangle) & \alpha_2 + \alpha_3 &  \alpha_1 + \alpha_3 &\alpha_2 + \alpha_3 & 2 (\alpha_1+\alpha_2+\alpha_3) \\
\hline
\end{array}
\end{align}
So, we have precisely one massless spacetime boson that is ``chiral" in the sense that 
it distinguishes $\tilde{\psi}$ from $\psi$.  

\subsubsection*{\underline{$\alpha_1 > \alpha_2 > 0 > \alpha_3 > \alpha_4$}}

Using the fact that $|\alpha_3| + |\alpha_4| = - \alpha_3 - \alpha_4 = \alpha_1 + \alpha_2$, we can rewrite the vacuum energy as 
\begin{align}
E_0 = - \frac{1}{2} + \alpha_1 + \alpha_2  \,.
\end{align}
The spectrum of first excited states is
\begin{align}
\begin{array}{|c|c|c|c|c|}
\hline
i & 1 & 2 & 3 & 4
\\
\hline
E(\tilde{\psi}^i_{-1/2-\alpha_i}|0\rangle) & 2 \alpha_1 + \alpha_2  &  \alpha_1 + 2\alpha_2 & |\alpha_4| & |\alpha_3| 
\\
\hline
E(\psi^i_{-1/2+\alpha_i}|0\rangle) & \alpha_2  &  \alpha_1 & \alpha_1 + \alpha_2 + |\alpha_3| & \alpha_1+\alpha_2 + |\alpha_4| \\
\hline
\end{array}
\end{align}
So, we have no massless spacetime boson. 

In the R-sector, the worldsheet vacuum energy is always zero. The vacuum states give spacetime fermions. A careful analysis of the GSO projection shows that the signs of the angles $\alpha_i$ are correlated with the chirality in the $2d$ gauge theory. The $(+,+,+,-)$ and $(-,-,-,+)$ cases (and their permutations) give a right-moving Fermion, while the $(+,+,-,-)$ case gives a left-moving Fermion.

We may summarize what we have learned about the open string spectrum as follows. Suppose the rotation from brane A to brane B is given by an $SU(4)$ matrix $U(\theta) = \mathrm{diag}(e^{i\theta_1}, e^{i\theta_2},e^{i\theta_3},e^{i\theta_4})$ with $|\theta_i| < \pi/2$ and $\theta_1+\theta_2+\theta_3+\theta_4 = 0$. Up to permutations, there are three distinct cases:

\begin{enumerate}

\item 
$(+,+,+,-)$ : chiral multiplet from A to B. 

\item 
$(-,-,-,+)$ : chiral multiplet from B to A. 

\item 
$(+,+,-,-)$ : Fermi multiplet between A and B. 

\end{enumerate}

Now, we would like to relax the restrictions $U(\theta) = \mathrm{diag}(e^{i\theta_1}, e^{i\theta_2},e^{i\theta_3},e^{i\theta_4})$ and $|\theta_i| < \pi/2$ and determine a covariant criteria for distinguishing the three cases. For the chiral versus Fermi distinction, the criterion is nothing but the orientation of the intersection. Let us assign to brane A a differential form 
\begin{align}
\alpha = dx^1 \wedge dx^2 \wedge dx^3 \wedge dx^4 \,. 
\end{align} 
For a diagonal $U$, the differential form for brane B is 
\begin{align}
\beta = (\cos\theta_1 dx^1 + \sin\theta_1 dy^2) \wedge \cdots \wedge (\cos\theta_4 dx^4 + \sin\theta_4 dy^4) \,. 
\end{align} 
The wedge product of the two gives 
\begin{align} 
\alpha \wedge \beta = \left(\prod_{i=1}^4 \sin\theta_i\right) \mbox{d(vol)} \,,
\label{intersection-chiral-vs-fermi}
\end{align}
where d(vol) is the oriented volume form of $\mathbb{C}^4$. So, when $s_1 = \mathrm{sgn}[\prod_i \sin\theta_i]$ is odd/even, the field type is chiral/Fermi, respectively. Covariantly, 
we may write 
\begin{align}
s_1 = \mathrm{sgn}[\det(U-U^*)] \,.
\label{ss1}
\end{align}
Note that $s_1$ is invariant under a transformation $U \rightarrow O_1 U O_2$, where 
$O_{1,2} \in SO(4)$. This means that, in going from brane A to brane B, 
the rotations within a 4-plane does not affect the result. Note also that $s_1$ is invariant under $U \leftrightarrow U^*$. 

Next we covariantize the distinction between the two orientations of chiral fields. Within the restrictions 
$U(\theta) = \mathrm{diag}(e^{i\theta_1}, e^{i\theta_2},e^{i\theta_3},e^{i\theta_4})$ and $|\theta_i| < \pi/2$, a good criterion is 
\begin{align}
s_2 = \mathrm{sgn}[\sin(\theta_1 + \theta_2) \sin(\theta_2 + \theta_3)\sin(\theta_3 + \theta_1)]\,.
\label{ss2}
\end{align}
To make it covariant, we start by noting that the three angles are the Cartan angles in the $SO(6)$ notation. We can covariantly switch from an $SU(4)$ basis to an $SO(6)$ basis by defining
\begin{align}
V_{ij,kl} = \epsilon_{ijpq} U^p{}_k U^q{}_l \,.
\end{align}
It follows from $U\in SU(4)$ that $V$ is Hermitian as a $6\times 6$ matrix. In terms of $V$, we can write the unrestricted and covariant form of $s_2$ as 
\begin{align}
s_2 = \mathrm{sgn}\left[ \mathrm{Pfaff} (\mathrm{Im}(V))\right] \,.
\end{align}
It is straightforward to show that $s_2$ is invariant under $U \rightarrow O_1 U O_2$ and antisymmetric under $U \leftrightarrow U^*$. 

Throughout this appendix, we assumed that $U$ is a generic element of $SU(4)$. If $U$ takes value in a proper subset of $SU(4)$ such as $SU(3)$ or $SU(2)\times SU(2)$, the spacetime supersymmetry is enhanced and the massless spectrum is enlarged accordingly. The covariant signs 
\eqref{ss1} and \eqref{ss2} may get flipped as $U$ passes through such walls of supersymmetry enhancement.  


\newpage
\bibliographystyle{JHEP}
\bibliography{mybib}

\end{document}